\documentclass[aps,prd,twocolumn,showpacs,nofootinbib,floatfix,superscriptaddress]{revtex4}  
\usepackage{graphicx}  
\usepackage{dcolumn}   
\usepackage{bm}        
\usepackage{amssymb}   
\usepackage{epsfig}    
\usepackage[abs]{overpic}

\newcommand{\met}{\mbox{$E\kern-0.57em\raise0.19ex\hbox{/}_{T}$}}

\begin{document}
\hspace{5.2in}\mbox{FERMILAB-PUB-11-547-E}

\title{\boldmath  
Measurement of the inclusive jet cross section in $p \bar{p}$ collisions at 
$\sqrt{s}=1.96$ TeV}
\date{October 17, 2011}
%
\affiliation{Universidad de Buenos Aires, Buenos Aires, Argentina}
\affiliation{LAFEX, Centro Brasileiro de Pesquisas F{\'\i}sicas, Rio de Janeiro, Brazil}
\affiliation{Universidade do Estado do Rio de Janeiro, Rio de Janeiro, Brazil}
\affiliation{Universidade Federal do ABC, Santo Andr\'e, Brazil}
\affiliation{Instituto de F\'{\i}sica Te\'orica, Universidade Estadual Paulista, S\~ao Paulo, Brazil}
\affiliation{Simon Fraser University, Vancouver, British Columbia, and York University, Toronto, Ontario, Canada}
\affiliation{University of Science and Technology of China, Hefei, People's Republic of China}
\affiliation{Universidad de los Andes, Bogot\'{a}, Colombia}
\affiliation{Charles University, Faculty of Mathematics and Physics, Center for Particle Physics, Prague, Czech Republic}
\affiliation{Czech Technical University in Prague, Prague, Czech Republic}
\affiliation{Center for Particle Physics, Institute of Physics, Academy of Sciences of the Czech Republic, Prague, Czech Republic}
\affiliation{Universidad San Francisco de Quito, Quito, Ecuador}
\affiliation{LPC, Universit\'e Blaise Pascal, CNRS/IN2P3, Clermont, France}
\affiliation{LPSC, Universit\'e Joseph Fourier Grenoble 1, CNRS/IN2P3, Institut National Polytechnique de Grenoble, Grenoble, France}
\affiliation{CPPM, Aix-Marseille Universit\'e, CNRS/IN2P3, Marseille, France}
\affiliation{LAL, Universit\'e Paris-Sud, CNRS/IN2P3, Orsay, France}
\affiliation{LPNHE, Universit\'es Paris VI and VII, CNRS/IN2P3, Paris, France}
\affiliation{CEA, Irfu, SPP, Saclay, France}
\affiliation{IPHC, Universit\'e de Strasbourg, CNRS/IN2P3, Strasbourg, France}
\affiliation{IPNL, Universit\'e Lyon 1, CNRS/IN2P3, Villeurbanne, France and Universit\'e de Lyon, Lyon, France}
\affiliation{III. Physikalisches Institut A, RWTH Aachen University, Aachen, Germany}
\affiliation{Physikalisches Institut, Universit{\"a}t Freiburg, Freiburg, Germany}
\affiliation{II. Physikalisches Institut, Georg-August-Universit{\"a}t G\"ottingen, G\"ottingen, Germany}
\affiliation{Institut f{\"u}r Physik, Universit{\"a}t Mainz, Mainz, Germany}
\affiliation{Ludwig-Maximilians-Universit{\"a}t M{\"u}nchen, M{\"u}nchen, Germany}
\affiliation{Fachbereich Physik, Bergische Universit{\"a}t Wuppertal, Wuppertal, Germany}
\affiliation{Panjab University, Chandigarh, India}
\affiliation{Delhi University, Delhi, India}
\affiliation{Tata Institute of Fundamental Research, Mumbai, India}
\affiliation{University College Dublin, Dublin, Ireland}
\affiliation{Korea Detector Laboratory, Korea University, Seoul, Korea}
\affiliation{CINVESTAV, Mexico City, Mexico}
\affiliation{Nikhef, Science Park, Amsterdam, the Netherlands}
\affiliation{Radboud University Nijmegen, Nijmegen, the Netherlands and Nikhef, Science Park, Amsterdam, the Netherlands}
\affiliation{Joint Institute for Nuclear Research, Dubna, Russia}
\affiliation{Institute for Theoretical and Experimental Physics, Moscow, Russia}
\affiliation{Moscow State University, Moscow, Russia}
\affiliation{Institute for High Energy Physics, Protvino, Russia}
\affiliation{Petersburg Nuclear Physics Institute, St. Petersburg, Russia}
\affiliation{Instituci\'{o} Catalana de Recerca i Estudis Avan\c{c}ats (ICREA) and Institut de F\'{i}sica d'Altes Energies (IFAE), Barcelona, Spain}
\affiliation{Stockholm University, Stockholm and Uppsala University, Uppsala, Sweden}
\affiliation{Lancaster University, Lancaster LA1 4YB, United Kingdom}
\affiliation{Imperial College London, London SW7 2AZ, United Kingdom}
\affiliation{The University of Manchester, Manchester M13 9PL, United Kingdom}
\affiliation{University of Arizona, Tucson, Arizona 85721, USA}
\affiliation{University of California Riverside, Riverside, California 92521, USA}
\affiliation{Florida State University, Tallahassee, Florida 32306, USA}
\affiliation{Fermi National Accelerator Laboratory, Batavia, Illinois 60510, USA}
\affiliation{University of Illinois at Chicago, Chicago, Illinois 60607, USA}
\affiliation{Northern Illinois University, DeKalb, Illinois 60115, USA}
\affiliation{Northwestern University, Evanston, Illinois 60208, USA}
\affiliation{Indiana University, Bloomington, Indiana 47405, USA}
\affiliation{Purdue University Calumet, Hammond, Indiana 46323, USA}
\affiliation{University of Notre Dame, Notre Dame, Indiana 46556, USA}
\affiliation{Iowa State University, Ames, Iowa 50011, USA}
\affiliation{University of Kansas, Lawrence, Kansas 66045, USA}
\affiliation{Kansas State University, Manhattan, Kansas 66506, USA}
\affiliation{Louisiana Tech University, Ruston, Louisiana 71272, USA}
\affiliation{Boston University, Boston, Massachusetts 02215, USA}
\affiliation{Northeastern University, Boston, Massachusetts 02115, USA}
\affiliation{University of Michigan, Ann Arbor, Michigan 48109, USA}
\affiliation{Michigan State University, East Lansing, Michigan 48824, USA}
\affiliation{University of Mississippi, University, Mississippi 38677, USA}
\affiliation{University of Nebraska, Lincoln, Nebraska 68588, USA}
\affiliation{Rutgers University, Piscataway, New Jersey 08855, USA}
\affiliation{Princeton University, Princeton, New Jersey 08544, USA}
\affiliation{State University of New York, Buffalo, New York 14260, USA}
\affiliation{Columbia University, New York, New York 10027, USA}
\affiliation{University of Rochester, Rochester, New York 14627, USA}
\affiliation{State University of New York, Stony Brook, New York 11794, USA}
\affiliation{Brookhaven National Laboratory, Upton, New York 11973, USA}
\affiliation{Langston University, Langston, Oklahoma 73050, USA}
\affiliation{University of Oklahoma, Norman, Oklahoma 73019, USA}
\affiliation{Oklahoma State University, Stillwater, Oklahoma 74078, USA}
\affiliation{Brown University, Providence, Rhode Island 02912, USA}
\affiliation{University of Texas, Arlington, Texas 76019, USA}
\affiliation{Southern Methodist University, Dallas, Texas 75275, USA}
\affiliation{Rice University, Houston, Texas 77005, USA}
\affiliation{University of Virginia, Charlottesville, Virginia 22901, USA}
\affiliation{University of Washington, Seattle, Washington 98195, USA}
\author{V.M.~Abazov} \affiliation{Joint Institute for Nuclear Research, Dubna, Russia}
\author{B.~Abbott} \affiliation{University of Oklahoma, Norman, Oklahoma 73019, USA}
\author{B.S.~Acharya} \affiliation{Tata Institute of Fundamental Research, Mumbai, India}
\author{M.~Adams} \affiliation{University of Illinois at Chicago, Chicago, Illinois 60607, USA}
\author{T.~Adams} \affiliation{Florida State University, Tallahassee, Florida 32306, USA}
\author{G.D.~Alexeev} \affiliation{Joint Institute for Nuclear Research, Dubna, Russia}
\author{G.~Alkhazov} \affiliation{Petersburg Nuclear Physics Institute, St. Petersburg, Russia}
\author{A.~Alton$^{a}$} \affiliation{University of Michigan, Ann Arbor, Michigan 48109, USA}
\author{G.~Alverson} \affiliation{Northeastern University, Boston, Massachusetts 02115, USA}
\author{G.A.~Alves} \affiliation{LAFEX, Centro Brasileiro de Pesquisas F{\'\i}sicas, Rio de Janeiro, Brazil}
% ...
%
%
\collaboration{The D0 Collaboration\footnote{with visitors from
$^{a}$Augustana College, Sioux Falls, SD, USA,
$^{b}$The University of Liverpool, Liverpool, UK,
$^{c}$UPIITA-IPN, Mexico City, Mexico,
$^{d}$SLAC, Menlo Park, CA, USA,
$^{e}$University College London, London, UK,
$^{f}$Centro de Investigacion en Computacion - IPN, Mexico City, Mexico,
$^{g}$ECFM, Universidad Autonoma de Sinaloa, Culiac\'an, Mexico,
$^{h}$Helsinki Institute of Physics, Helsinki, Finland,
and 
$^{i}$Universit{\"a}t Bern, Bern, Switzerland.
$^{\ddag}$Deceased.
}} \noaffiliation
\vskip 0.25cm

\begin{abstract}
We present a measurement of the inclusive jet cross section using the 
Run II cone algorithm and data collected by the D0 experiment
in $p \bar{p}$ collisions at a center-of-mass energy $\sqrt s=$1.96~TeV,
corresponding to an integrated luminosity of 0.70~fb$^{-1}$.
The jet energy calibration and
the method used to extract the inclusive jet cross section are described. 
We discuss the main uncertainties, which are dominated by the jet 
energy scale uncertainty. The results cover jet transverse momenta 
from 50~GeV to 600~GeV with jet rapidities in the range $-2.4$ to 
2.4 and are compared to predictions using recent proton parton distribution 
functions. Studies of correlations between systematic uncertainties in 
transverse momentum and rapidity are presented.
\end{abstract}

\maketitle

\section{Introduction and motivation}
\begin{figure}
\includegraphics[width=\columnwidth]{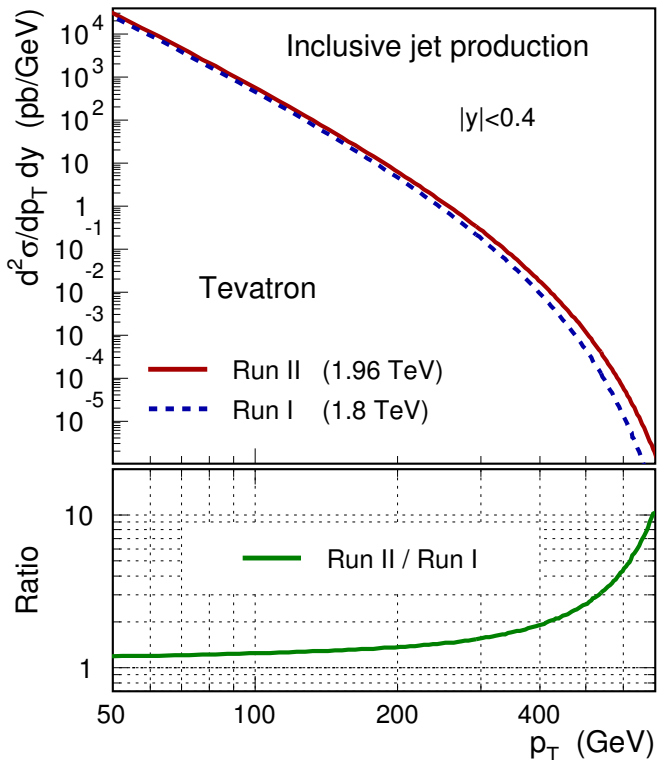}
\caption{\label{run1run2} (color online)
Inclusive jet production cross section for central jets
($|y|<0.4$) for Run I and Run II energies at the 
Tevatron obtained using NLO QCD as implemented in
{\sc{nlojet++}}. The ratio of the
two curves is shown in the bottom panel. 
We note an increase of the Run II cross section with respect 
to Run I of up to a factor 10 at highest jet $p_T$.}
\end{figure}

\begin{figure}
\includegraphics[width=\columnwidth]{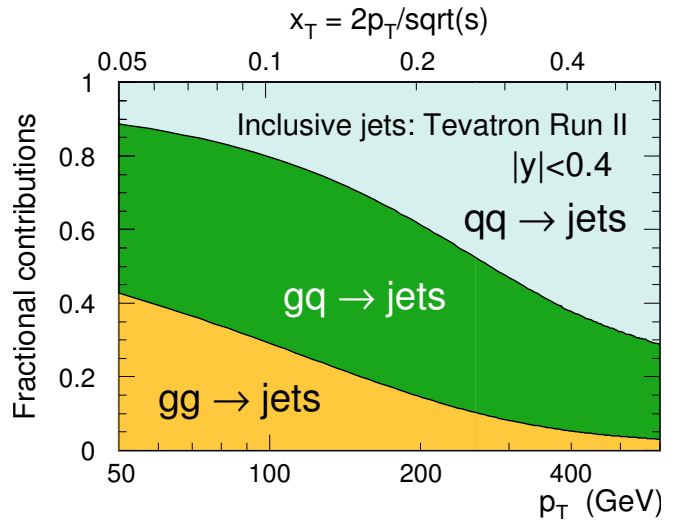}
\caption{\label{fig1b} (color online) Fractional contributions of the
$qq$, $qg$ and $gg$ sub-processes to the inclusive jet
cross section for central jets as a function of
jet $p_T$ and of the fraction of the beam energy carried
by the jet, $x_T$.}
\end{figure}

The measurement of the cross section for inclusive production of hadronic 
jets in hadron collisions provides stringent tests of quantum chromodynamics 
(QCD).  The inclusive jet cross section in $p\bar{p}$ collisions 
for jets with large momentum transverse to the beam axis ($p_T$) is 
directly sensitive to the strong coupling constant ($\alpha_s$)~\cite{alphas} 
and the parton distribution functions (PDFs) of the proton~\cite{pdfreview}. 
At the Tevatron $p\bar{p}$ collider, data are divided into two sets 
corresponding to Run~I (1992---1996) and Run~II (2002---2011). 
The increased $p \bar{p}$ 
center-of-mass energy between Run~I ($\sqrt{s}=1.8$~TeV) and Run~II 
($\sqrt{s}=1.96$~TeV) leads to a
significant increase in the cross section at large $p_T$ -- a factor of three 
at $p_T \approx 550$~GeV, as shown in Fig.~\ref{run1run2} obtained using the 
next-to-leading order (NLO) QCD calculation as implemented in
{\sc{nlojet++}}~\cite{nlojet}. This increases the
sensitivity to potential new observations such as quark compositeness
and extra dimensions~\cite{composite}. 
The integrated luminosity of the inclusive jet cross section measurement 
discussed in this Article exceeds the Run I luminosity by more than a factor of 
five, allowing for more stringent constraints on the PDFs. 
In Fig.~\ref{fig1b} we show 
the different subprocesses that contribute to the inclusive jet cross section.
In particular the gluon density can be further constrained using 
these data, since the $gg$ and $qg$ initial states contribute significantly 
to the cross section across almost the full $p_T$ range of the measurement. 
The gluon distribution is still poorly known, especially for gluons
carrying a large momentum fraction $x$.  In contrast, the quark PDFs 
are already well constrained by fixed target and electron-proton 
collider experiments~\cite{pdfreview}. 

In this Article, we report measurements by the D0 collaboration of the 
inclusive jet cross section in $p\bar{p}$ collisions at a center-of-mass 
energy of $\sqrt{s}=1.96$~TeV. We give details of the analysis leading to 
the results published in Ref.~\cite{prl,mikko}, with particular attention 
to the jet energy scale determination.
The precision achieved for the jet energy scale in 
the D0 experiment is unprecedented for any hadron collider experiment
to date, and the methods applied to reach this precision will be 
useful for future hadron collider experiments.  

The data sample, collected with the D0 detector during 2004--2005 in 
Run~II of the Fermilab Tevatron, corresponds to an integrated 
luminosity of ${\cal L}=0.70$~fb$^{-1}$~\cite{lumi}. The cross section is 
presented in six bins of jet rapidity ($y$), extending to $|y|=2.4$, as a 
function of jet $p_T$ starting at $p_T=50$~GeV. The rapidity is related to 
the polar scattering angle $\theta$ with respect to the beam axis 
by $y=0.5 \ln [(1+ \beta \cos \theta)/(1-\beta \cos \theta)]$ 
with $\beta = |\vec p|/E$.
The measurement also extends the kinematic reach of earlier 
measurements of the inclusive jet cross section by the CDF and D0 
Collaborations~\cite{Aaltonen:2008eq,cdf,d0_runI,cdf_run1}. 

This Article is organized as follows. After a brief description of the D0 
detector in Sec.~\ref{sec:detector}, we discuss the jet 
algorithm used in Run II in Sec.~\ref{sec:algos}.  Section~\ref{sec:theory} 
describes the theoretical predictions for the inclusive jet cross section 
before the D0 measurement in Ref.~\cite{prl}. 
Section~\ref{sec:jes} gives an extensive description of the methods used to 
measure the jet energy scale and to determine the corresponding uncertainty.
This is the leading uncertainty for the measurement of the 
inclusive jet cross section.
Sections~\ref{sec:trigger}--\ref{sec:unfolding} describe the 
jet triggers, event and jet selection criteria, determination 
of the jet $p_T$ resolution and the unfolding method. In
Sections~\ref{sec:results}--\ref{sec:conclusions}, we describe our results 
and compare them with predictions using recent PDF parameterizations.

\section{Detector}\label{sec:detector}

In this section, we briefly describe the Run II D0 detector~\cite{d0det} 
and the main components used in the measurement of the inclusive jet 
cross section. 

\subsection{Calorimeter}

\begin{figure*}[t!]
\includegraphics[width=0.7\textwidth]{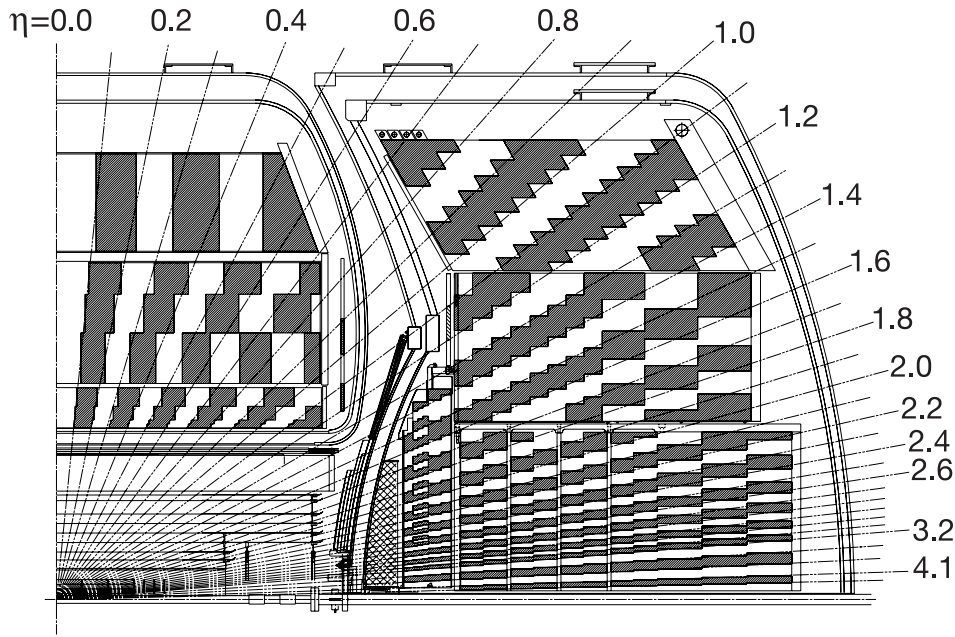}
\caption{\label{calo} 
Side view of a quadrant of the D0 calorimeters (CC, EC and ICR) showing the 
transverse and longitudinal segmentation~\cite{d0det}. The shading pattern 
indicates the cells for signal readout. The lines indicate the pseudorapidity 
intervals defined from the center of the detector. The CC covers the region
$|\eta| < 1.2$ and the EC extends the coverage up to $|\eta| \sim$4.2. 
The inter-cryostat detector is visible as a thin dark shaded tile 
between the cryostats, within $1.1<|\eta|<1.4$, and 
the massless gap detectors are inside the cryostats,
within $0.8<|\eta|<1.2$ (in the CC) and $1.0<|\eta|<1.3$ (in the EC).}
\end{figure*}

The calorimeter and the tracking detectors, used
to measure the position of the interaction point, are the most important 
detector components used to measure 
the jet $p_T$. An accurate and stable energy response 
is required for reliable measurements of the cross section
for jet production. The calorimeter consists of the following subdetectors:
the uranium/liquid argon calorimeter divided into a central (CC) 
and two end (EC) sections, the plastic scintillator inter-cryostat
detector (ICD), and the massless gap (MG) detectors. Both the CC and ECs
are segmented longitudinally into electromagnetic (EM), fine hadronic 
(HAD), and coarse hadronic (CH) sections. A schematic view of the 
calorimeter showing its projective tower geometry as a function of 
pseudorapidity $\eta = - \ln \tan (\theta/2)$, where $\theta$ is the 
polar angle from the beamline, is given in 
Fig.~\ref{calo}. The choice of binning in the inclusive jet cross section 
measurement closely follows the structure of the calorimeter: $|\eta|<0.8$ 
is well-contained within the CC, $1.6<|\eta|<2.4$
within the EC, whereas the more challenging inter-cryostat region (ICR) 
$0.8<|\eta|<1.6$ has energy sharing between the four sub-detectors.

\subsubsection{Central and end calorimeters}

The CC covers detector pseudorapidity $|\eta| < 1.2$, and the two ECs 
extend the range up 
to $|\eta| = 4.2$. Both the electromagnetic and fine hadronic calorimeters 
are sampling calorimeters with an active medium of liquid argon and absorber 
plates of nearly-pure depleted uranium. Incoming particles traversing the 
uranium absorber plates initiate showers of secondary particles that ionize 
the argon in the gaps between the absorber plates. A high-voltage electric 
field collects the free electrons on resistively-coated copper pads that 
act as signal boards~\cite{d0det,run1det}. The outer part of the calorimeter, 
the coarse hadronic section, uses copper in the CC and 
stainless steel in the EC for the absorber plates. The calorimeter is 
transversely segmented into cells in pseudorapidity
and azimuthal angle of $0.1 \times 0.1$ (and $0.05 \times 0.05$ in the third 
layer of the EM calorimeter) for $|\eta|<3.2$ to allow for more precise 
location of EM shower centroids.  At $|\eta|>3.2$, the cell size grows to 
0.2 or more for both $\eta$ and the azimuthal angle $\phi$. 
These high pseudorapidities are not 
used for the jet cross section measurements since the jet triggers are 
limited to $|\eta|<3.2$.  The total depth of the EM calorimeter is about 
20 electromagnetic radiation lengths, and the combined thickness of the 
electromagnetic and hadronic calorimeters is about 7 nuclear interaction 
lengths~\cite{run1det}.

A typical calorimeter cell consists of an absorber plate and a liquid argon
gap. The metal plate is grounded, while the resistive plate of the signal board
located in the liquid argon gap is kept at a high voltage of $\sim$2.0 kV. 
The drift time of the electrons across
the typical 2.3 mm gap is 450 ns, longer than the
separation between two subsequent Tevatron bunch
crossings of 396 ns. To minimize the effect of pile-up
from interactions from different bunch crossings,
only two-thirds of the charge collected is used
in the shaper circuits and then provided to baseline subtraction boards. 
To remove the baseline, the signal corresponding to a sampling occurring 
396 ns earlier (the time between two bunch crossings) is subtracted.
Only cells with a signal at least 2.5 times the standard deviation of the
electronic noise after baseline subtraction are kept in nominal
conditions of data taking. This defines the on-line
zero-suppression mode of the calorimeter.

\subsubsection{Inter-cryostat detector and massless gaps}
 
The regions between the CC and the ECs are instrumented with the inter cryostat 
detector and massless gaps. The ICD and MG detectors provide energy 
measurement for the otherwise poorly instrumented regions located 
at roughly $0.8<|\eta|<1.4$, where the depth of the passive material coming from 
cryostat walls, stiffening rings and cables varies rapidly with rapidity. The 
ICD relies on photomultipliers to record the signals from plates of 
scintillating plastic and covers the region $1.1<|\eta|<1.4$. The 
signal from the ICD is stretched in time to match that of the EM calorimeter 
and augments the EM calorimetry that is absent in the region
$1.2<|\eta|<1.35$. The ICD is supplemented by the MG detectors that are placed 
inside the cryostat walls in the CC and the ECs from $0.8<|\eta|<1.2$ and 
$1.0<|\eta|<1.3$, respectively. Unlike typical calorimeter cells, the 
massless gap detectors do not have absorber plates, but they sample the 
showers that develop in the cryostat walls, calorimeter support structures, 
and other calorimeter cells.

In addition to the CC, ECs, and ICD, preshower detectors are located in the 
central and forward regions, but they are not used in this analysis.

\subsection{Tracking detectors}

The tracking detectors are not used directly in jet reconstruction
since the jet finding algorithms in D0 use only energy deposits in the
calorimeter towers. However, the tracking detectors are used to reconstruct 
the position of the primary vertex of the $p \bar{p}$ interaction, 
which is necessary to precisely measure the jet rapidity and transverse 
momentum. The position of the primary vertex is typically distributed as 
a 20 cm-wide Gaussian distribution along the beamline
direction around the nominal interaction point of $(x,y,z)=(0,0,0)$ located
in the center of the detector.  In the detector description and 
data analysis, we use a right-handed coordinate system in which the $z$-axis 
is along the proton direction and the $y$-axis is upward.
The inner tracking system, consisting of the silicon microstrip tracker, 
provides a 35~$\mu$m vertex resolution along the beam line and 15~$\mu$m 
resolution in the $r-\phi$ plane for tracks with a minimum $p_T$ of 10~GeV 
at $\eta=0$. The outer tracking system, consisting of the central fiber 
tracker, uses scintillating fiber technology to complement the silicon tracker. 
Both detectors are located in the 2~T magnetic field of the 
superconducting solenoidal magnet to allow measurements of the momentum of 
charged particles.

\subsection{Muon detector}

The muon detector is composed of a combination of proportional drift tubes in 
the central region ($|\eta|\lesssim 1.0$), and smaller, faster mini drift 
tubes in the forward region ($1.0 \le |\eta| \le 2.0$). Both are separated 
in three layers (A, B, C).  Toroidal magnets are located between 
the A and B layers of the muon detector in the central and forward regions
to allow reconstruction of the muon momentum.
The muon system is not used directly in our analysis (we do
not correct for muons in jets), but very high energy jets can leak 
outside the calorimeter and show some hits in the A layer. We do not include 
these hits in jet reconstruction, but instead correct the jet cross sections 
for asymmetries introduced in the jet energy resolution (described in
Sec~\ref{finalres}).

\subsection{Luminosity detector}

The luminosity monitor (LM) is constructed of scintillating tiles on both 
sides of the interaction point that detect the particles coming from 
inelastic collisions. The luminosity $\mathcal{L}$ is determined from 
the average number of observed interactions $\bar{N}_\mathrm{LM}$ using 
the formula
\begin{equation}
\mathcal{L} = \frac{f\bar{N}_\mathrm{LM}}{\sigma_\mathrm{LM}},
\end{equation}
where $f$ is the $p \bar{p}$ bunch crossing frequency, and 
$\sigma_\mathrm{LM}$ is the effective cross section for inelastic collisions 
measured by the LM that takes into account event losses due to inefficiencies 
and geometric acceptance~\cite{lumi}. In practice, $\bar{N}_\mathrm{LM}$ is 
calculated by inverting the expression for the Poisson probability of 
observing zero LM hits in either of the two arrays
\begin{equation}
\label{eq:P0}
P(0) = {\textrm e}^{-\sigma_\mathrm{LM}\mathcal{L}/f} \times 
\left(2{\textrm e}^{-\sigma_\mathrm{SS}\mathcal{L}/(2f)} - 
{\textrm e}^{-\sigma_\mathrm{SS}\mathcal{L}/f}\right).
\end{equation}
The right-most term of Eq.~\ref{eq:P0} accounts for the possibility of 
producing double-sided LM hits from a combination of single-sided (SS) 
LM hits, where $\sigma_\mathrm{SS}$ is the effective cross section for 
only one of the arrays to show hits. The uncertainty on the luminosity 
determination is estimated to be 6.1\%~\cite{lumi}. This uncertainty is 
dominated by the 5.4\% uncertainty coming from the determination of 
$\sigma_\mathrm{LM}$, roughly half of which is due to acceptance and 
efficiency of the LM detectors with the remainder due to the 
uncertainty in the total 
inelastic cross section at 1.96~TeV described in~\cite{lumi,sigma_inelastic}.

\section{Jet reconstruction}\label{sec:algos}

Jets are reconstructed using the Run II midpoint cone 
algorithm~\cite{run2cone}, which is an iterative cone algorithm that 
considers energy deposits as four-vectors to construct the jet four-momentum. 
The same algorithm is used with different inputs in data and Monte Carlo (MC).
It is used to build jets from energy deposits in the calorimeter in data
or in fully simulated MC events, out of stable particles in simulation,
and out of partons produced either in a parton shower simulation
or from a next-to-leading order theoretical calculation.

In data and in MC events processed through a
simulation of the response of the D0 detector, the first step is to define the 
seeds for jet reconstruction. Pseudoprojective
towers, as illustrated in Fig.~\ref{calo}, are built by adding the 4-momenta 
of the calorimeter cells.  The 4-momentum associated with the energy
deposit in each cell of the calorimeter is computed using
the direction defined by the reconstructed $p \bar{p}$ interaction
vertex and the center of the cell and assuming $E=|p|$.
All non-zero-suppressed cells are used in jet reconstruction.
Calorimeter towers are ordered in decreasing transverse momentum and 
are used as seeds to form preclusters using a simple cone algorithm of 
radius 0.3 in ($\eta$, $\phi$) plane, starting with the tower having
the highest $p_T$ and then descending the list until no towers remain above a 
minimum threshold of $p_T > 500$~MeV.  All towers added to a precluster 
are removed from the list, avoiding overlaps between preclusters.  
Preclusters with $p_T>1$ GeV are used as seeds for the jet clustering 
algorithm.  The goal of preclustering in data is to reduce the number of 
seeds and the computing time to reconstruct 
jets. As verified by MC studies~\cite{run2cone}, the low value of the $p_T$ 
threshold on the jet seeds ensures that there are no significant variations
in the jet observables for the $p_T$ range considered in this measurement 
($p_T>$ 50 GeV).

The seeds -- preclusters in data and in MC events processed through a 
simulation of the response of the D0 detector, or stable particles in MC, 
or partons from NLO calculation -- are used as center points for proto-jets. 
All calorimeter towers, particles or partons within 
$\Delta R=\sqrt{(\Delta y)^2 +(\Delta\phi)^2} \le R_{\text{cone}}$, 
where $R_{\text{cone}}=0.7$, are added to the proto-jet. The 
four-momentum of the proto-jet is the sum of the four-momenta of all included 
calorimeter towers, particles or partons. The direction of the 
resulting four-vector is used as the center point for a new cone. 
When the proto-jet four-momentum does not coincide with the cone axis, 
the procedure is repeated using the new axis as the center point
until a stable solution is found.  The maximum number of iterations is 50 
and the solution is considered to be stable if the difference in $\Delta R$ 
between two iterations is smaller than 0.001.  In the rare cases of bistable 
solutions the last iteration is retained.  Any protojets falling below a 
threshold,
$p_{T,\textrm{jet}}<p_{T,\textrm{min}}$, with $p_{T,\textrm{min}}=3$~GeV, are 
discarded.

The presence of a threshold requirement on the
cluster seeds introduces a dependency on
infrared and collinear radiation. In order to reduce the sensitivity to soft
radiation, $p_T$-weighted mid-points between pairs of 
proto-jets are used as additional seeds 
if the distance between pairs, $\Delta R$ in the ($y$, $\phi$) plane
to the proto-jet, is between 0.7 and 1.4. The list of stable 
proto-jets obtained from this procedure may contain many overlapping and 
identical jet candidates. To resolve these ambiguities the proto-jets are 
sorted in order of decreasing $p_T$ and processed through a split-and-merge 
procedure to remove overlaps.  If two proto-jets have overlapping cones, 
they are merged if the overlap region contains more than 50\% of the 
transverse momentum of the lower 
$p_T$ jet. Otherwise, the jets are split with calorimeter cells or particles 
in the overlap region being assigned to the nearest jet in ($y$, $\phi$). 
In both cases, the jet four-momenta are recomputed after this reassignment.  
In case of multiple overlaps, the algorithm always starts with the highest 
$p_T$ proto-jet to redistribute the shared towers.  As mentioned above, 
the jet four-momentum is computed as the sum of the four-momenta 
of the (massless) calorimeter energy deposits included in the jet, 
and consequently the calorimeter jets are massive by construction if the 
jet cone contains cells with different locations in the ($\eta$, $\phi$) plane.
The variables used to characterize the jets are 
the jet $p_T$ and $y$. The split-and-merge procedure may modify the cone axis 
and jet four-momentum for the final jets, and include towers outside the 
initial 0.7 cone.

\section{Theoretical predictions}\label{sec:theory}

\begin{figure*}
\includegraphics[width=0.8\textwidth]{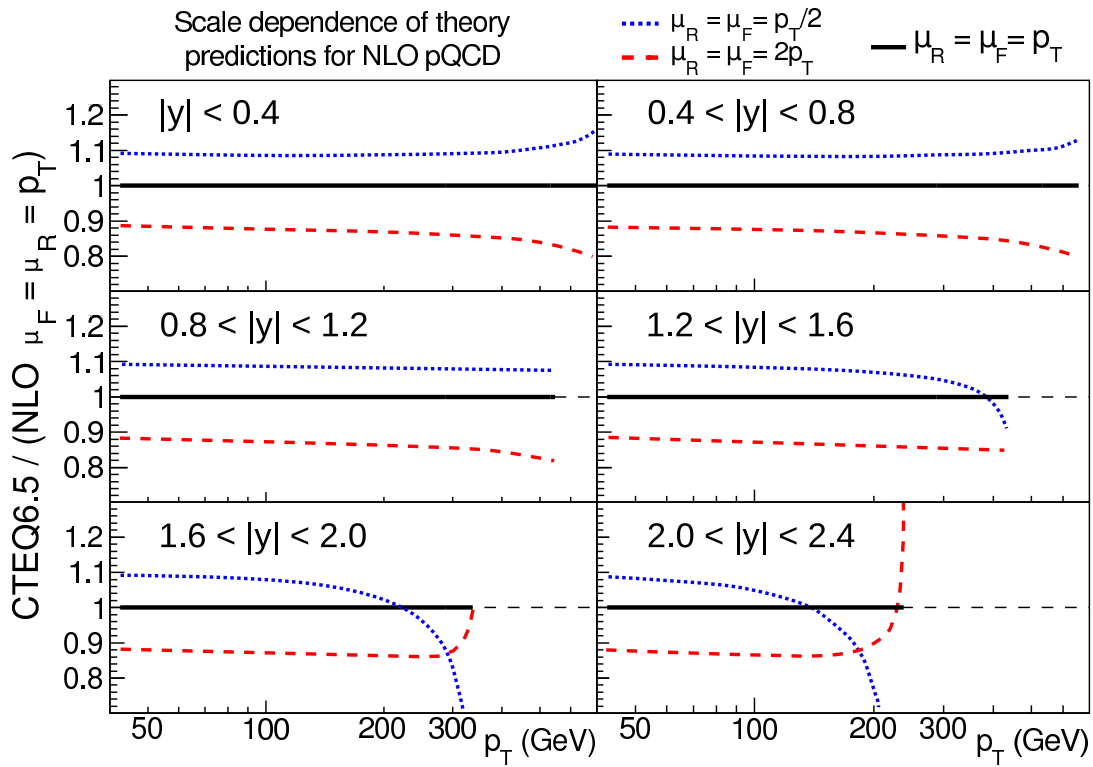}
\caption{\label{renorm} (color online)
Uncertainty on the inclusive jet cross section due to the choice of the 
renormalization and factorization scales $\mu_R$ and $\mu_F$ in the
NLO QCD calculation using {\sc{nlojet++}}.}
\end{figure*}

In this section, we describe how we compute the predictions of the 
inclusive jet cross sections that are later compared to our measurements.

\subsubsection{Jet cross section at NLO}

We use the program {\sc{FastNLO}}~\cite{fastnlo}, which is based on the
matrix elements implemented in {\sc{nlojet++}}, to calculate the inclusive 
cross sections to next-to-leading order precision and to evaluate the 
effects of the choice of proton PDFs, such as CTEQ6 or 
MRST2004~\cite{cteq6,mrst2004}, in a computationally efficient manner.
Perturbative QCD (PQCD) requires the specification of the renormalization 
scale $\mu_R$ and the factorization scale $\mu_F$.   Typical choices 
set both $\mu_R=\mu_F$ to the $p_T$ of each of the individual jets, 
with half and twice this scale used to estimate the theoretical scale 
uncertainty. The uncertainty on the NLO prediction of the inclusive jet cross 
section due to the choice of renormalization and factorization scales is 
given in Fig.~\ref{renorm} and is about 10--20\%.

\begin{figure}[b!]
\includegraphics[width=0.95\columnwidth]{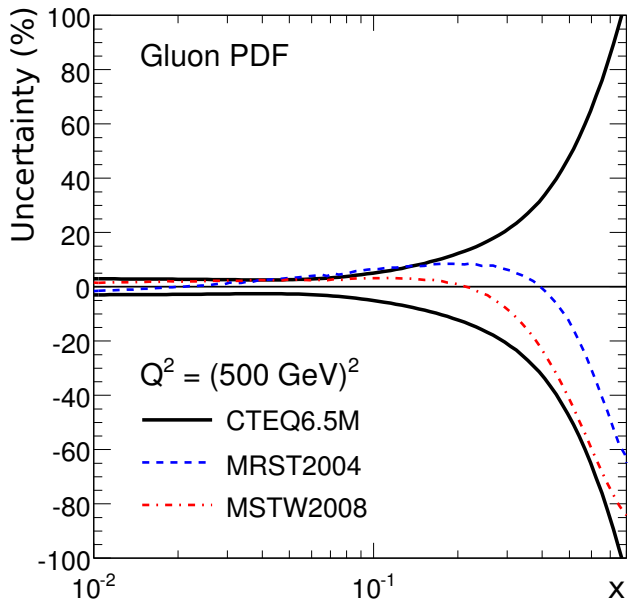}
\caption{\label{pdf} (color online)
Uncertainty of the CTEQ6.5M gluon PDF (solid lines) in percent compared to 
the differences between CTEQ6.5M  and MRST2004 (dashed line), 
MSTW2008 (dash-dotted line) central values.}
\end{figure}

\subsubsection{Parton distribution functions}

A discussion of the different PDFs and methods to reduce their 
uncertainties using various measurements at the Tevatron and the LHC can be 
found in reference~\cite{pdfreview}. In this paper, we briefly describe 
the PDFs used in the comparison between the measurements and the theoretical 
predictions.

One of the PDF sets used in this analysis is provided by the CTEQ 
Collaboration. This most recent global fit from the CTEQ Collaboration 
performed prior to the measurement described in this paper, called 
CTEQ6.5M~\cite{cteq6}, utilizes D0 and CDF Run I measurements, as well 
as the most recent deep inelastic scattering (DIS) data from the HERA 
collider at DESY and existing fixed target 
DIS and Drell-Yan data. The central prediction of the CTEQ6.5M PDF 
is supplemented with the provision of 20 eigenvector basis PDF sets
to estimate the PDF uncertainty, 
representing independent variations of the PDFs within the 
90\% C.L. of the data sets used in the fit.

Another widely used PDF parameterization is provided by the MRST 
Collaboration~\cite{mrst2004}. Our measurements are compared to the 
MRST2004 parameterization, which does not include our results. A third
PDF parameterization is MSTW2008~\cite{mstw2008} which uses our results.
The differences with respect to CTEQ6.5M are mainly in the 
description of the gluons at high-$x$ and are
within the CTEQ6.5M uncertainty band, as shown in Fig.~\ref{pdf}. 
We also note that the uncertainty on the gluon density calculated by the
CTEQ6.5M parameterization is larger than 40\% for $x \ge 0.5$ and 
squared four-momentum transfer $Q^2=500^2$~GeV.  
Comparisons between our data and NLO calculations
using these and other PDF parameterizations are given in Sec.~\ref{sec:results}.

\section{Jet energy scale measurement}\label{sec:jes}

In this section we describe the method used to obtain the jet energy scale 
(JES) applied in the measurement of the inclusive jet cross section as a 
function of jet $p_T$.  To compare the theoretical predictions to data, 
both need to be corrected to a common reference-level, chosen here to be the 
``particle-level jets."
We correct the calorimeter jet energies to the particle level, and apply 
non-perturbative corrections (hadronization and underlying event) to 
theoretical NLO cross sections to move from the parton to the particle level. 
Particle jets~\cite{SMWG} are clustered from stable particles after 
fragmentation, including particles from the true underlying event, but 
excluding undetected 
energy from muons and neutrinos.
The JES procedure provides a correction  factor that translates on average 
the energy of jets measured in the calorimeter to the energy of 
the corresponding particle jets.  The jet energy scale is determined from data 
acquired during the same running period as used in the measurement of the
inclusive jet cross section.

The main effects that need to be considered when correcting jet energies
from the calorimeter measurement $E_\mathrm{meas}$ to 
the particle level $E_\mathrm{particle}$ are the offset energy ($O$), 
calorimeter response ($R$), and 
detector showering ($S$). These corrections can be expressed as a simple 
formula
\begin{equation}
\label{eq:simple_jes}
E_\mathrm{particle} = \frac{E_\mathrm{meas} - O}{R \cdot S}.
\end{equation}
The offset energy $O$ originates from electronics noise, calorimeter noise from 
uranium decays, residual energy from previous bunch crossings (``pile-up"), 
and energy from multiple $p \bar{p}$ collisions during a bunch crossing. 
The underlying event energy corresponding to multiple parton interactions
in a single $p \bar{p}$ collision is not considered as part of the offset 
energy since it is included in the jet energy at the particle level. This 
also avoids correcting the data with model dependent offset corrections. 
The calorimeter response $R$ is the average fraction of the energy measured 
in the calorimeter for the particles inside the jet cone. The detector 
showering is the net flow of energy 
in and out of the jet cone due to detector effects, such as the magnetic field, 
scattering from passive material, and shower development in the calorimeter. The 
correction $S$ is defined as the ratio of the response-corrected calorimeter 
jet energy, in the absence of offset, and the particle jet energy. 
The correction does not include the effects of real QCD emissions, 
which arise from partons that shower outside the jet cone. We discuss 
each correction in turn below.

\subsection{Determination of the offset energy\label{sec:offset}}

The offset energy consists of the energy in the jet that is not related to the 
primary $p\bar{p}$ collision (hard scatter and underlying event). The 
offset energy is divided into two distinct categories, noise and pile-up (NP), 
and multiple $p \bar{p}$ interactions (MI).
The noise component corresponds to the contributions of calorimeter and 
electronics noise, as well as the decay of the uranium nuclei in the 
calorimeter. The pile-up energy corresponds to the energy left in the 
calorimeter from previous or next collisions because of the long 
integration time of 
the calorimeter electronics. The typical value of the NP offset
in a cone, $R=0.7$, is 0.2 GeV in the CC and ECs and 0.5 GeV in 
the ICR for the instantaneous luminosities considered in this analysis. 

The MI offset is the energy deposited by additional collisions during 
the bunch crossing. The value of the MI offset increases 
linearly with the number of additional interactions, which is characterized
by the number of reconstructed $p\bar p$ interaction vertices in a given event. 
A typical value of MI is of the 
order of 0.5 GeV in the CC per additional interaction.

The offset energies are measured directly from data using ``zero bias" and 
``minimum bias" data collected at a constant rate of about 0.5 Hz during data 
taking. The only requirement for zero bias events is coincident timing with 
the beam crossing; minimum bias events additionally require energy depositions 
above thresholds in coincidence in the two luminosity monitors, 
indicating that an inelastic collision took place. The 
offset is estimated from the average energy density in all calorimeter towers 
within detector rings of fixed pseudorapidity. 
The offset energy for a given jet cone is then calculated by 
summing the average offset in towers within the cone radius around the jet 
center. The NP offset energy is measured using zero bias data with a veto on 
the luminosity monitor (no interaction occurred), and the MI energy for a given 
number $N$ of interactions is the difference in the energy in minimum bias 
events with $(N+1)$ vertices and with a single vertex. 

\begin{figure}
\includegraphics[width=\columnwidth]{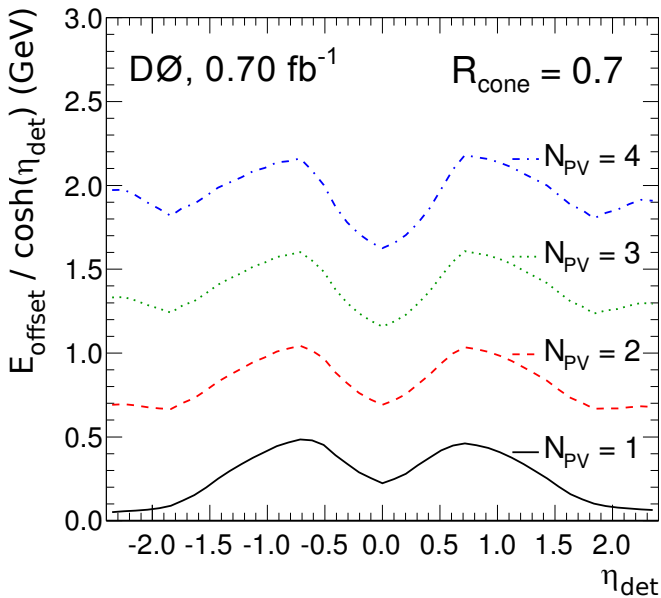}
\caption{\label{offset} (color online)
Offset corrections as a function of the jet pseudorapidity in the detector 
(without taking into account the vertex position) for different numbers of 
reconstructed primary vertices $N_\mathrm{PV}$. The 
special case $N_\mathrm{PV}=1$ includes only the noise contribution 
to the offset.}
\end{figure}

The offset energy for different numbers of $p \bar{p}$ interactions (measured 
by the number of reconstructed vertices) is displayed in 
Fig.~\ref{offset} and is found to depend linearly on the number of
interactions within a 5\% uncertainty.  The average vertex multiplicity 
in the sample used to measure 
the inclusive jet $p_T$ cross section is $\sim$1.5 -- 2.0, hence the 
average offset correction to jet $p_T$ is $\sim$0.5 GeV in the CC 
and EC and $\sim$0.7 GeV in the ICR. 
The uncertainties on the offset corrections are of the order of 1\% of 
the overall energy correction at low jet $p_T$ and
are negligible
for jet $p_T$ above $\sim$100 GeV. They are significantly smaller than the
total jet energy scale uncertainties. 

\subsection{Determination of the jet energy response}

The jet energy response, $R$, can be factorized into two parts 
$R = R_\mathrm{cc}(E)\cdot F_\eta(\eta,E)$. The $R_\mathrm{cc}$ term
uses the $p_T$ balance between the $\gamma$ and the jet in 
$\gamma +$jet events with a high (photon) purity in 
the CC region to determine an absolute response correction,
while the second term $F_\eta$ normalizes the response of the calorimeter 
as a function of jet pseudorapidity.

\subsubsection{Jet response in the CC}

The missing transverse energy (\met ) projection fraction (MPF) 
method~\cite{mpf} is applied in $\gamma + {\textrm {jet}}$ events to 
measure the response for jet energies in the CC region.  Use of the MPF 
reduces the sensitivity of the measurement to showering and additional
unreconstructed jets.  We project the vector sum of all calorimeter 
tower energies transverse to the beam (including those of the photon), 
which equals the opposite of the $\met$ in the event, onto the photon 
transverse momentum vector $\vec{p}_{T,\gamma}$. 
At the particle level, the photon is balanced against the hadronic recoil, 
$\vec{p}_{T,\gamma} + \vec{p}_{T,\mathrm{had}}=0$, 
where $\vec{p}_{T,\gamma}$
and $\vec{p}_{T,\mathrm{had}}$ are the transverse momentum
of the photon and the hadronic recoil system, respectively. 
The measured jet $p_T$ will be affected by the energy response of 
the calorimeter causing an imbalance in the jet and 
photon transverse momenta, resulting in a non-zero $\met$, 
\begin{equation}
R_{\mathrm{em}}\cdot\vec{p}_{T,\gamma} + 
R_{\mathrm{had}}\cdot\vec{p}_{T,\mathrm{had}}= -
\vec{\met},
\label{good}
\end{equation}
where $R_{\text{em}}$ and $R_{\text{had}}$ are the electromagnetic and hadronic
calorimeter responses, respectively. 

The MPF method necessitates a precise energy 
calibration for electrons and photons. The electron energy scale is determined 
from data using $Z\rightarrow e^+e^-$ decays~\cite{wmass}. MC 
simulations tuned to reproduce the response for electrons
in data are used to derive the response difference between photons and 
electrons. The leading uncertainty in this simulation is caused by limited
knowledge of the number of radiation lengths of material in front 
of the calorimeter.

Using the corrected photon energy scale ($R_{\text{em}}=1$),
$R_{\text{had}}$ is determined after projecting all terms in
Eq.~\ref{good} on the photon $p_T$ unit vector $\hat{n}_{\gamma}$.
In the MPF method, the jet response $R_{\text{had}}$ is 
thus directly defined through the $\met$
\begin{equation}
R_\mathrm{had} = 1 + \frac{\vec{\met}\cdot \hat{n}_{\gamma}}{|\vec{p}_{T,\gamma}|},
\end{equation}
where we use $|\vec{p}_{T,\gamma}| = - \hat{n}_\gamma \cdot \vec{p}_{T,\mathrm{had}}$.
When the jet is required to 
be back-to-back with the photon (difference in azimuthal angle
larger than 2.9 radians) and no additional jets are allowed in 
events with a single $p \bar{p}$ interaction, the hadronic recoil 
response $R_\mathrm{had}$ can be identified with the jet response 
$R_{\text{jet}}$.  The impact of the proton remnants is small on average.
The jet energy response depends on the particle jet energy and the results 
are usually binned in jet $p_T$.  However, the measured jet energy has poor 
resolution and can lead to a large bias in the measurement of the response. 
To avoid this resolution bias, the jet energy response is measured as a 
function of the estimator
\begin{equation}
E' = p_{T,\gamma}\cdot\cosh \eta_\mathrm{jet}.
\label{Eprime}
\end{equation}
$E'$ is strongly correlated to the particle level jet 
energy and has a better resolution than the measured jet energy.
We parameterize all corrections as a function of $E'$ and map back to the
measured jet energy $E_{\mathrm{meas}}$ on a jet-by-jet basis 
by inverting the equation
\begin{equation}
E_\mathrm{meas} - O = R_\mathrm{had}(E') S_\mathrm{phys}(E') E',
\end{equation}
where $O$ is the offset contribution, $R_\mathrm{had}(E')$ contains all jet
energy corrections back to particle level, and 
$S_\mathrm{phys}(E')=E_\mathrm{jet}^\mathrm{ptcl}/E'$
contains the additional corrections for particle showering, causing
energy to flow out of or into the jet cone. The latter component accounts for 
energy loss from out-of-cone
radiation (physics showering), leading to a correction of 0.90--1.00 at jet 
$p_T>$50
GeV and $|y|<$3.0. The equation is iteratively solved using Newton's method.
The resulting estimate of the jet energy is observed to
agree with the true $E'$ to better than 2\% at jet $p_T>50$~GeV,
resulting in less than 0.2\% uncertainty on the jet response $R_{\text{had}}$.

\begin{figure}
\includegraphics[width=\columnwidth]{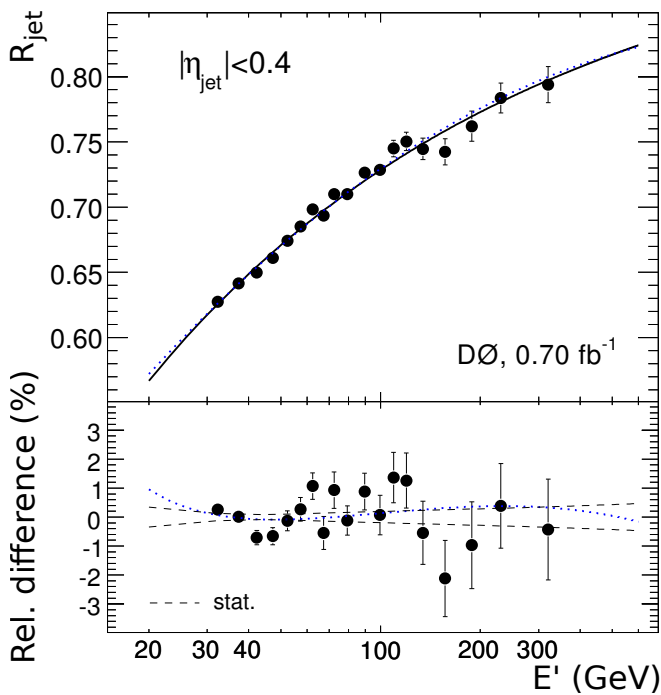}
\caption{\label{extrapol} (color online)
Extrapolation of the jet energy response in the CC at high $E'$ using the 
rescaled MC (see the main text) and a fit to the MC points. 
The dotted line shows a simple quadratic logarithmic fit to 
data for comparison with the tuned MC results displayed by the solid line. 
We also display in the bottom panel the relative difference between both 
curves and the statistical uncertainty on the fit to the rescaled MC in 
dashed lines.}
\end{figure}

Another issue in using the MPF is related to photon identification. 
To have a clean $\gamma+$jet sample in data, only  CC photons are used 
with tight selection criteria. 
However, in some jets a large fraction of their transverse momentum
is carried by photons from $\pi^0$, $\eta$, or $K^0_s$ decays, which
form a sample of ``electromagnetic'' jets (``EM-jets'').
If these photons are sufficiently close together, 
and there is little activity around the photons, the jet can mimic an isolated 
single photon typical for $\gamma$+jet events. Because the cross section for
$\gamma$+jet events is $\sim$3 -- 4 orders of magnitude lower than that of 
dijet events, these EM-jets contribute a significant background for true 
$\gamma$+jet events. An artificial neural network (ANN) is trained to 
discriminate between photon and EM-jets~\cite{photonANN} using input 
variables based on the shape of the 
calorimeter shower and measurements of charged particle tracks in the 
vicinity of the photon candidate.  The distribution of the photon ANN output for
the simulated photon signal and for the EM-jet background samples are
fitted to the data for each $E'$ and $\eta$ bin using a maximum likelihood
optimization to obtain the fractions of signal events in the data.
To reduce the 
uncertainty in the jet energy scale due to contamination from background in the 
$\gamma$+jet events, the difference in the response determined from 
real $\gamma+$jet and dijet events, where one of the jets is misidentified
as a prompt photon, is estimated using MC and applied as a correction based 
on the estimated purity of the selected photons in data.   
The jet energy response after 
all corrections as a function of $E'$ in the CC is given in 
Fig.~\ref{extrapol}. The main uncertainty is due to the uncertainty on 
the photon energy scale, which is on the order of 0.5\% 
at $E'\approx 20$~GeV and 0.8\% at $E'\approx 500$~GeV. 
The choice of fragmentation model used in {\sc{Pythia}}~\cite{pythia} 
was an additional source of systematic uncertainty on the photon 
purity~\cite{gamma}. 

The statistics of the $\gamma$+jets sample limits the direct response 
measurements in the CC to $E'<350$~GeV. The measured energy response 
in this region must 
be extrapolated to the highest jet energies at $\approx 600$~GeV. To avoid 
a statistical uncertainty of more than 2\% at high-$p_T$ in the CC, 
MC models are used to constrain the high-$p_T$ response. For this 
purpose, the measurement of the response in $\gamma+$jet events in the MC is 
rescaled to the measurement in data by modifying the response of the calorimeter
for single pions in MC. 
Figure~\ref{extrapol} shows the measured response for jets in data compared to
the rescaled MC prediction and to a quadratic fit in $\log E'$.
The uncertainty in the fragmentation model for the 
high $E'$ extrapolation is estimated using the differences between the 
{\sc{Pythia}} and {\sc{Herwig}}~\cite{herwig} generators after 
turning off the underlying event modeling. This leads to a systematic 
uncertainty of about 0.8\% at $E'=600$~GeV. 
The systematic uncertainties related to PDFs (especially
due to the uncertainty on the gluon fraction in the 
proton) are about 0.2\%.

\begin{figure}
\includegraphics[width=0.95\columnwidth]{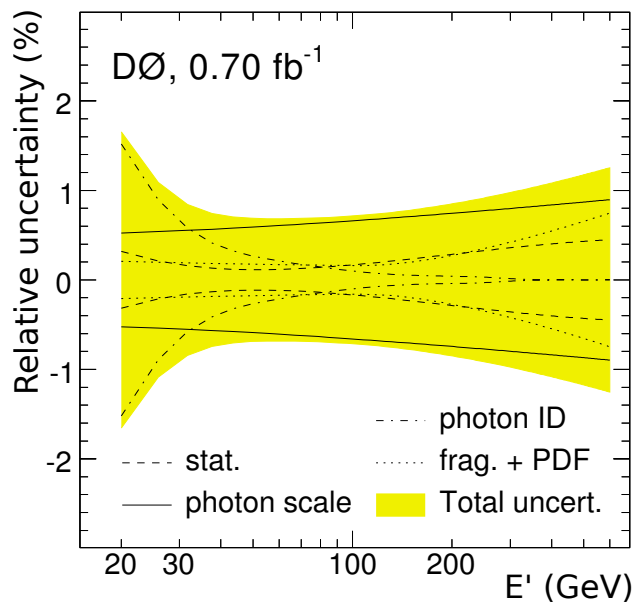}
\caption{\label{respuncert} (color online)
Different sources of uncertainty on the jet $p_T$ response in the CC: photon 
energy scale, photon identification, fragmentation, and PDF.}
\end{figure}

The total uncertainty on the jet $p_T$ response as a function of $E'$ is given 
in Fig.~\ref{respuncert}. The dominant uncertainty comes from the 
photon energy scale. The uncertainty due to photon identification is 
related to the uncertainty on the sample purity and contributes mainly at 
$E'$ energies below 50 GeV.

\subsubsection{Pseudorapidity dependent corrections}

\begin{figure*}
\vspace{0.5cm}
\begin{overpic}[width=0.48\textwidth]
{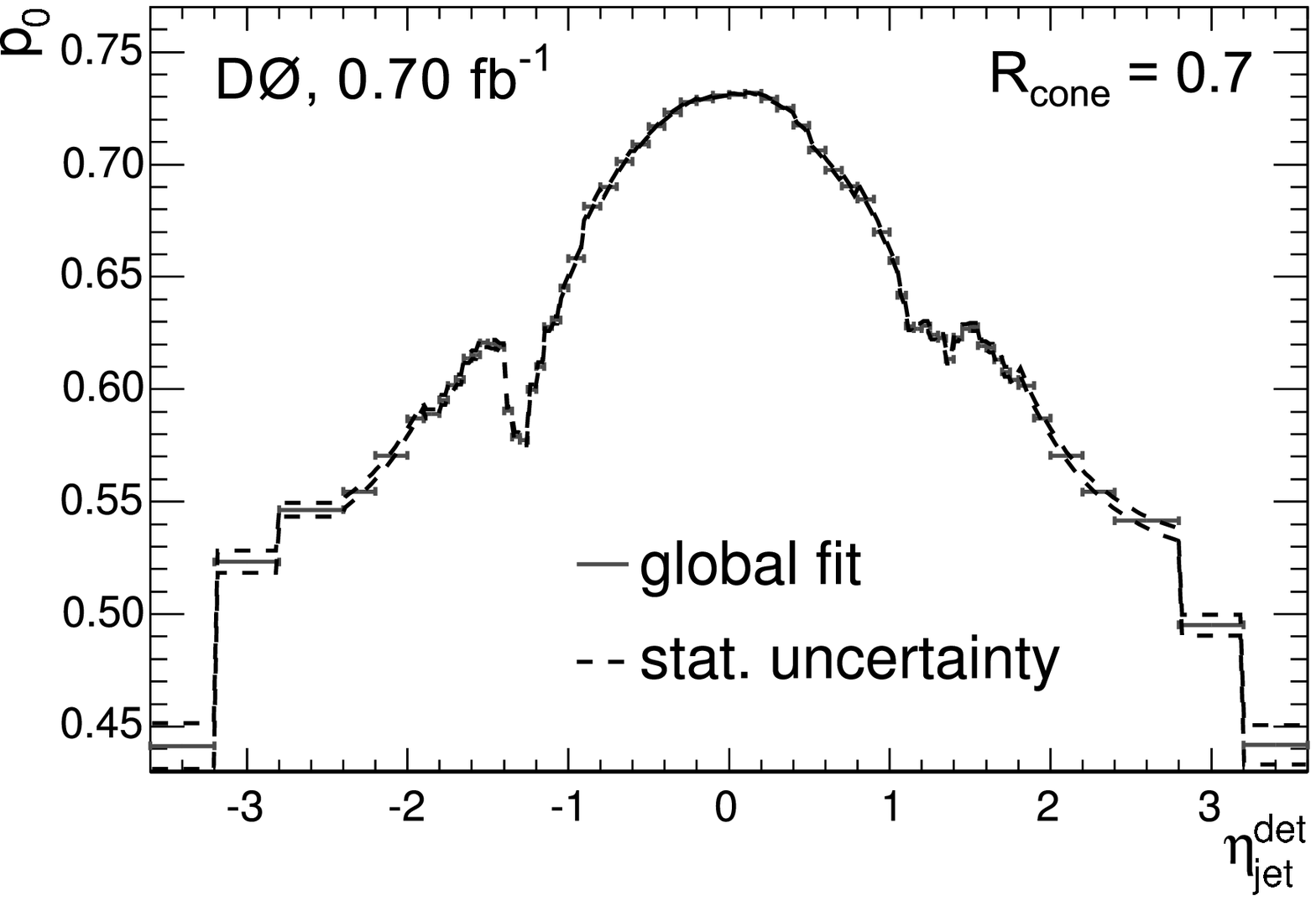}
\put(220,100){\textsf{(a)}}
\end{overpic}
\begin{overpic}[width=0.48\textwidth]
{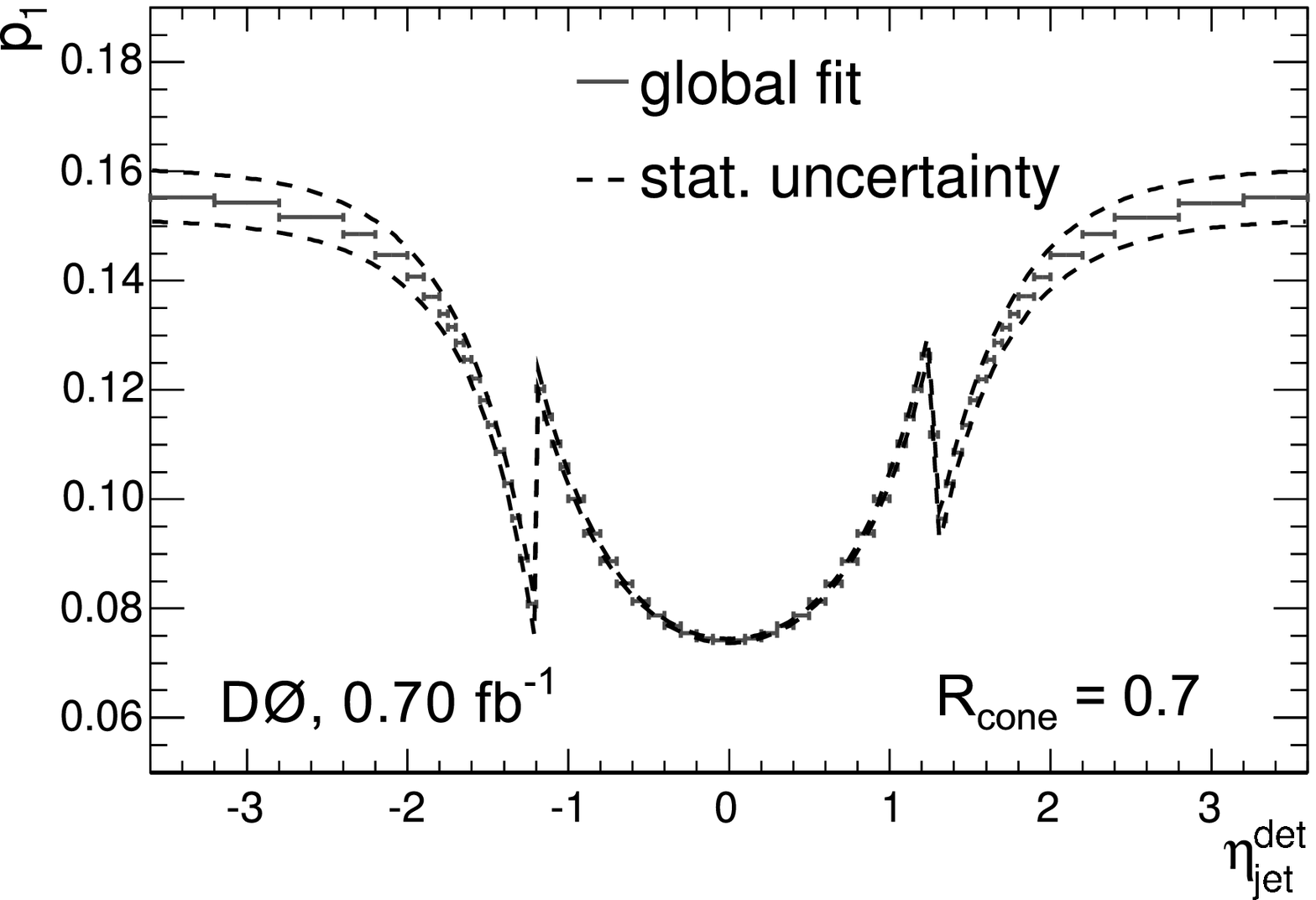}
\put(220,100){\textsf{(b)}}
\end{overpic}
\vspace{0.5cm}
\begin{overpic}[width=0.48\textwidth]
{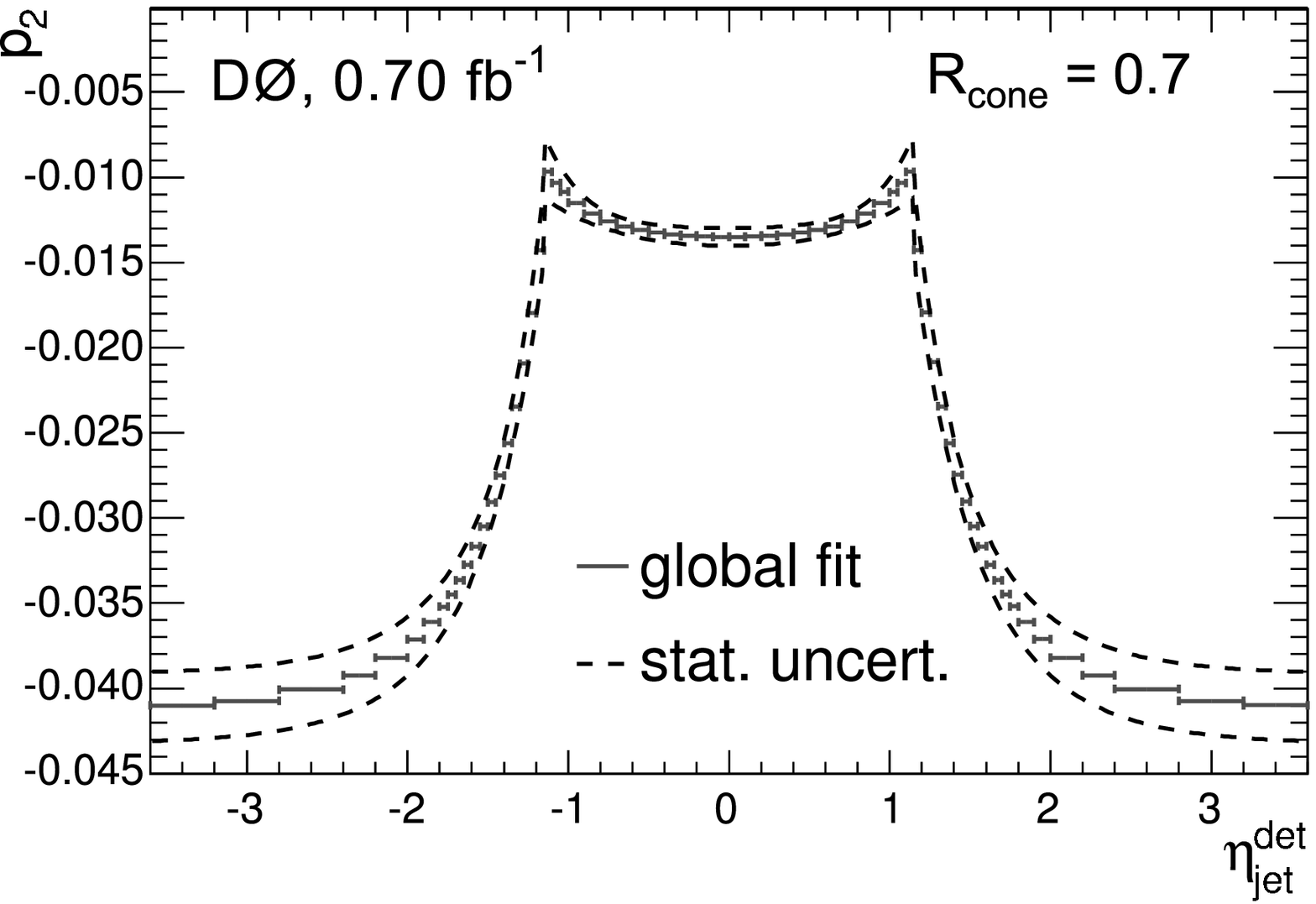}
\put(220,100){\textsf{(c)}}
\end{overpic}
\begin{overpic}[width=0.48\textwidth]
{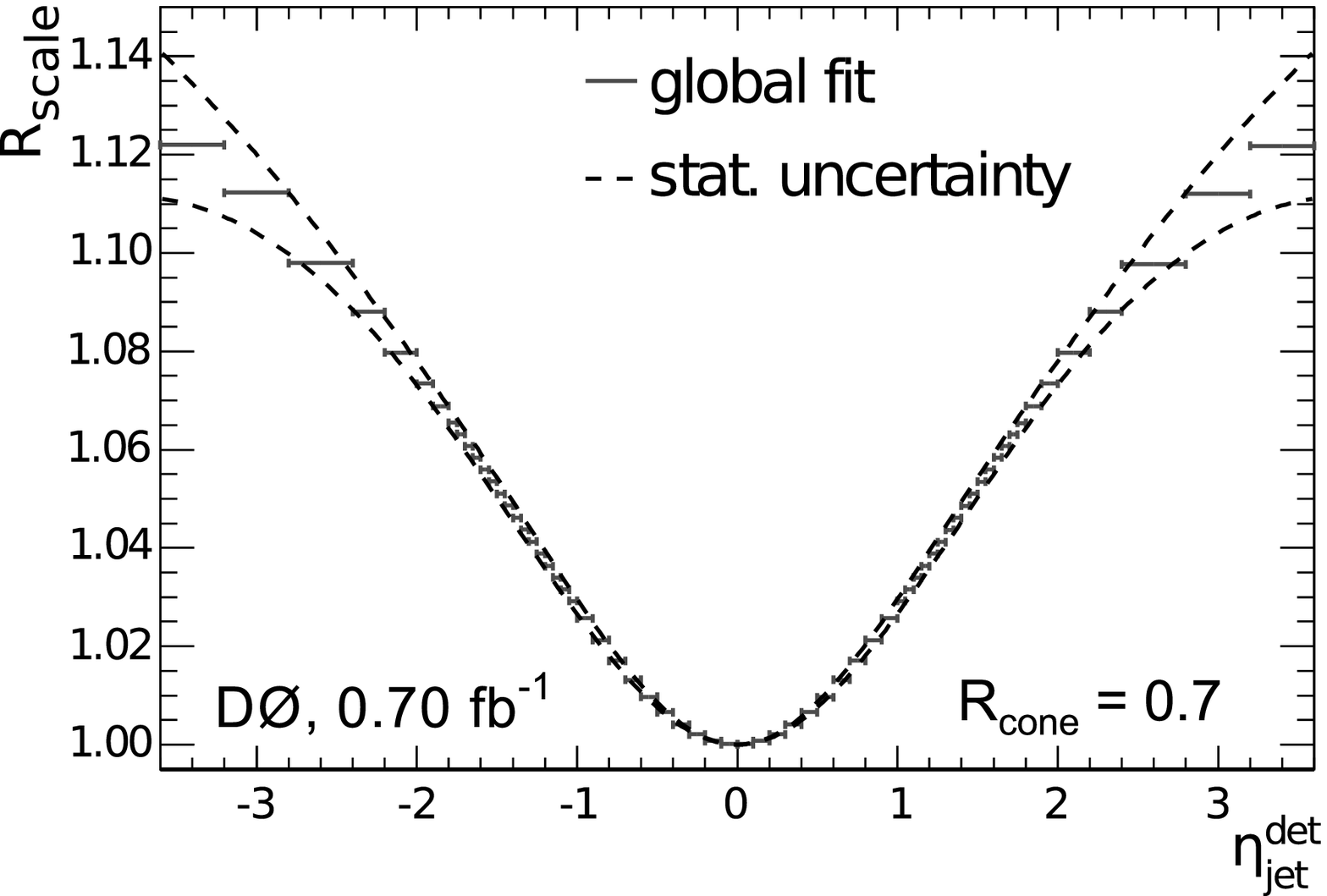}
\put(220,100){\textsf{(d)}}
\end{overpic}
\caption{\label{figpi}  The (a-c) parameters of the $\eta$-dependent 
correction and (d) the $\eta$-dependent scaling factor applied to the 
dijet samples.
The sharp features are due to changes in the detector structure, 
moving from the central to forward calorimeters.}
\end{figure*}

The purpose of the $\eta$-dependent corrections is to equalize the jet 
response everywhere as a function of pseudorapidity in the calorimeter after 
the jets are corrected for offset effects. The D0 calorimeter is 
inter-calibrated at the cell level as a function of the azimuthal 
angle $\phi$ by equalizing the response of the calorimeter in dedicated
$\phi$-symmetric data samples.  This yields a jet response that is 
independent of $\phi$, so only the $\eta$ dependence of the response 
needs to be corrected.  The $\eta$ dependence 
of the response is mostly due to the changing calorimeter detector elements, 
especially in the ICR, different amounts of passive material and the varying 
angle of incidence with jet $\eta$.  The $\eta$-dependent 
corrections $F_{\eta}(E,\eta)$ normalize the response at forward 
pseudorapidities to that measured in the CC ($R_{CC}$).
This leads to the definition
\begin{equation}
F_{\eta}(E, \eta) \equiv R(E, \eta) / R_\mathrm{CC}(E),
\end{equation}
where $R(E,\eta)$ is the response of the detector for a jet of energy $E$, 
located at detector pseudorapidity $\eta$.
We use both dijet and $\gamma$+jet samples
to determine $F_{\eta}$.  The dijet sample provides high statistics 
and high reach in jet energy for the forward region. One of the 
jets is required to be central and the response measurement is binned in 
terms of the $p_T$ of the central jet (using the dijet $E'$, defined 
as in Eq.~\ref{Eprime} where the photon is replaced by the central jet) 
after correcting for the offset and calorimeter response. This binning 
leads to a resolution bias, which is later corrected.

\begin{figure*}
\begin{overpic}[width=0.49\textwidth]
{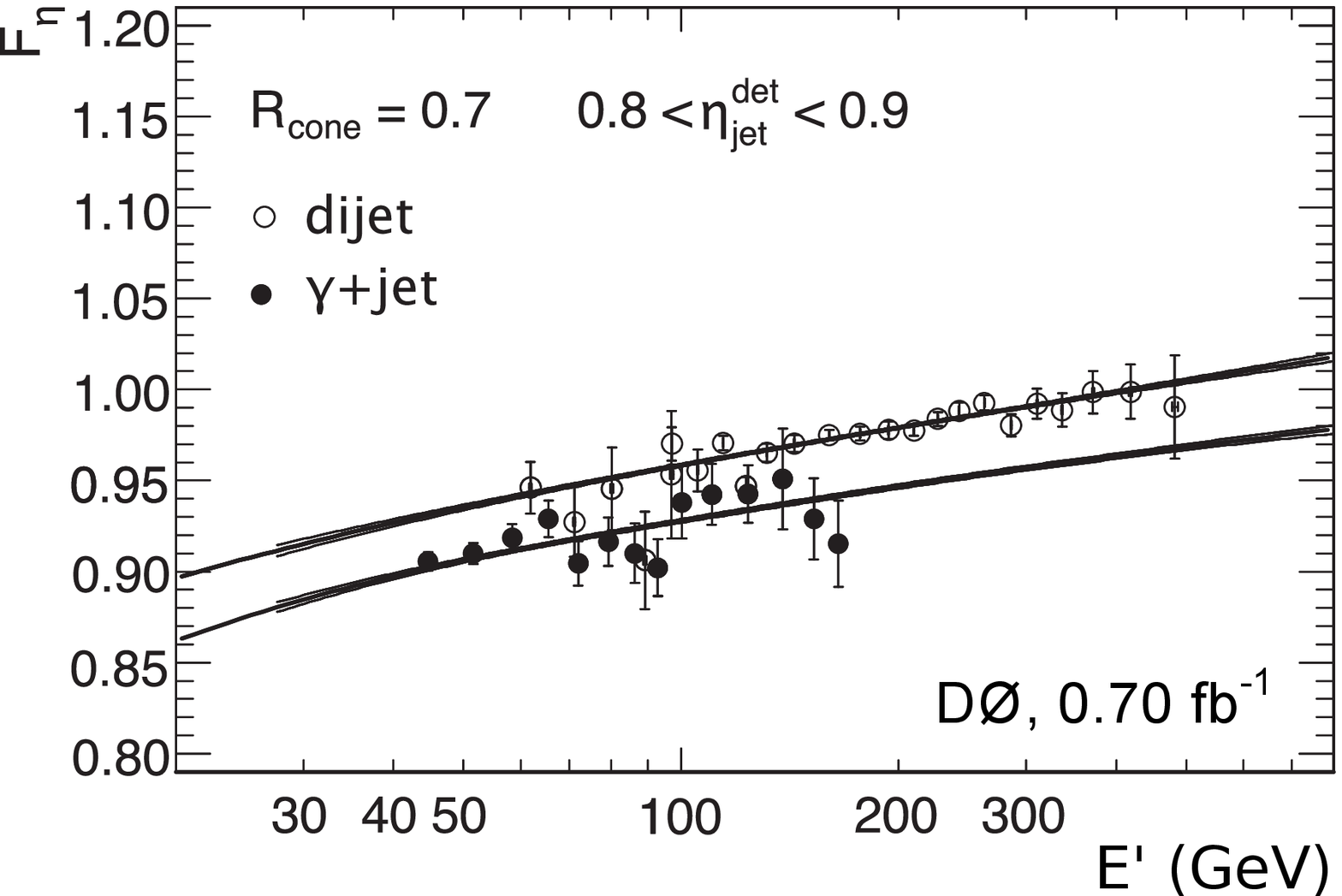}
\put(220,140){\textsf{(a)}}
\end{overpic}
\begin{overpic}[width=0.49\textwidth]
{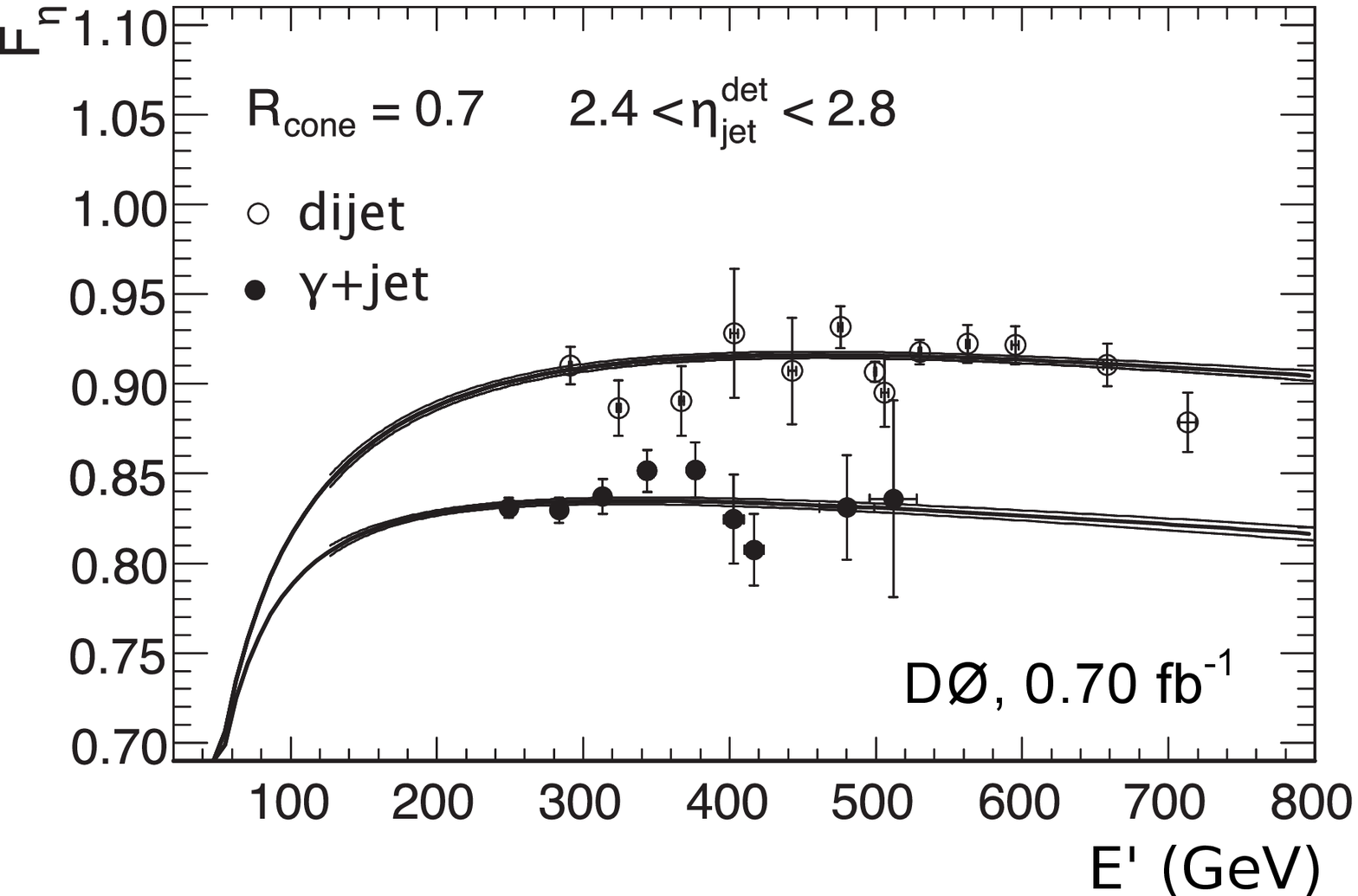}
\put(220,140){\textsf{(b)}}
\end{overpic}
\caption{\label{etadep}Fits of $F_\eta$ in $\gamma$+jet and dijet data for two 
different regions in $\eta$ as a function of $E'$. The central fit values and
the uncertainty band are displayed on the figure.}
\end{figure*}

The $\eta$-dependence of the response, $F_{\eta}$, is fitted using a 
quadratic-logarithmic function of $E'$
\begin{equation}
F_\eta(E',\eta) = \frac{p_0(\eta) + p_1(\eta)\ln(E') + p_2(\eta)\ln^2(E')}
{R_\mathrm{CC}(E')},
\end{equation}
where the $p_i$ are fitted as a function of detector $\eta$. The $F_{\eta}$
and $p_i$'s are given in Fig.~\ref{figpi}(a-c).
As an example of the data used in this fit, we give in 
Fig.~\ref{etadep} the $\eta$-dependent corrections for two bins in $\eta$ for 
the dijet and $\gamma +$jet samples. Although the correction factors 
depend on the sample ($\gamma+$jet, dijets), we can remove this 
dependency by scaling 
the dijet correction in the overlap region between the CC and the EC by an 
energy-independent factor 
$R_\mathrm{scale}$:
\begin{equation}
R_{\mathrm{scale}}(\eta) = 1 + q_1\ln[\cosh(\eta)] + q_2\ln^2[\cosh(\eta)],
\end{equation}
where $q_1$ and $q_2$ are two parameters fitted to data and the result is
given in  Fig.~\ref{figpi}(d).
This functional form is motivated by phenomenological studies of 
the difference in the jet responses measured in $\gamma +$jet and dijet 
samples, as discussed in the next section.

The jet $p_T$ resolution 
is worse than the $\gamma$ $p_T$ resolution. Due to the steeply
falling inclusive jet cross 
section, more jets migrate into a given $p_T$ bin from lower $p_T$ than from 
higher $p_T$, giving rise to a $p_T$ bias compared to the particle level. 
The effect of this resolution bias is taken into account in the final 
measurement of the jet $p_T$ response versus $\eta$ using the CC jet $p_T$ 
resolutions obtained from dijet events as described
in Sec.~\ref{sec:resolutions}. In 
particular, the jet transverse momenta in dijet events 
in the CC are {\textit {a priori}} perfectly balanced on 
average by definition $[F_\eta(E',\eta=0)=R_\mathrm{scale}(\eta=0)=1]$, which 
provides a strong constraint for the bias correction.

\begin{figure}
\vspace{1cm}
\includegraphics[width=\columnwidth]{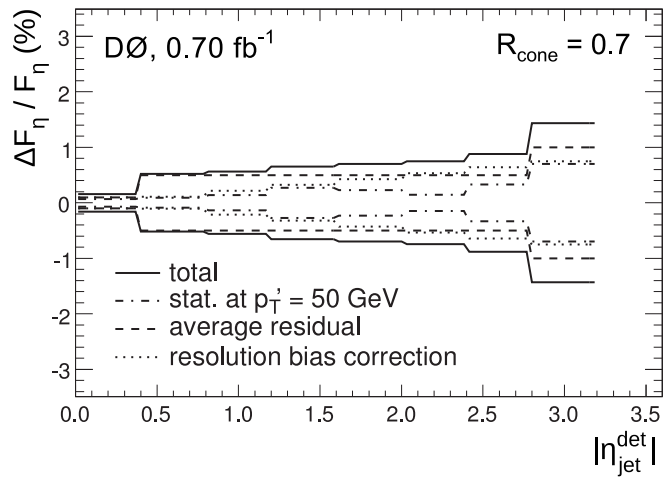}
\caption{\label{etadepuncert} 
Relative uncertainties on the $\eta$-dependent corrections as a function of 
jet detector rapidity.}
\end{figure}

With the application of the dijet-specific scale factor and resolution bias 
corrections we obtain systematic uncertainties in the $\eta$-dependent 
corrections that are less than 1\% for $|\eta|<2.8$ as illustrated in 
Fig.~\ref{etadepuncert}. The leading systematic uncertainty is from the average
residuals of the fits for $F_\eta$ and 
is estimated to be 0.5\% for $0.4<|y|<2.4$ and constant versus energy. 
This residual accounts for the scatter of the data points around the central 
fit and covers possible variation in the shape of the fit function. The 
uncertainty due to the resolution bias correction is of the order of $0.5$\% 
at $|\eta|=2.0$ and reduces to zero at $\eta=0$.

\subsubsection{Dijet specific response}

\begin{figure}
\includegraphics[width=\columnwidth]{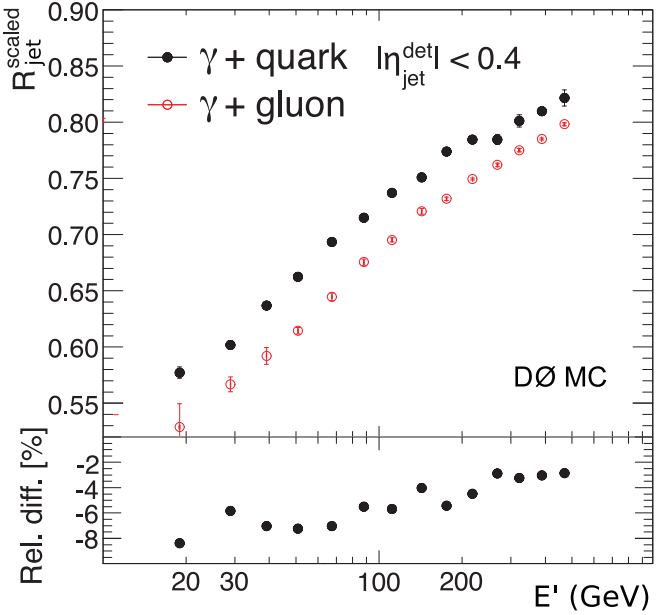}
\caption{\label{jetresp} (color online)
Quark- and gluon-initiated jet responses and their relative differences for CC 
jets as a function of $E'$.}
\end{figure}

\begin{figure}
\includegraphics[width=\columnwidth]{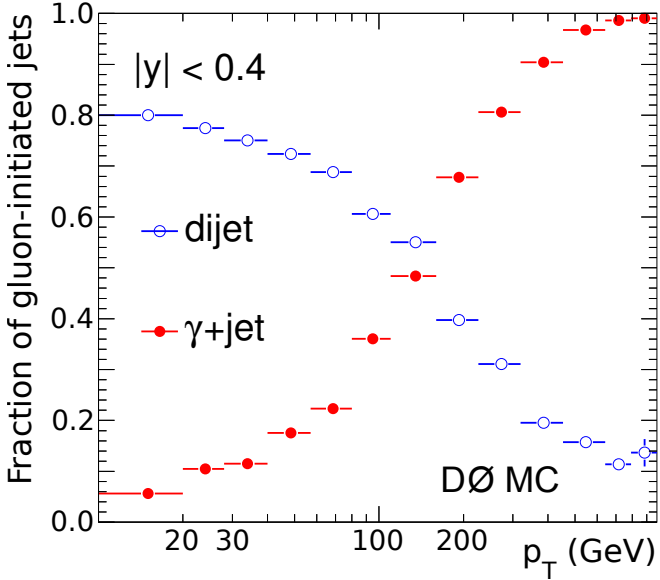}
\caption{\label{fraction} (color online)
Fraction of gluon initiated jets in $\gamma +$jet and dijet events in the CC.}
\end{figure}

The methods presented so far allow for a precise measurement of the MPF 
response in the CC for the $\gamma$+jet sample. However, the response for 
dijet and $\gamma +$jet events is different. Figure~\ref{jetresp}  
displays response for the quark and gluon initiated jets measured in MC 
simulations after rescaling the single pion response to data. The 
gluon-initiated jets have a lower response than quark-initiated jets because 
they have on average higher particle multiplicity with softer particles. 
The soft particles lead to a lower jet response due to the  
falling single pion response at low energy. Figure~\ref{fraction} displays 
the fraction of gluon-initiated jets in MC for 
$\gamma +$jet and dijet events. The $\gamma +$jet jet energy 
scale cannot be used directly for the measurement of the inclusive jet cross 
section,  because this sample is strongly dominated by dijets. 
This effect also explains 
the differences we observe in Fig.~\ref{etadep} for the $\eta$-dependent 
corrections in $\gamma +$jet and dijet samples. The difference observed in 
$F_{\eta}$ versus $E'$ at fixed $\eta$ is due to the different amounts of quark 
and gluon jets in the samples. The gluon versus quark fractions depend 
primarily on energy (not $p_T$ or $\eta$) which leads to a correction 
factor dependent on 
$\cosh(\eta)$. Once this difference is taken into account, it is 
possible to combine both samples to fit $F_{\eta}$.

To calculate the relative difference in response between $\gamma +$jet 
and dijet samples in the CC, we first scale the single pion response in MC
to reproduce the measured jet response in the $\gamma +$jet data.  
The measurement from data of the absolute jet response in $\gamma+$jet events 
in the CC is then scaled to its dijet equivalent. The dijet $\eta$-dependent 
corrections are obtained from a global fit to $\gamma +$jet and dijet data, 
which accounts for the sample-dependent response.

The differences between the dijet response used in this analysis and the
$\gamma +$jet response used in most other analyses are contained in the
$\eta$-dependent scale factor $F_\eta$ and the ratio of tuned MC responses 
$R_{\mathrm{dijet}/\gamma+\mathrm{jet}}$ at $\eta=0$
\begin{eqnarray}
R_\mathrm{dijet}(E',\eta) &=& R_{\mathrm{dijet}/\gamma+\mathrm{jet}}(E')\cdot
F_\eta(\eta) \cdot \nonumber \\
&~& R_{\gamma+\mathrm{jet}}(E',\eta).
\end{eqnarray}
The ratio between the dijet and $\gamma+$jet responses $R_{\mathrm{dijet}/\gamma+
\mathrm{jet}}$
is in practice given by the information presented in Figs.~\ref{jetresp} and 
\ref{fraction} and can be expressed using the responses for 
the gluon- and quark-jets\footnote{We note that nearly all the quark-initiated 
jets come from light quarks.} ($R_{\text{gluon}}$ and $R_{\text{quark}}$) and 
the fractions of gluon-jets
in the dijet and $\gamma+$jet samples ($f_{\text{gluon}}^{\text{dijet}}$ and 
$f_{\text{gluon}}^{\gamma +\text{jet}}$)
\begin{eqnarray}
R_{\text{dijet}/\gamma +\text{jet}} = \frac{(R_{\text{gluon}}
f_{\text{gluon}}^{\text{dijet}} +
 R_{\text{quark}}
(1-f_{\text{gluon}}^{\text{dijet}}))} 
{(R_{\text{gluon}} f_{\text{gluon}}^{\gamma +\text{jet}} + R_{\text{quark}} 
(1-f_{\text{gluon}}^{\gamma +\text{jet}}))}.
\end{eqnarray}

\subsection{Showering correction}

Jets are extended objects and deposit their energy over a wide area in the 
calorimeter. When the cone algorithm is used, some of this energy is  
deposited outside the jet cone due to interactions with the 
magnetic field and passive material. This is called 
detector showering and needs to be taken into account in the jet energy scale 
determination. In addition, part of the energy of the incident parton is 
lost outside the jet cone because of hadronization and the finite size 
of the jet cone. This is called physics showering and is taken into 
account in the energy scale correction to the particle level.

\begin{figure}
\includegraphics[width=\columnwidth]{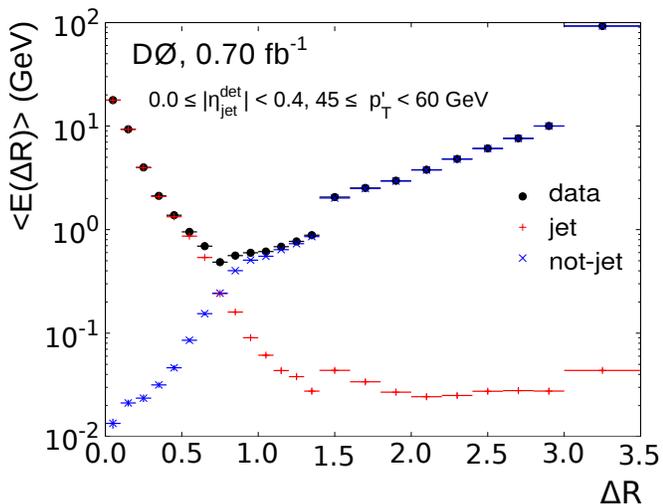}
\caption{\label{show} (color online)
Jet energy profiles as a function of distance from the jet axis $\Delta R$
for MC and data used to compute showering corrections.
The data are corrected for offset energy from noise and additional $p\bar{p}$
collisions and are compared to MC jets without offset (jet) and contributions
from the underlying event (not-jet).
We note the good agreement between data and the sum of energy contributions
from inside and outside the jet in MC.} 
\end{figure}

The determination of the showering corrections requires a good understanding of 
the transverse jet energy profile. In a dedicated study, the cell-level 
information from MC is kept to generate energy density profiles as a 
function of the distance 
$\Delta R=\sqrt{(y_{particle}-y_{jet})^2+(\phi_{particle}-\phi_{jet})^2}$ 
between the particle and the jet axis for 
particles originating from inside the particle jet, from outside the 
jet, and from offset due in particular to pile up or additional 
interactions in one bunch crossing.  The sum of these 
profiles is fitted to the measured energy profile in data to account for 
possible response differences between data and MC. The energy profiles 
are created by summing the energy in the cells at a given radius from the cone 
axis. The profiles are calculated for back-to-back $\gamma$+jet events and 
show the jet core at $\Delta R$ around 0 and the photon contribution at 
$\Delta R\approx \pi$. The energy density in the range 
$R_\mathrm{cone}<\Delta R< \pi$ is primarily offset energy. 
Figure~\ref{show} shows an example of the showering profiles in MC without 
any zero bias event overlay (i.e. with only the underlying event and 
no offset). It gives the average energy in a given rapidity and 
transverse energy bin coming from inside and outside the jet as a function 
of the distance $\Delta R$ in rapidity and
azimuthal angle from the center of the jet. The MC describes the data 
when both the energies inside and outside the jet are considered.

The estimate of the showering correction $\hat{S}$ for $\gamma +$jet events in 
MC and data is obtained by comparing the energy deposited by all particles  
inside the calorimeter jet cone $\sum_{\Delta R=0}^{R_\mathrm{cone}} E_\mathrm{in} +$ 
$\sum_{\Delta R=0}^{R_\mathrm{cone}} E_\mathrm{out}$ originating from inside or
outside the particle jet to that from the original particle jet 
$\sum_{\Delta R=0}^\mathrm{\infty} E_\mathrm{in}$ using the 
fit-weighted templates
\begin{equation}
\hat{S} = \frac{\sum_{\Delta R=0}^{R_\mathrm{cone}} E_\mathrm{in} + 
\sum_{\Delta R=0}^{R_\mathrm{cone}} E_\mathrm{out}}
{\sum_{\Delta R=0}^\mathrm{\infty} E_\mathrm{in}},
\end{equation}
where $E_{\text{in}}$ and $E_{\text{out}}$ are the energies coming from 
inside and outside the jet. To take into account any potential bias in the 
method, the final value of the showering correction in data is computed as
\begin{equation}
S_\mathrm{data} = \hat{S}_\mathrm{data}\cdot
\frac{S_\mathrm{MC}^\mathrm{true}}{\hat{S}_\mathrm{MC}},
\end{equation}
where the true showering $S_\mathrm{MC}^\mathrm{true}$ is directly 
available in MC. This bias correction amounts to less than 0.3\%.

While the showering templates are measured in energy, the applicable quantity 
for the cross section measurement is jet $p_T$.  When mapping the showering
templates to $p_T$ the deposits in rapidity are weighted by 
$\cosh(y_0)/\cosh(y_i)$, where $y_0$ is the cone axis and $y_i$ is the 
rapidity of the energy deposit. As a result of this weighting, the effects 
of showering in $p_T$ are generally suppressed relative to energy showering.  
This can also tilt the jet toward $y=0$ and cause a net increase in the jet 
$p_T$, leading to $S_\mathrm{data}>1$.  The differences between energy 
and $p_T$ showering can be up to (1--2)\% over the kinematic region of the 
cross section measurements.

\begin{figure}
\includegraphics[width=\columnwidth]{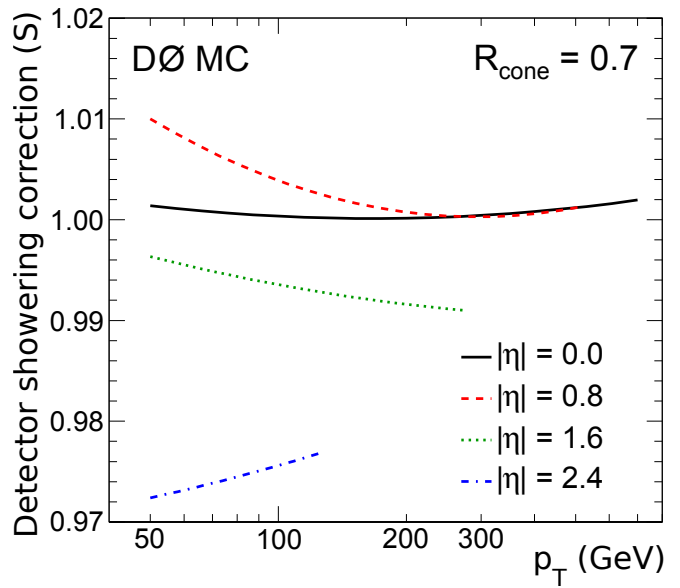}
\caption{\label{shower} (color online) Jet showering corrections shown 
as a function of jet $p_T$ for different regions of jet pseudorapidity.}
\end{figure}

The last step of the showering correction is to make the transition
from $\gamma +$jet to dijet events. This remaining correction is computed 
directly using the differences 
in showering in $\gamma +$jet and dijet MC. The final jet $p_T$ 
showering corrections are given in Fig.~\ref{shower}.

The uncertainties on the showering correction are less than 1\% 
of the overall correction factor at $p_T>50$~GeV. The main sources of 
uncertainty come from the difference between data and MC in the 
single pion response at low 
$p_T$, the quality of the fits of MC templates to data, and the description of 
the underlying event determined by varying {\sc{Pythia}} tunes for Tevatron data
at higher $p_T$.

\subsection{Potential biases in the method}

\subsubsection{Topology bias of the MPF method}

The MPF method balances a photon or a central jet against the full 
remaining hadronic 
recoil, but the measured MPF response is interpreted as the response of the 
probe jet. The precision of this interpretation may be biased
because the hadronic recoil includes particles not related to the probe jet, 
for example, particles coming from soft gluon radiation. These additional 
particles are generally softer than those in the core of the jet and are 
expected to lower the response of the recoil with respect to that of the 
jet core.

In the case of the energy measurement, an additional bias is caused by the 
systematic mismeasurement of the jet rapidity, because the MPF method is 
inherently based on balancing $p_T$. As we will see in the following, the 
rapidity bias is particularly large in the ICR, where the absolute rapidity is 
systematically underestimated and causes a corresponding increase in the MPF 
response: the same calorimeter energy now corresponds to higher $p_T$. Since 
the bias versus energy 
has a non-trivial rapidity dependence and
the cross section measurement is performed as a function of $p_T$, we derive
and apply topology bias corrections as a function of jet $p_T$.

\begin{figure}
\includegraphics[width=\columnwidth]{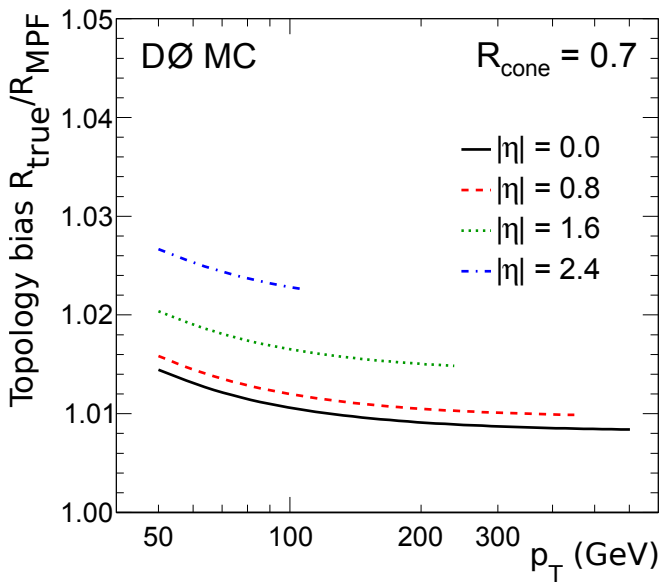}
\caption{\label{biasMPF} (color online)
Topology bias in the MPF method for jet $p_T$ response.}
\end{figure}

The bias of the MPF response is determined in tuned MC by comparing the MPF 
response to the true response defined at the particle level. The result for 
the $p_T$ response is shown in Fig.~\ref{biasMPF}. This bias is about 1--2.5\%
for the different rapidity regions with little $p_T$ dependence ($<0.5\%$)
for $p_T>50$~GeV. The 
MPF response bias is quite small since the method is based on the $p_T$ 
balance and the cone size of $R=$ 0.7 is large enough to contain most of the 
hadronic recoil in the absence of additional soft non-reconstructed jets. The 
bias is significantly larger, 2--4\%, for $R=$ 0.5 jet cones. The systematic 
uncertainty on the MPF method bias is computed as the difference between 
the $\gamma +$jet and dijet samples and found to be of the order of 0.1\%.

\subsubsection{Zero suppression bias}

An off-line zero suppression further suppresses the energies of
calorimeter cells in order to reduce the amount of noise, in 
particular in the coarse hadronic section, that can contribute 
to jet energies.  The algorithm used for this zero-suppression 
retains calorimeter cells if their energy exceeds the average
baseline noise by $4\sigma$, where $\sigma$ represents the 
measured standard deviation of the noise for a given cell.
Neighboring cells are also retained if their energy exceeds a 
threshold of $2.5\sigma$.

The zero-suppression algorithm produces a small positive noise
offset contribution because of the asymmetric zero suppression
(negative energies are never kept). For cells with high enough
real energy deposits, as within the jet core, the zero-suppression
produces no net offset, and positive and negative noise offset
contributions are expected to cancel.  Conversely, the energies measured   
for particles incident on the calorimeter, including those from uranium
decay, are reduced by the zero-suppression when cells are below
threshold.  Therefore the average offset within a jet is different from
the offset outside of a jet which we measure using zero bias and minimum
bias events.

\begin{figure}
\includegraphics[width=\columnwidth]{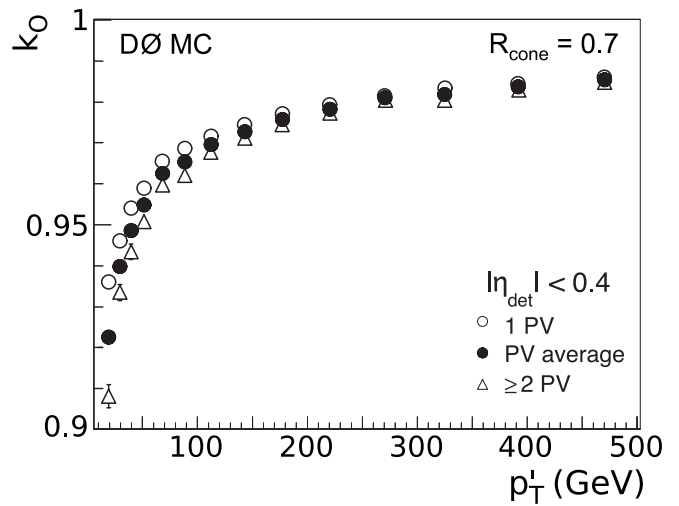}
\caption{\label{zerobias} 
Bias in the measurement of jet energy, $k_O$, due to zero suppression effects
on the offset correction, shown as a function of $p'_T$ for
central jets and different number of 
reconstructed vertices.}
\end{figure}

The true offset is increased inside the jet environment compared 
to the average energy density measured outside jets in zero bias and 
minimum bias events. The correction factor for the true offset 
inside a jet, $k_O$, is defined as
\begin{equation}
k_O = \frac{\left<E_\mathrm{meas}^\mathrm{jet}(\mathrm{no~ZB~overlay})\right>}
{\left<E_\mathrm{meas}^\mathrm{jet}(\mathrm{ZB~overlay}) - O_\mathrm{meas}\right>},
\end{equation}
where $E_\mathrm{meas}^\mathrm{jet}$ is the energy of a reconstructed jet and
$O_\mathrm{meas}$ is the measured offset correction described in
Sec.~\ref{sec:offset}. The same MC events are reconstructed with and 
without zero bias event overlay 
(offset). The zero bias event sample was collected without any calorimeter zero 
suppression so that its effect can be studied in detail. Figure~\ref{zerobias} 
shows the effect of zero-suppression on the offset correction
for jets in the CC.  For jet $p'_T>50$ GeV, where 
$p'_T= E'/\mathrm{cosh}~ \eta_{\mathrm{jet}}$, the resulting bias on jet energy
varies between 5\% at low $p'_T$ and 2\% at higher $p'_T$.

\begin{figure}
\includegraphics[width=\columnwidth]{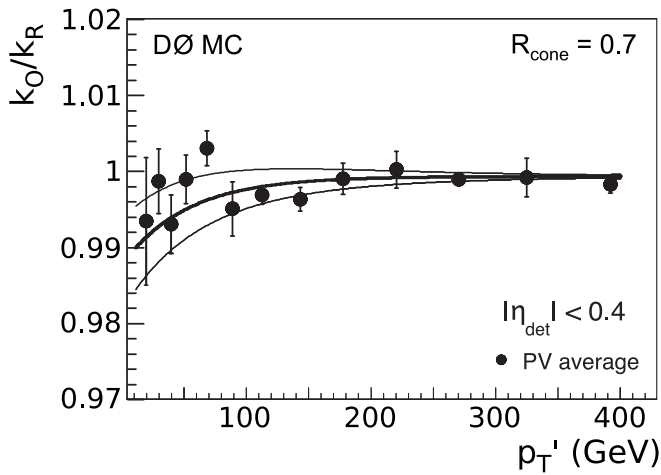}
\caption{\label{zerobiasb} 
Zero suppression bias $k_O/k_R$ in CC. The outer solid lines show the 
uncertainty attributed to the bias correction and the 1$\sigma$ contours.}
\end{figure}

The bias in offset is almost fully canceled by an opposite bias in the 
MPF response, defined as
\begin{equation}
k_R = \frac{\left<R_\mathrm{had}(\mathrm{no~ZB~overlay})\right>}
{\left<R_\mathrm{had}(\mathrm{ZB~overlay})\right>},
\end{equation}
because the increased offset inside the jet increases the $\met$ in the 
photon direction. This artificially increases the estimated MPF response
(see Eq.~5). 
Only the ratio $k_O/k_R$ is therefore relevant for the final bias correction 
due to the zero suppression bias. The combined bias is found to be less 
than 0.5\% for jet $p_T>50$~GeV in all rapidity bins, largely cancelling 
the topological bias, and approaches zero at high $p_T$, as shown in 
Fig.~\ref{zerobiasb}.

\subsubsection{Rapidity bias}

\begin{figure}[t]
\includegraphics[width=\columnwidth]{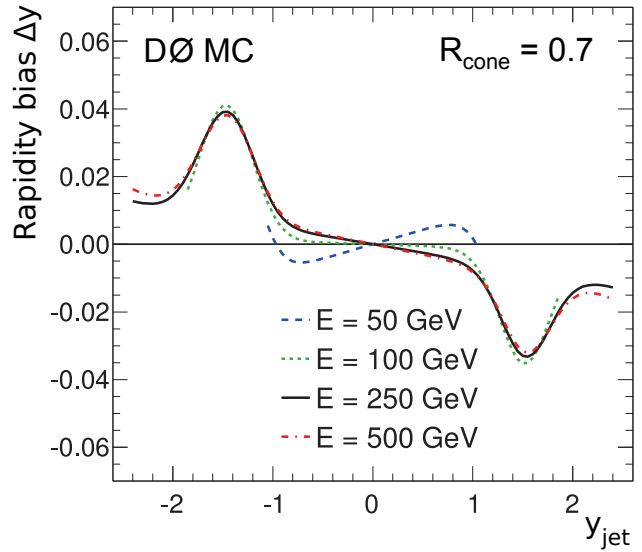}
\caption{\label{rapbias} (color online)
Rapidity bias obtained in MC for different jet energies $E$. The curves cover the range 
$p_T>30$~GeV and $|y_{\mathrm{jet}}|<2.4$.}
\end{figure}

Since the inclusive jet cross section is measured in bins of rapidity,
we checked for any potential bias in the reconstruction of jet rapidity
using the simulation, as shown
in Fig.~\ref{rapbias}. The rapidity is generally biased 
towards the central calorimeter, with the largest deviations observed in the 
ICR. This is attributed to detector effects in the ICR in addition to
the jet cone algorithm itself. The absolute effect on the inclusive 
jet measurement is small compared to the effect of jet $p_T$ calibration.

\subsection{Final jet energy scale corrections and uncertainties}

\begin{figure*}
\begin{overpic}[width=0.49\textwidth]
{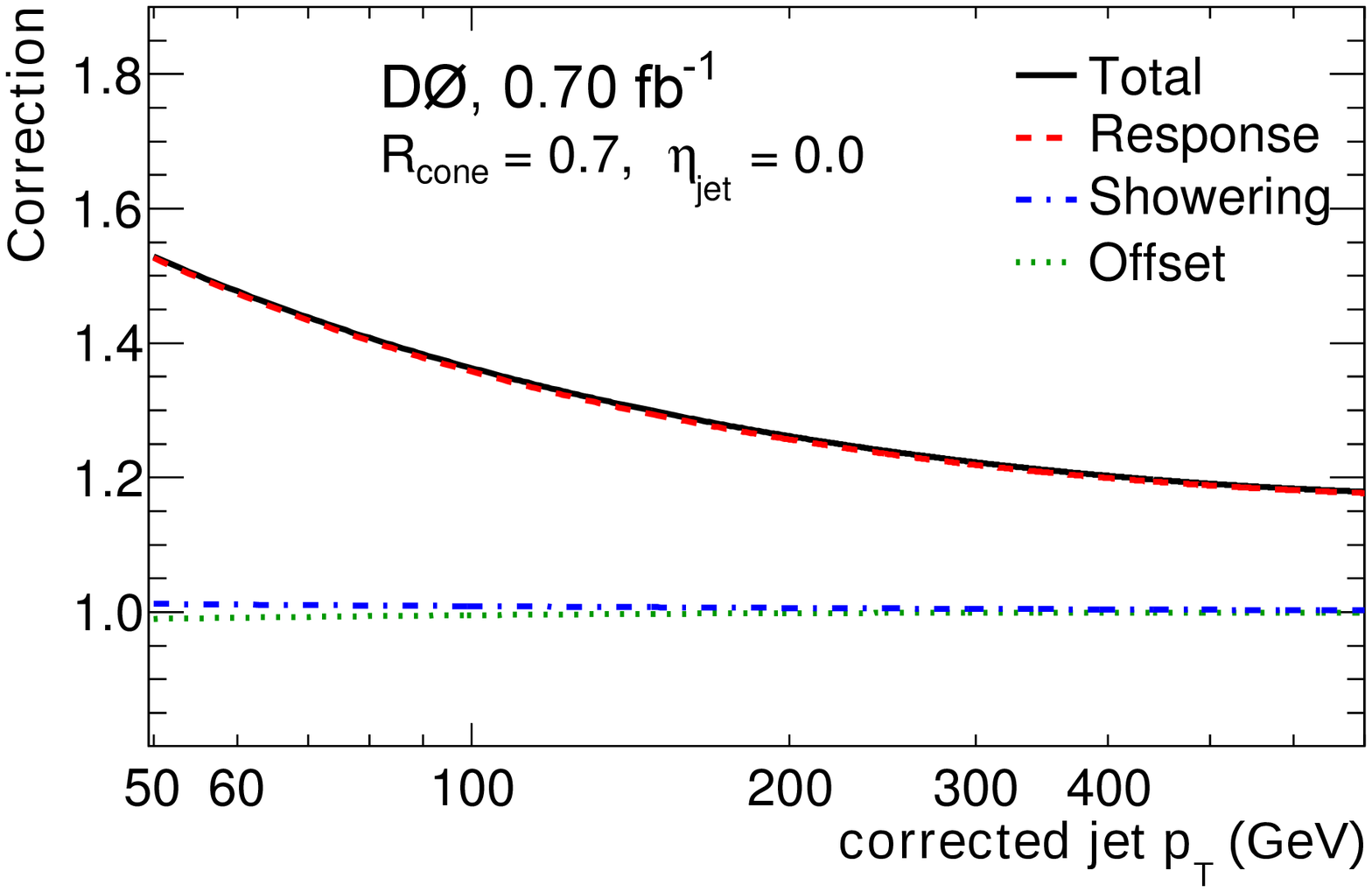}
\put(40,140){\textsf{(a)}}
\end{overpic}
\begin{overpic}[width=0.49\textwidth]
{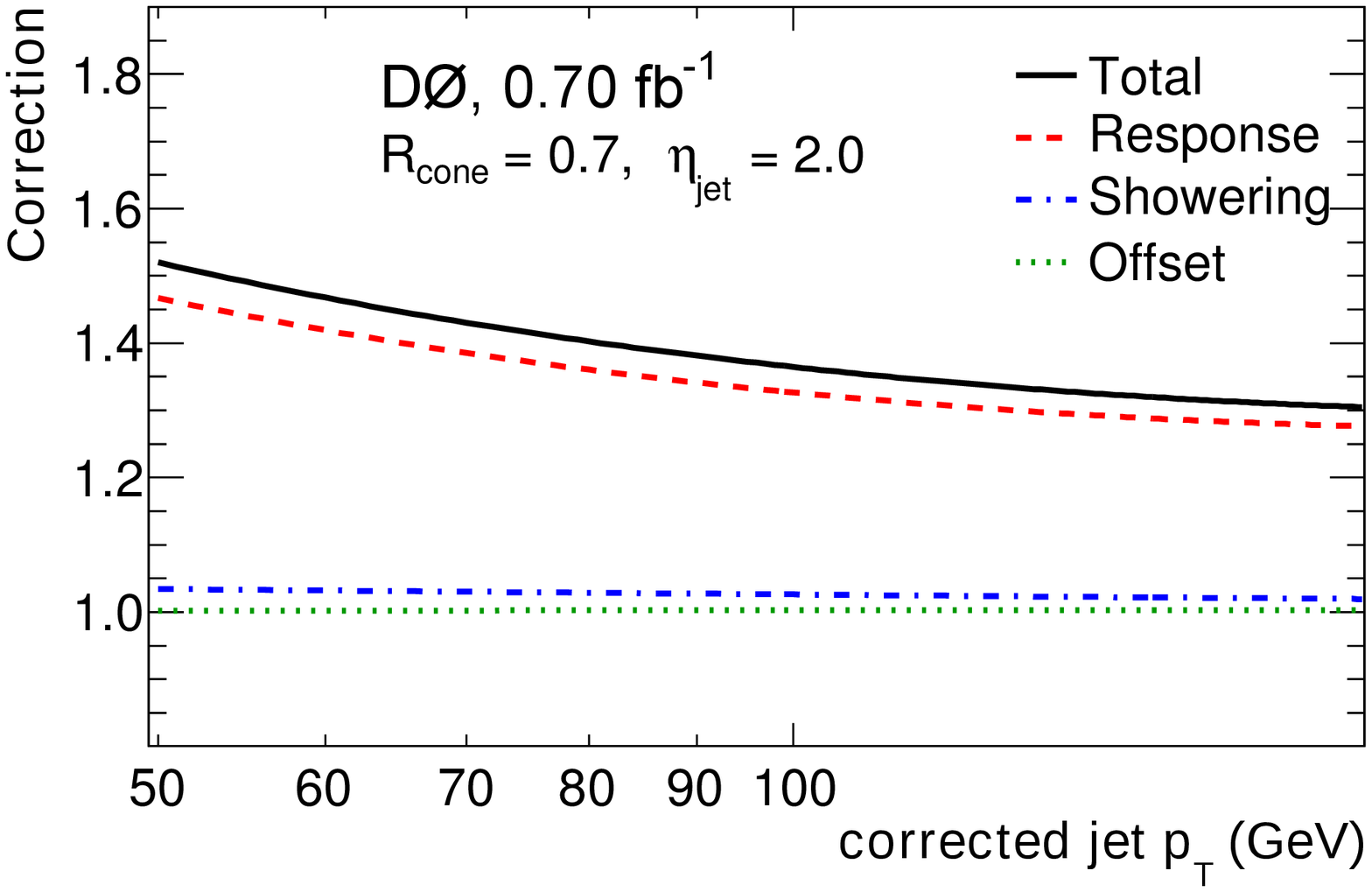}
\put(40,140){\textsf{(b)}}
\end{overpic}
\hspace{0.47\textwidth}\\
\begin{overpic}[width=0.49\textwidth]
{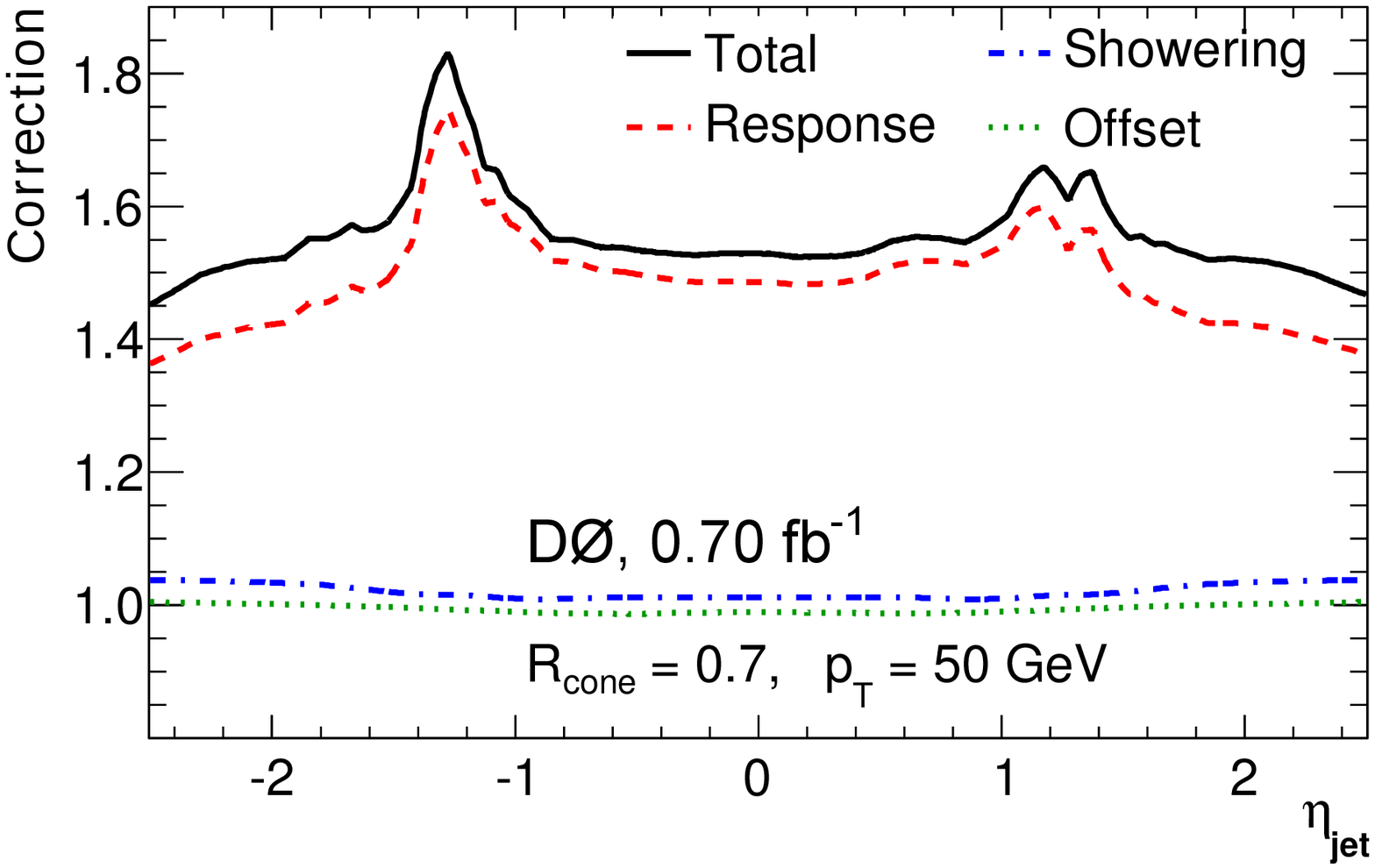}
\put(40,135){\textsf{(c)}}
\end{overpic}
\begin{overpic}[width=0.49\textwidth]
{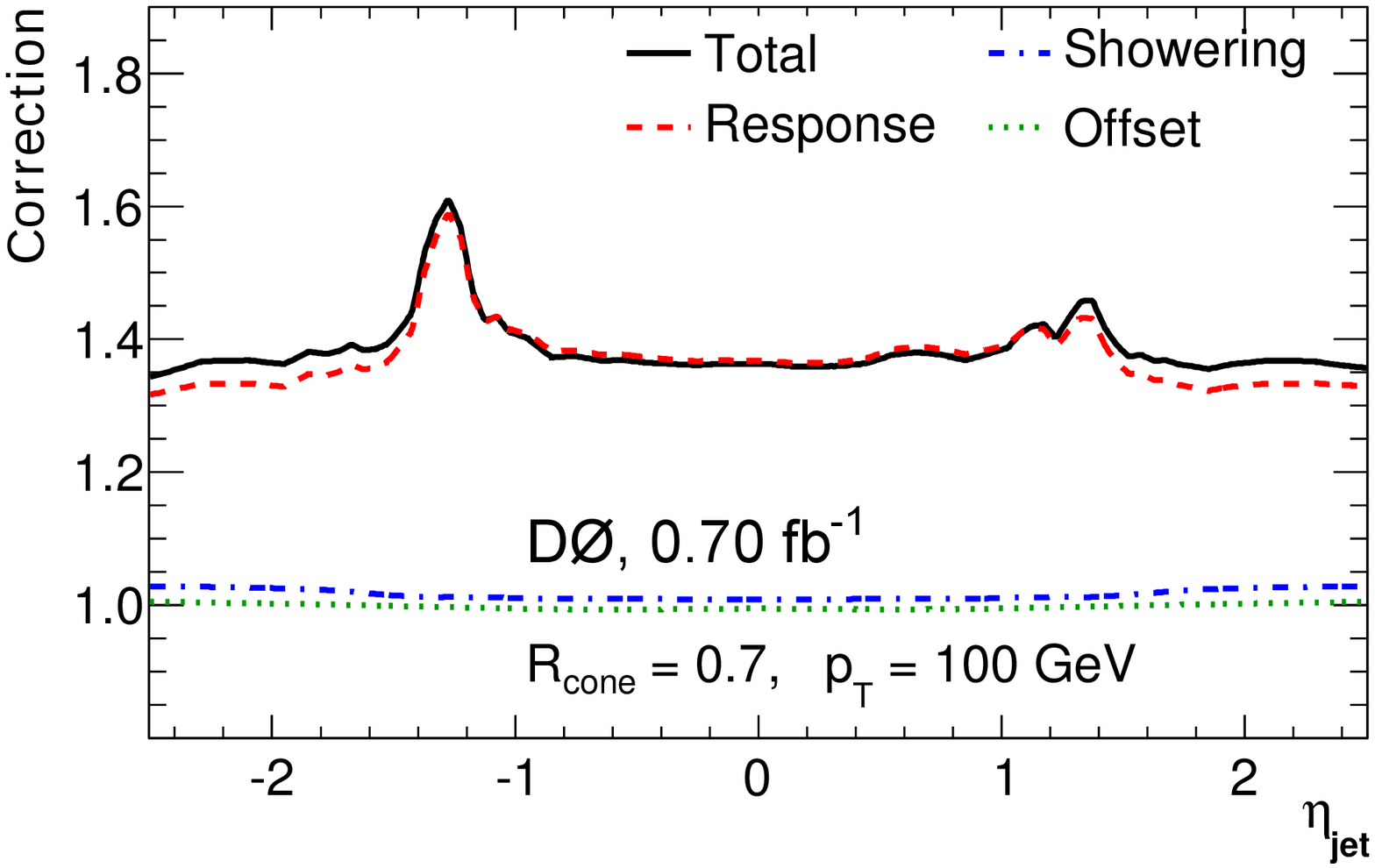}
\put(40,135){\textsf{(d)}}
\end{overpic}
\hspace{0.47\textwidth}\\
\caption{\label{jes} (color online) Jet energy scale corrections 
as a function of jet 
$p_T$ for (a) central and (b) forward rapidity, and as a function of jet 
rapidity for (c) low and (d) high jet $p_T$.}
\end{figure*}

\begin{figure*}
\begin{overpic}[width=0.49\textwidth]
{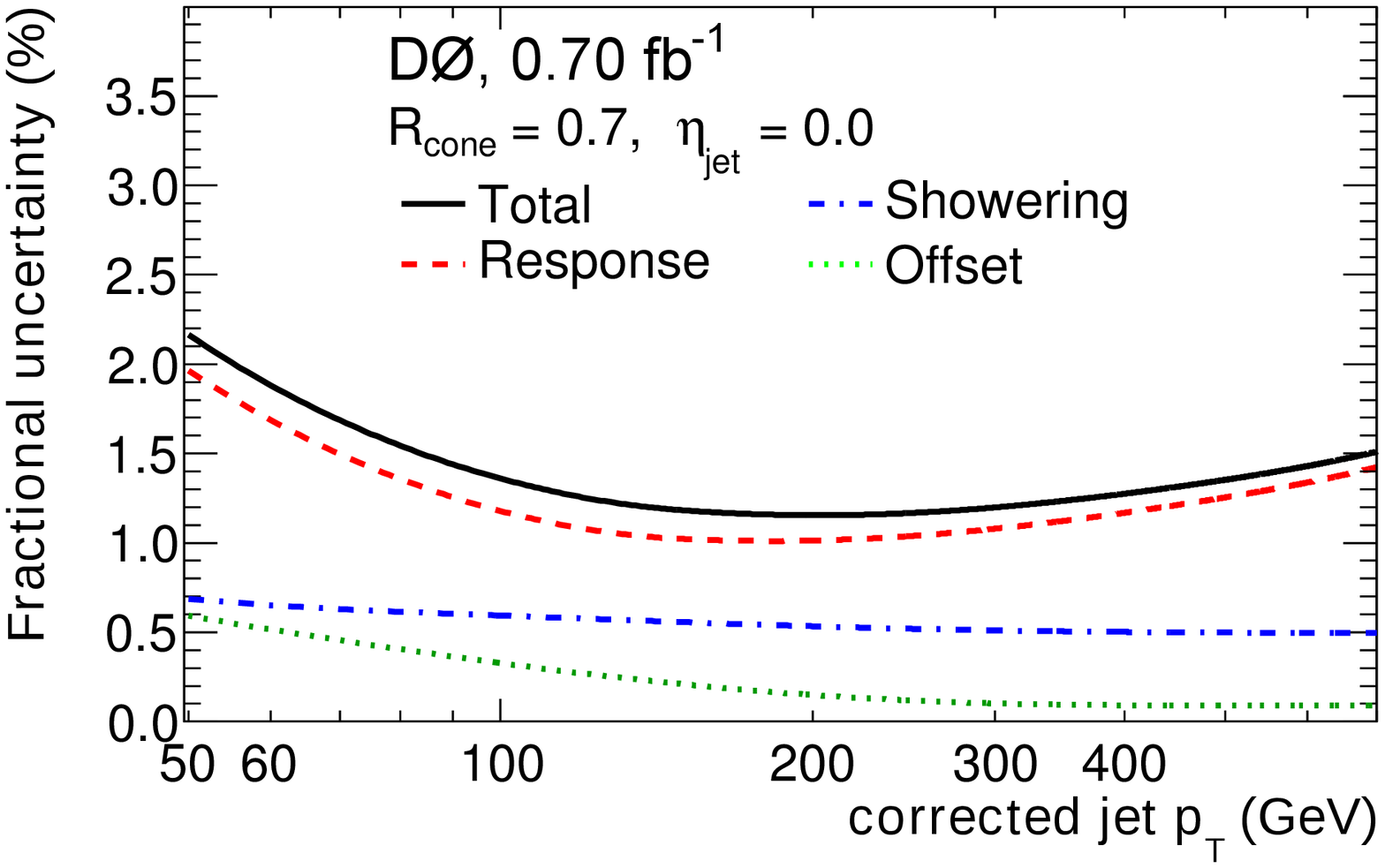}
\put(40,135){\textsf{(a)}}
\end{overpic}
\begin{overpic}[width=0.49\textwidth]
{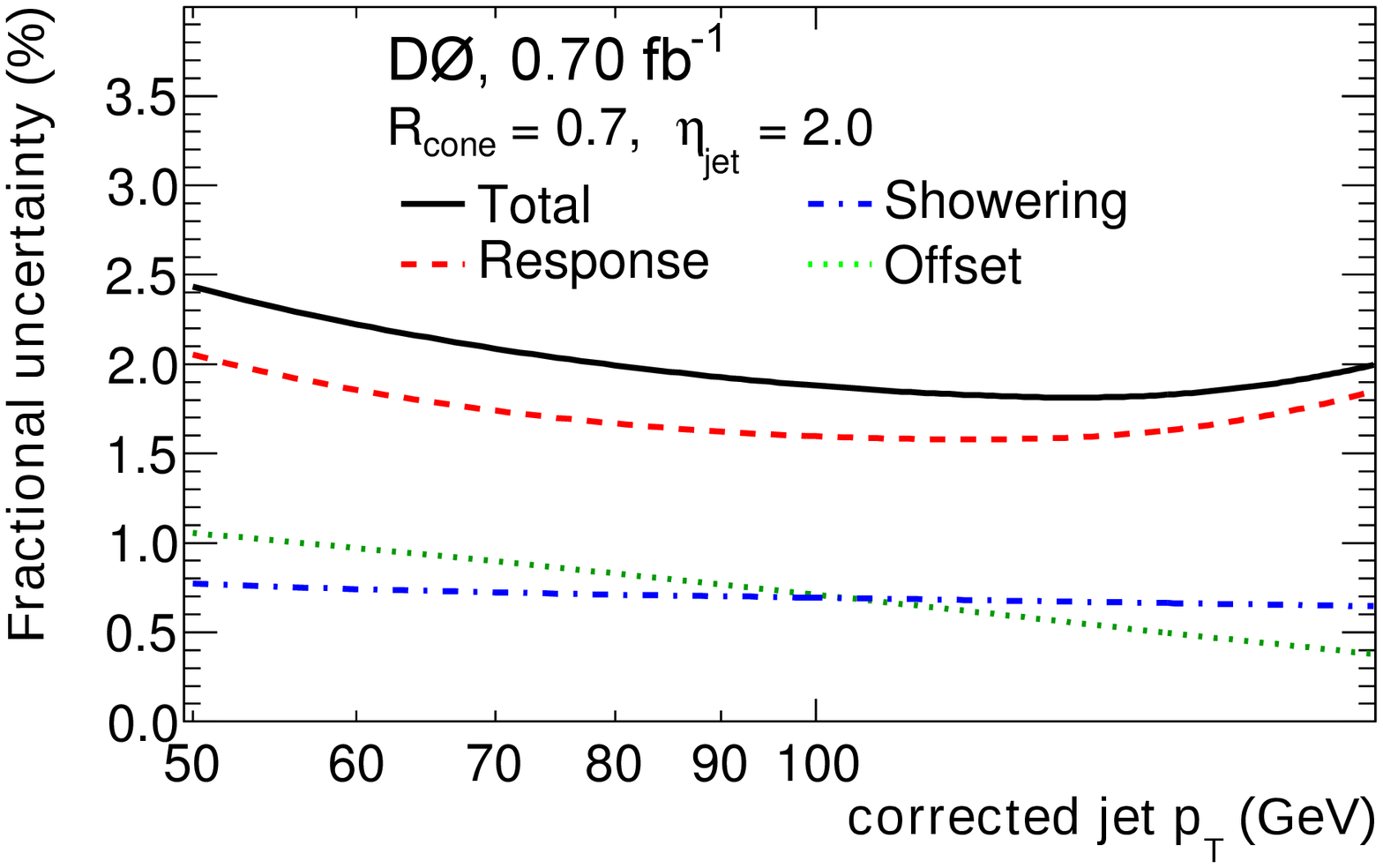}
\put(40,135){\textsf{(b)}}
\end{overpic}
\\
\begin{overpic}[width=0.49\textwidth]
{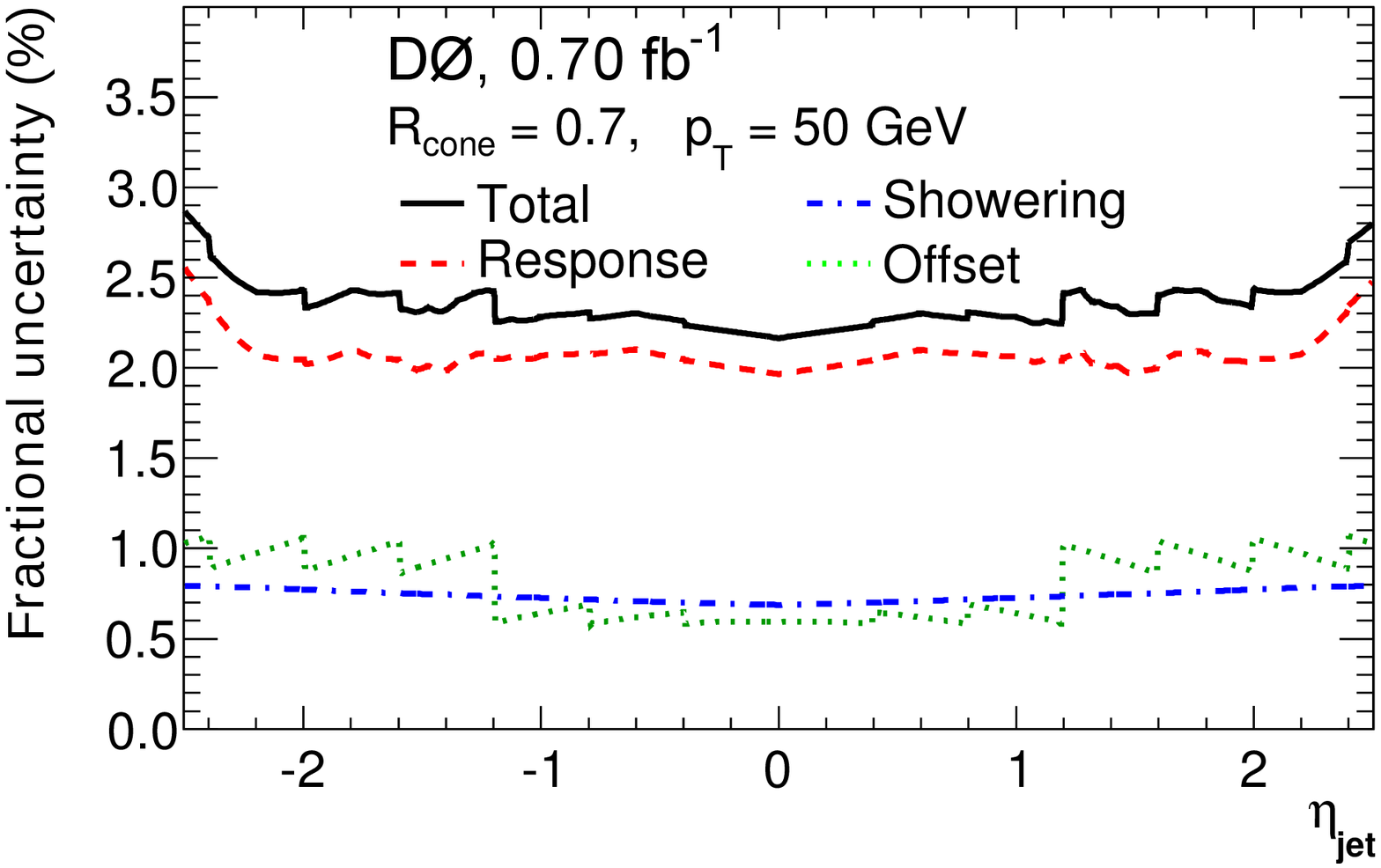}
\put(40,135){\textsf{(c)}}
\end{overpic}
\begin{overpic}[width=0.49\textwidth]
{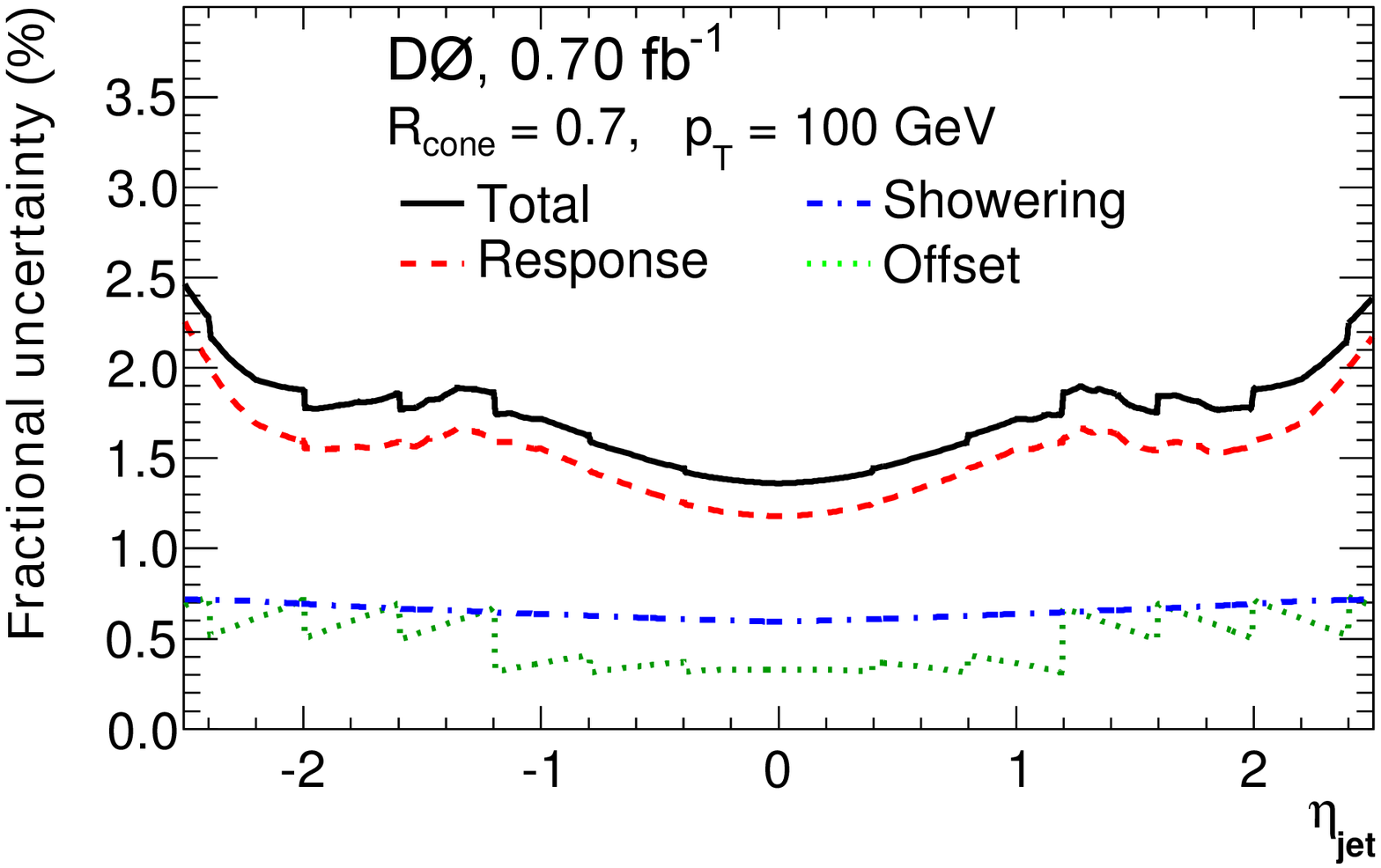}
\put(40,135){\textsf{(d)}}
\end{overpic}
\hspace{0.47\textwidth}\\
\caption{\label{jeserror} (color online) Jet energy scale uncertainties 
as a function of jet 
$p_T$ for (a) central and (b) forward rapidity, and as a function of jet 
pseudorapidity for (c) low and (d) high jet $p_T$.}
\end{figure*}

Figure~\ref{jes} shows the jet energy scale corrections as a function of jet 
$p_T$ for central and forward rapidity, and as a function of jet rapidity at 
low and high jet $p_T$. The corrections range between 1.2 and 1.8 for the 
kinematic range of the cross section measurement. The response correction is 
by far the largest one, while the showering correction starts to be 
noticeable at large rapidity. At
high rapidity, the actual angular distance for each $\Delta \eta$  bin is small,
while the radius of the showering is slightly increasing
due to the increasing energy of the jet at fixed $p_T$ as one goes forward.
The total correction is computed using Eq. 3.  The combined effects of
the uncertainties associated with each component of the correction are
summarized in Fig.~\ref{jeserror} as a 
function of jet $p_T$ for central and forward rapidity, and as a function of 
jet rapidity for low and high jet $p_T$ --- high jet energy corresponds
to low $p_T$ at high rapidity. 

The corrections do not show a significant dependence as a function 
of jet rapidity except in the region of the ICR. The uncertainties 
vary between 1.2 and 2.5\% for the kinematic range of the cross 
section measurement and are dominated by the uncertainties of the 
jet response. The uncertainties obtained in the CC and for 
jet $p_T\approx$ 100--500 GeV are the smallest ones obtained by any 
experiment operating at a hadron collider. These uncertainties 
do not depend strongly on jet pseudorapidity and $p_T$.

\subsection{Closure tests}

\begin{figure*}
\begin{overpic}[width=0.49\textwidth]
{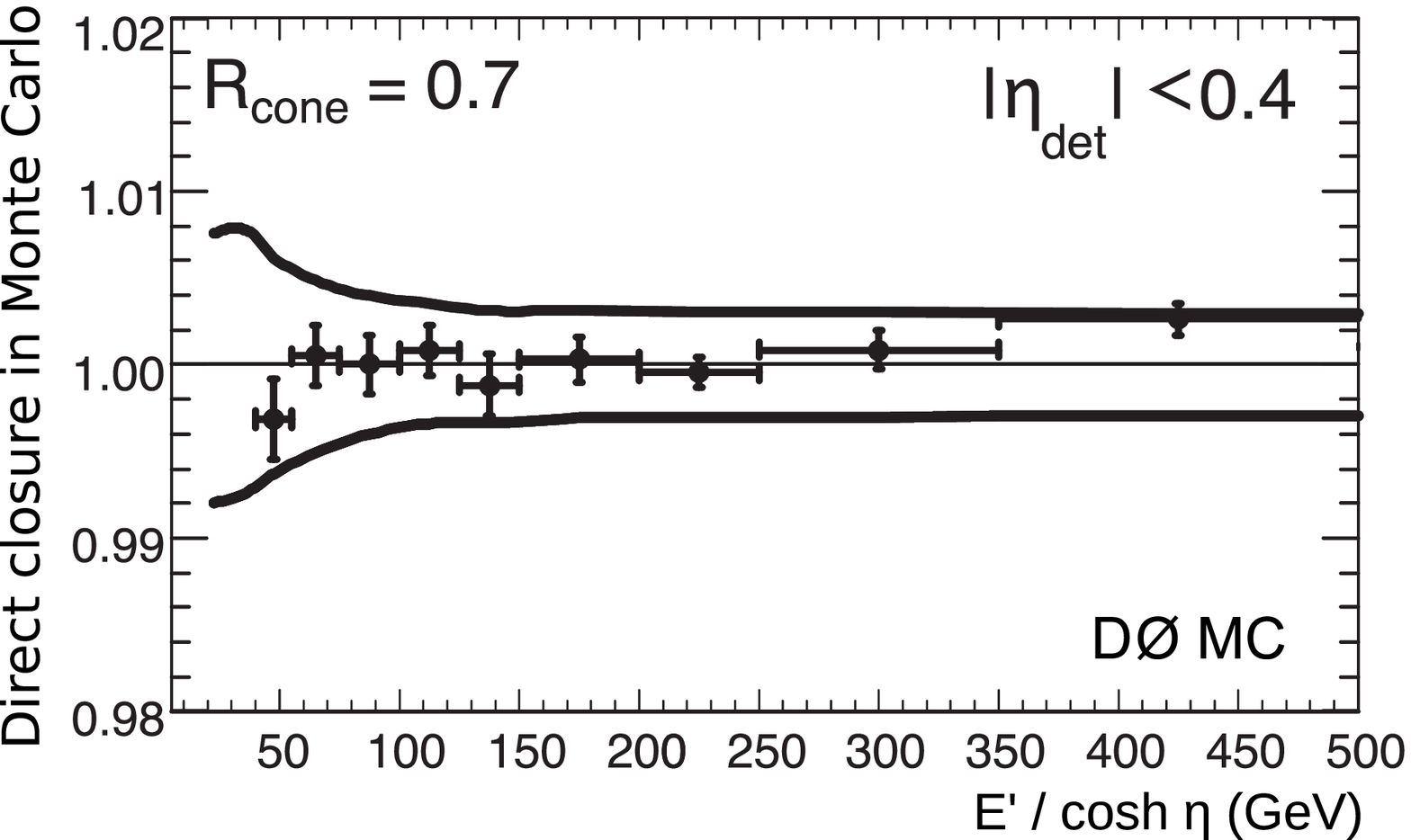}
\put(40,40){\textsf{(a)}}
\end{overpic}
\begin{overpic}[width=0.49\textwidth]
{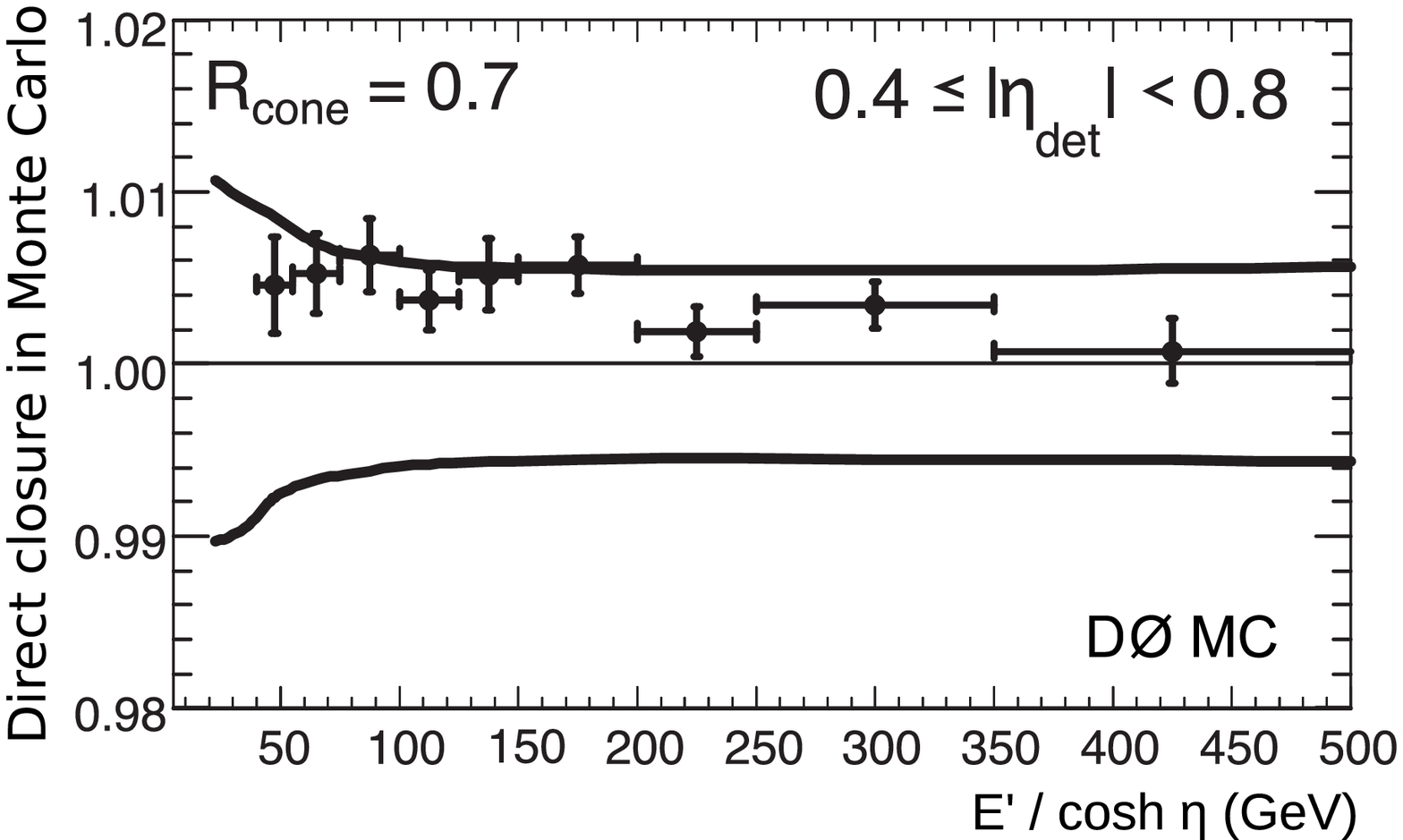}
\put(40,40){\textsf{(b)}}
\end{overpic}
\caption{\label{jesclosure} Closure test of jet energy scale in MC 
for (a) $|\eta_{det}|<0.4$ and (b) $0.4<|\eta_{det}|<0.8$. The band 
outlined by the solid curves corresponds to the uncertainties
in the extraction of jet energy scale in MC which are mainly statistical.}
\end{figure*}

The aim of the closure tests is to verify the accuracy of the 
jet energy scale correction using either MC or data and to evaluate the 
remaining difference as an additional systematic uncertainty related to the 
method. As an example, one test is to use the full method in MC and 
to compare the results with the particle level jet energy. The direct 
closure variable is defined as
\begin{equation}
\label{eq:direct}
D = \frac{\left<E^{\mathrm{corr}}_{\mathrm{jet}}\right>}
{\left<E^{\mathrm{particle}}_{\mathrm{jet}}\right>},
\end{equation}
where $E^{\mathrm{corr}}_{\mathrm{jet}}$ is the corrected jet energy and 
$E^{\mathrm{particle}}_{\mathrm{jet}}$ is the energy of the closest particle jet 
matching the reconstructed jet within $\Delta R < R_\mathrm{cone}/2$. Results
from the direct closure test are shown in Fig.~\ref{jesclosure} in two 
regions of jet rapidity. We note that we obtain consistency of the method
within statistical uncertainties (D is close to unity within less than 1\%) 
and no additional systematic uncertainty is introduced.
Closure tests using data are performed relative to MC by comparing 
ratios of fully corrected jet energies 
$<E_{data}^{corrected}>/<E_{MC}^{corrected}>$ in fixed regions of $E'$ and $\eta$.
Again we find good agreement within the expected uncertainties of the jet 
energy scale.

\section{Triggering on jets}\label{sec:trigger}

\begin{figure*}
\begin{overpic}[width=0.49\textwidth]
{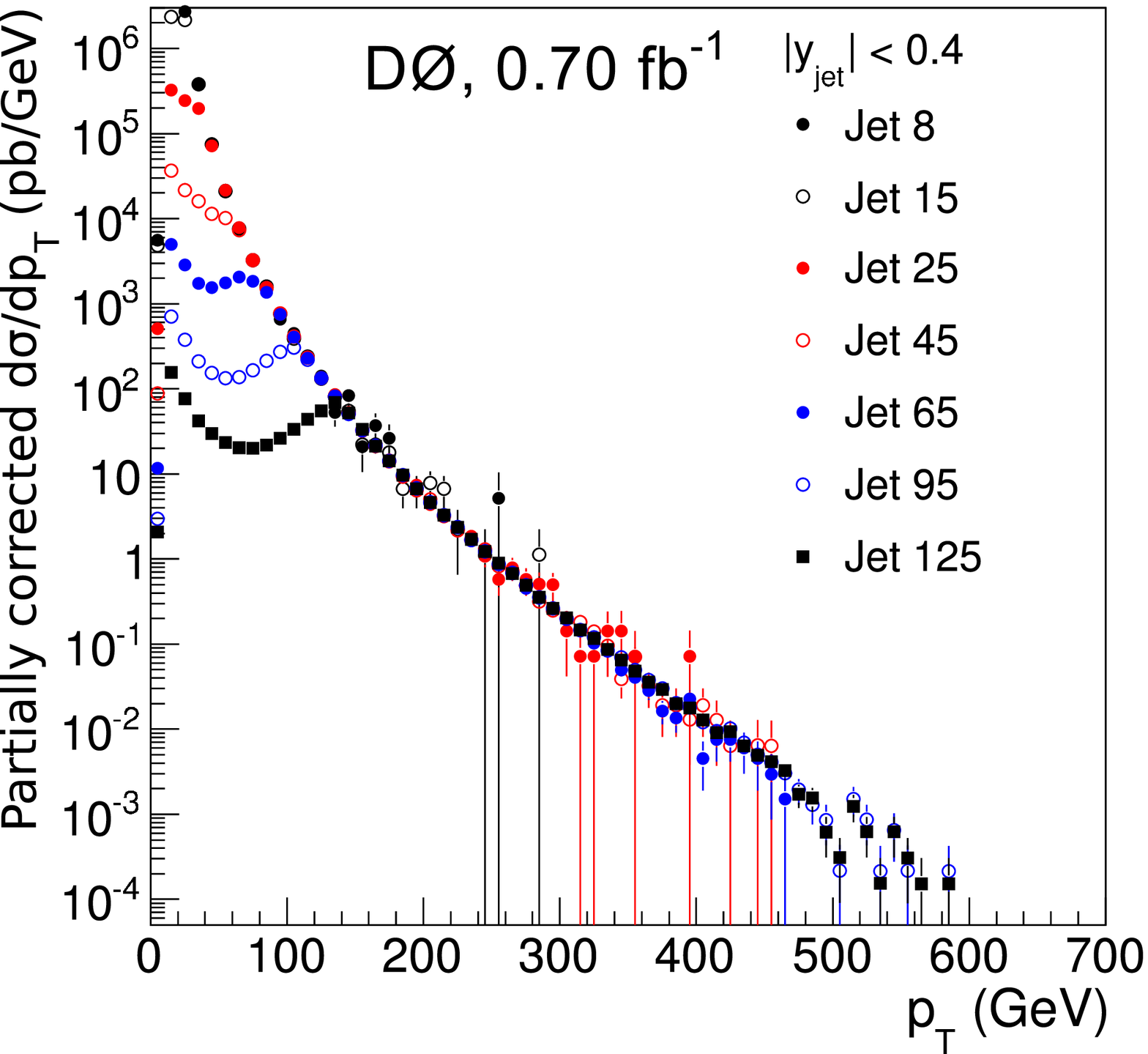}
\put(40,40){\textsf{(a)}}
\end{overpic}
\begin{overpic}[width=0.49\textwidth]
{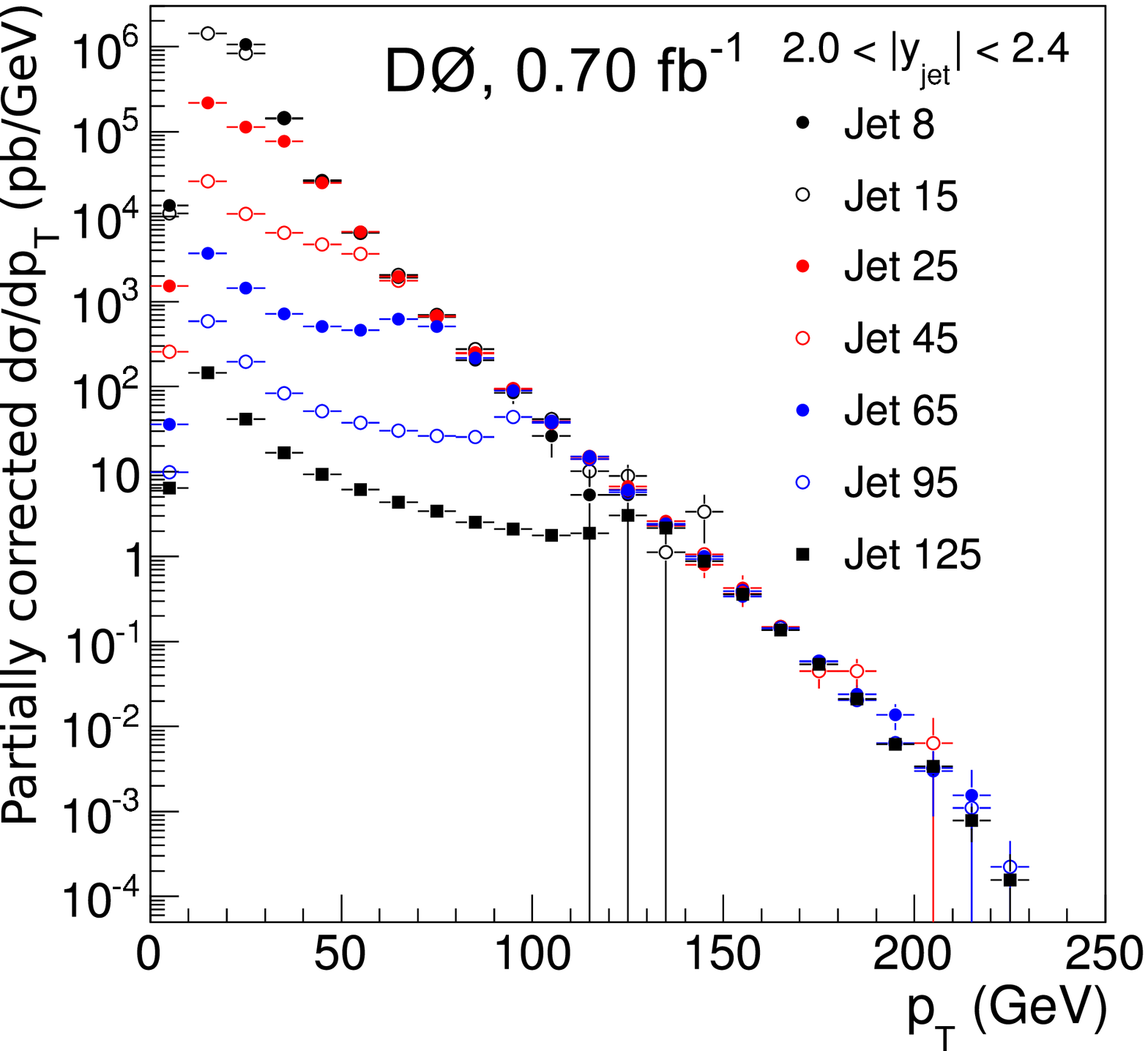}
\put(40,40){\textsf{(b)}}
\end{overpic}
\caption{\label{fig4} (color online) Inclusive jet $p_T$ cross section 
without unfolding corrections for the different single jet triggers as a 
function of jet $p_T$ for (a) $|y|<0.4$ and (b) $2.0<|y|<2.4$. The 
average prescales are 34000, 7100, 460, 41, 9.6,
1.4 and 1 for the 8, 15, 25, 45, 65, 95, and 125 GeV triggers, respectively.}
\end{figure*}

In this section, we briefly describe how we determine the absolute jet 
trigger efficiencies. Two different samples based on jet or muon triggers are 
used. The D0 trigger system is composed of three consecutive levels called 
L1, L2, and L3.  At L1, a single jet trigger typically requires $n$ 
calorimeter trigger towers above a given threshold, where a trigger 
tower is defined by the hardware summation of energies in $2\times 2$ 
calorimeter towers.  The trigger towers are read out separately from 
the precision calorimeter electronics via a fast digitizer and are 
used in both L1 and L2 triggers. 
All events used in this analysis are required to pass a trigger designed 
to fire if a single jet with $p_T>$ 50 GeV is in the event.
For instance, the 65 GeV single jet trigger requires 
the presence of three calorimeter towers with a transverse momentum
above 5 GeV. 
This requirement is often satisfied by the presence of trigger towers 
belonging to different jets, ensuring high trigger efficiency.  In most of 
these events, there are two high-$p_T$ jets in the event or more 
than two low-$p_T$ jets, which ensures that the event passes the L1 threshold. 
A detailed analysis shows that the L1 single jet efficiency is more than 
98\% for the full kinematic range of our measurement, which is corrected
for the residual inefficiency.  At L2 we perform a clustering of the trigger 
tower energies and apply a threshold based on the $p_T$ of highest energy 
cluster.  Seven L3 triggers corresponding to uncorrected L3 
jet $p_T$ thresholds of 8, 15, 25, 45, 65, 95, and 125 GeV are used in the 
analysis. The highest-$p_T$ L3 trigger was never prescaled during data 
collection. In 
Fig.~\ref{fig4}, we show the jet cross section before any unfolding 
corrections as a function of jet $p_T$ for the different jet triggers for 
two domains in jet rapidity $|y|<0.4$, and $2.0<|y|<2.4$.

\begin{figure}
\includegraphics[width=0.95\columnwidth]{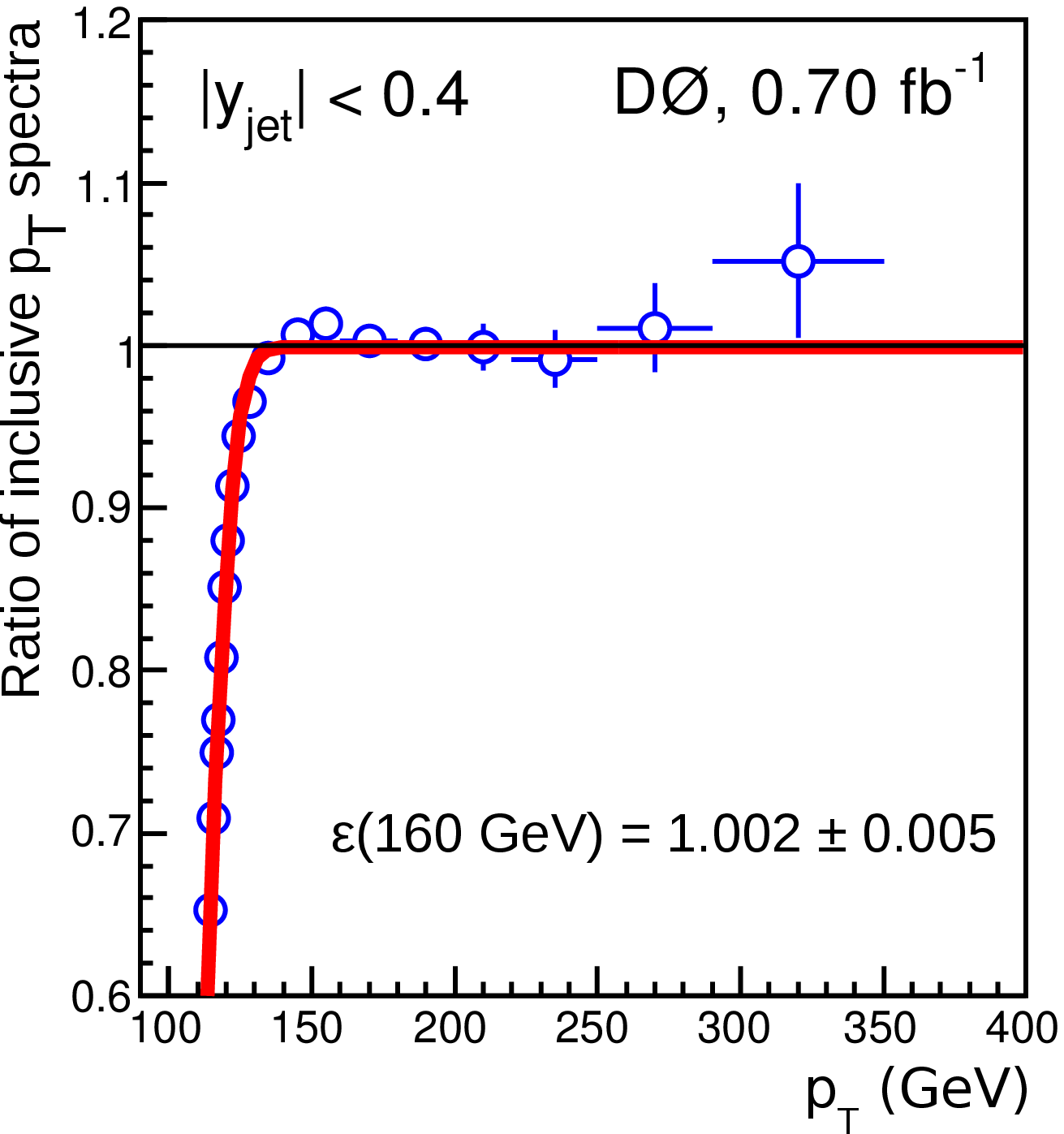}
\caption{\label{fig5} (color online)
Ratio of two consecutive jet triggers used to obtain the relative jet 
trigger efficiency for the 95 and 65 GeV single jet triggers. The fit of the 
turn-on curve determines the 95 GeV jet $p_T$ 99\%-threshold to be 
130 GeV. A 
higher threshold of 160 GeV that is consistent with an 
efficiency $\epsilon=1.00$ is used in the final analysis.}
\end{figure}

The first method used for computing the jet efficiency is to obtain 
the relative jet 
trigger efficiency with respect to the lower $p_T$ jet trigger. For instance, 
the ratio of the 95 and 65 GeV triggers is shown in Fig.~\ref{fig5}. For 
this purpose, we plot the ratio of the number of events that pass the 95 GeV 
trigger to those that pass the 65 GeV threshold as a function of jet 
$p_T$ after jet energy scale and vertex efficiency corrections to cancel known 
luminosity dependencies as discussed in Sec.~VII. 
The ratio is scaled by the relative integrated 
luminosities of these triggers to account for the different prescales. When 
this ratio reaches 1, the 95 GeV threshold trigger is 100\% efficient with 
respect to the 65 GeV one. A fit to this ratio gives the different 
thresholds for which the triggers are fully efficient ($>$99.9\%). 
The jet energy scale corrected $p_T$ at which each trigger becomes fully
efficient is given in Table~\ref{threshold}. These thresholds take into 
account the $p_T$ binning used in the 
analysis and can be significantly higher than the minimum usable threshold. 
We note that this method does not allow us to obtain the absolute trigger 
efficiency since it gives all efficiencies with respect to the lowest 
8 GeV $p_T$ trigger as a reference. 

A second method is used to measure the absolute single jet trigger efficiency.
It uses 
independent muon and minimum bias triggers. The minimum bias trigger 
only requires energy deposits in the luminosity monitors. As its name 
indicates, it shows very little selection bias and is ideal for trigger 
studies. 
Unfortunately, the sample collected during all of Run II at the Tevatron at 
0.5 Hz only yields statistics adequate to study jets below 70 GeV using 
this trigger, and this method does not allow exploration of the 
high $p_T$ jet trigger efficiency.  For this 
reason, inclusive muon triggers without any calorimeter requirements are also 
used. This allows us to check what fraction of the offline reconstructed
jets in muon triggered events pass the calorimeter jet
trigger requirement, providing a direct estimate of trigger efficiencies up to
400 GeV in the CC.
The conclusion of this study is that all jet triggers are more 
than 98\% efficient above the thresholds defined above, 
and the residual inefficiency is determined to a precision of better than 1\%. 
Both methods to obtain the trigger efficiencies are 
useful since the muon triggers have a tendency to enrich the inclusive jet 
samples in $b$ and $c$-jets where the $b$ and $c$ quarks decay leptonically, 
which might lead to different trigger efficiencies as a function of jet $p_T$. 

\begin{table}[htb!]

\begin{tabular}{|ccccccc|}
\hline \hline
Rapidity / L3 trigger  & 15 & 25 &  45 &  65 &  95 & 125 \\ 
$|y|<0.4$     & 50 & 60 & 100 & 120 & 160 & 200 \\
$0.4\le|y|<0.8$ & 50 & 60 & 100 & 120 & 160 & 200 \\
$0.8\le|y|<1.2$ & 50 & 90 & 110 & 140 & 190 & 230 \\
$1.2\le|y|<1.6$ & 50 & 80 &  90 & 140 & 190 & 240 \\
$1.6\le|y|<2.0$ & 50 & 70 &  90 & 110 & 160 & 190 \\
$2.0\le|y|<2.4$ & 50 & 70 &  90 & 120 & 160 & 200 \\
\hline \hline
\end{tabular}
\caption[Trigger $p_T$ thresholds in GeV used in the final analysis]
{\label{threshold}
Jet energy scale corrected $p_T$ in GeV at which each L3 trigger becomes 
fully efficient in different jet $y$ bins.}
\end{table}

\section{Event selections and efficiencies}

In this section, we discuss the selections that are used to remove 
background events in the sample. The selections fall into three different 
categories. The event quality flags remove events 
suffering from diverse calorimeter noise issues. The vertex requirement 
selects events with a high quality vertex 
close to the center of the calorimeter to improve the jet $p_T$ and $y$ 
measurements and to reduce the background from cosmic ray
events. The $\met$ requirement is designed to remove the remaining 
cosmic ray background, especially at high jet $p_T$.

\subsection{Event quality flags}

Event quality flags ensure that the subdetectors used in the analysis were 
working properly when the data were collected. Calorimeter event quality 
flags allow removal of events showing coherent pedestal shifts in the 
analog-to-digital converters, parts of the calorimeter not correctly read out, 
or high coherent noise. This is especially important for high-$p_T$ jets which 
can originate artificially from noisy towers in the calorimeter. Note that 
the vertex and $\met$ requirements also remove most of these events. 
The inefficiency induced by the calorimeter event quality flag rejection is 
estimated using an independent sample whose trigger is known to be 
unaffected by the calorimeter problems, the zero bias trigger. The 
inefficiency is calculated to be (3.2 $\pm$ 1.0)\% where the 1.0\% 
uncertainty covers the time and luminosity dependence of the inefficiency.

\subsection{Reconstructed vertex requirement}

The vertex selection is based on three different requirements: there must 
be at least one reconstructed vertex, the $z$-position along 
the beam line of the primary reconstructed vertex must be within 50 cm 
of the detector center
($|z_{\mathrm{vertex}}|<50$ cm), and the number of tracks fitted to the 
vertex has to be at least three to ensure an accurate measurement. 
The $z$-vertex position requirement ensures that the vertex is in 
the high efficiency tracking region.  The third requirement rejects 
vertices originating from fake high $p_T$ tracks. To each reconstructed 
vertex is assigned a probability that it comes from a minimum 
bias interaction based on the $\ln (p_T)$ distributions of the tracks 
with $p_T >$ 0.5 GeV pointing to the vertex.
The vertex with the lowest minimum bias probability is selected
as the primary vertex.

The efficiency of reconstructing a vertex with at least three tracks pointing 
to it (without the requirement on the $z$-vertex position) is found to 
be (99.6 $\pm$ 0.4)\%, independent of jet $p_T$ and $y$. The 
observed 0.4\% inefficiency is 
consistent with about 0.6\% of the primary vertices not being reconstructed 
because of tracking inefficiencies, and 0.2\% being replaced by a minimum 
bias vertex.

\begin{figure}
\includegraphics[width=\columnwidth]{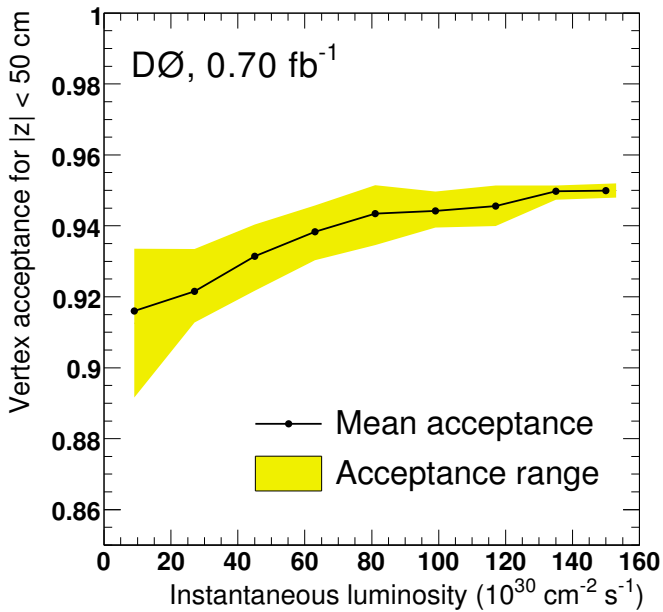}
\caption{\label{fig6} (color online)
Vertex acceptance for the requirement on the $z$-vertex position 
$|z_{\mathrm{vertex}}|<50$~cm as a function of instantaneous luminosity. 
The shaded band indicates the variation for different running periods.}
\end{figure}

The leading inefficiency comes from the requirement on the vertex position 
along the $z$-axis. The 
fraction of events rejected by this requirement is of the order of 7\%. To  
determine the efficiency of this requirement we take into account the shape 
of the luminous region. The longitudinal shape of the luminous region is 
approximated by the expression
\begin{equation}
\frac{d\mathcal{L}(z)}{dz} = N_pN_{\bar p}\frac{1}{\sqrt{2\pi}\sigma_z}
\frac{e^{-(z-z_{0z})^2/2 \sigma_z^2}}{4\pi\sigma_x(z)\sigma_y(z)},
\end{equation}
where the overlap of the proton and antiproton beam bunches having $N_p$ and 
$N_{\bar p}$ particles is described with a Gaussian distribution of width 
$\sigma_z$ in the $z$ direction, with a possible offset $z_{0z}$ relative to 
the nominal interaction point. $\sigma_x(z)$ and $\sigma_y(z)$ represent 
the transverse size of the  beam spot and vary as a function of $z$:
\begin{equation}
\sigma_T^2(z) = \frac{1}{6\pi\gamma}\epsilon_T\beta_T^*\left[1+
\frac{(z-z_{0T})^2}{\beta_T^{*2}}\right].
\end{equation}
Here $T$ is either $x$ or $y$, $z_{0T}$ is the minimum of the 
$\beta$ function describing the beam dimensions near the interaction 
point in direction $T$ and any offset in the $x$ and $y$ 
directions with respect to the nominal interaction point, $\gamma$ is the 
Lorentz factor of the beam  particles. The emittance $\epsilon_T$ and beta 
parameter $\beta_T^*$ describe the beam dimensions at the interaction 
point. The parameterization can be integrated to yield
\begin{equation}
\epsilon_{|z_\mathrm{vertex}|<50\mathrm{~cm}} =
 \frac{\int_{-50\mathrm{~cm}}^{50\mathrm{~cm}} f(z_\mathrm{vertex},\mathrm{run},
 \mathcal{L})}{\int_{-140\mathrm{~cm}}^{140\mathrm{~cm}} f(z_\mathrm{vertex},
 \mathrm{run},\mathcal{L})},
\end{equation}
where the limits of integration in the denominator come from the requirements 
used in the luminosity determination.
This parameterization is fitted to minimum bias data in the high tracking 
efficiency region ($|z_\mathrm{vertex}|<$ 40--60~cm) in bins of instantaneous 
luminosity for several run ranges
(the changes in beam optics as a function of time affect the beam shape as 
described by the $\beta^*$ parameter). The changes as a function of 
instantaneous luminosity are primarily due to the variations in the beam 
parameters during a 
store. The vertex efficiency varies by up to 6\% as a function of 
instantaneous luminosity and by up to 4\% as a function of the period of data 
taking for a fixed value of luminosity. The parameterizations have been 
determined as a function of time and instantaneous luminosity, and are applied 
as such on a per-event basis.  Figure~\ref{fig6} shows the mean vertex 
efficiency as a function of instantaneous luminosity, with the range of 
efficiencies overlaid. The uncertainty on the vertex acceptance is estimated 
to be 0.5\% by comparing results from fits to minimum bias data at 
$|z_\mathrm{vertex}|<60$~cm and $|z_\mathrm{vertex}|<40$~cm. In addition, an 
increased uncertainty of 0.4\% added in quadrature at high $|y|$ is introduced to 
account for the possibility of a lower vertex reconstruction efficiency.

\subsection{Missing transverse energy requirement}

\begin{figure}
\includegraphics[width=\columnwidth]{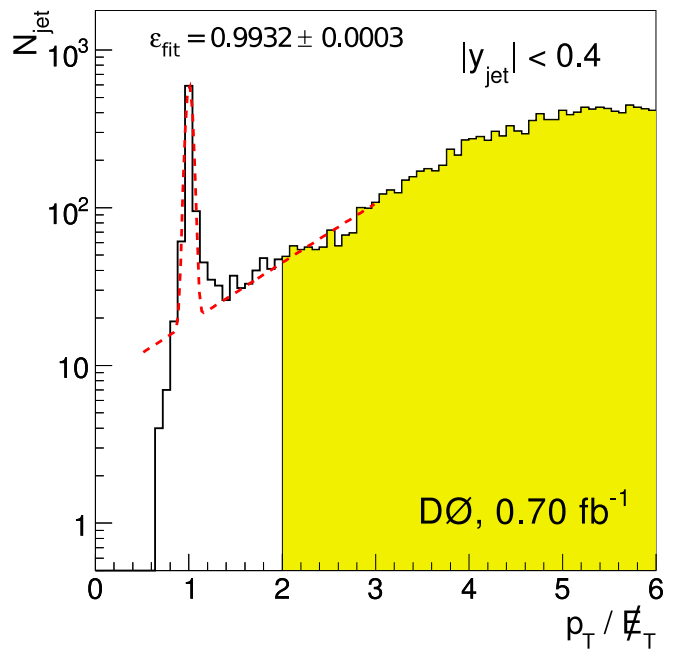}
\caption{\label{fig7} (color online) 
Distribution of $p_{T,\mathrm{lead}}/\met$ for 
jet events with leading jet $p_T>200$ GeV. A peak from cosmic ray background is 
visible around 1. The shaded region shows jets passing the $p_{T,\mathrm{lead}}/\met$  
requirement.}
\end{figure}

A requirement on the missing transverse energy in an event is applied to remove 
the remaining background from cosmic rays that induce showers in the 
calorimeter.  The cross section for these cosmic ray interactions falls 
much less steeply versus $p_T$ than the inclusive jet cross section, and is 
typically comparable at $p_T \approx$ 400 GeV. The issue of background
from cosmic rays is thus more important for high-$p_T$ jets. Fortunately, 
cosmic ray showers deposit most of their energy on one side of the 
calorimeter, have no reconstructed vertex, and produce high uncorrected 
$\met$ that peaks at $p_{T,\mathrm{lead}}/\met\approx 1$ where 
$p_{T,\mathrm{lead}}$ is the uncorrected $p_T$ of the leading 
jet of the event. These events are fully and efficiently removed by requiring
$p_{T,\mathrm{lead}}/\met>1.4$, when $p_{T,\mathrm{lead}}<100$~GeV and 
$p_{T,\mathrm{lead}}/\met>2.0$, when $p_{T,\mathrm{lead}} \ge 100$~GeV. 
Figure~\ref{fig7} shows the distribution of $p_{T,\mathrm{lead}}/\met$ for 
the high-$p_T$ jet trigger with $p_T>200$~GeV, with the selected events at 
$p_{T,\mathrm{lead}}/\met>2.0$ shown by the shaded region. 
A spike coming from cosmic ray events 
is visible at 1. An upper limit of 0.4\% is estimated on the inefficiency
of the $\met$ requirement 
and used as an uncertainty, but no correction is applied. This 
upper limit is based on studies of fits of distributions like the one in 
Fig.~\ref{fig7}, 
and track-matching inefficiency for jets since cosmic ray events are 
usually out-of-time with the tracking read-out.

\section{Jet identification requirements and efficiencies}

The jet identification requirements are designed to remove instrumental 
backgrounds such as jets formed from sources of transient noise in the 
calorimeter and also physics background from electrons and photons. 
The jet requirements are based on the 
fractions of jet energy deposited in the electromagnetic calorimeter (EMF) 
and in the coarse hadronic calorimeter (CHF). EMF$<$ 0.95 is 
required to remove overlaps between jets and electromagnetic objects, i.e. 
electrons and photons. This retains true jets with a 99\% efficiency. 
A lower limit on EMF (either 0 or varying between 0.03 and 0.05 depending 
on the pseudorapidity region in the calorimeter) as well as an upper limit 
on CHF (varying between 0.4 and 0.6) removes jets that are formed 
predominantly out of 
noise in the hadronic calorimeter. An additional requirement, L1 confirmation, 
is based on the ratio of the $p_T$ as measured by the L1 trigger system and 
as measured by the precision read-out. It is required to be above 0.5 for 
jet $p_T<$ 80 GeV, and there is no requirement for higher $p_T$ jets. This 
removes jets formed out of noise, for example due to coherent noise in the 
precision readout electronics.

\begin{figure}
\includegraphics[width=\columnwidth]{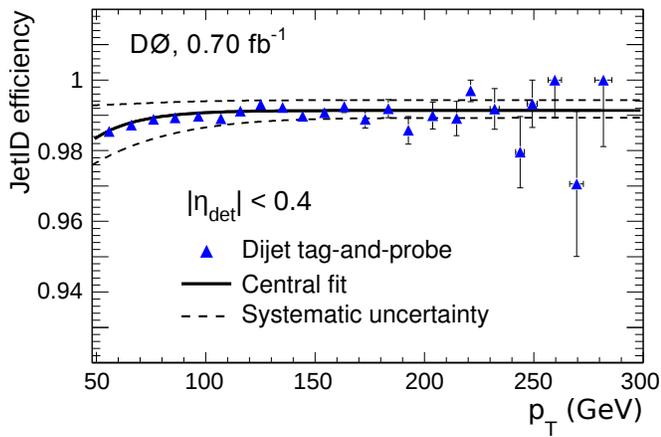} 
\caption{\label{fig8} (color online) Jet identification efficiencies 
obtained for the dijet sample. Dashed lines indicate the systematic 
uncertainty.}
\end{figure}

The jet identification efficiencies are determined using a
data driven method.
This method uses track jets which are jets built 
with a cone algorithm using charged particle tracks instead of calorimeter 
energy clusters. We select a leading $p_T$ tagged object, which in this 
case is a photon or a track jet associated with a good calorimeter jet, 
and a probe object, which is the leading track jet that is back-to-back 
in $\phi$ with the tag object.  
Events with additional track jets are
vetoed to ensure that the leading objects are balanced in $p_T$. The 
reconstruction efficiency is defined as 
the fraction of probe objects with a calorimeter jet found within the 
0.7 jet cone, and the jet identification efficiency is the fraction of those 
calorimeter jets passing the jet identification requirements. The data 
driven method has been used for three different samples: dijet, $\gamma+$jet 
and, $Z+$jet, which all lead to the same result. The central value for the 
jet identification efficiency shown in Fig.~\ref{fig8} is taken from the 
dijet sample. The efficiency for $p_T>50$ GeV, where we perform the 
measurement of the inclusive jet $p_T$ cross section, is 99\% in all 
calorimeter regions except in the region $0.8  <|y|< 1.2$ where it 
is about 98\%. 

Because the data driven method is used for calorimeter jets that are 
independently identified as track jets, we also 
directly measure the efficiencies by computing the fraction of events removed 
by each jet identification requirement individually after applying all other 
requirements in the inclusive jet sample. This method assumes that each 
jet identification cut removes only good jets. The efficiencies described
above are found to be in good agreement with those from the tag-and-probe 
method.

\section{Jet transverse momentum resolution}\label{sec:resolutions}

In this section, we discuss the determination of the jet $p_T$ resolution, 
which is needed for the unfolding of the inclusive jet $p_T$ cross section. 
The jet $p_T$ resolution is determined from data using the dijet asymmetry 
distribution, which can be obtained with minimal input from MC. This 
method requires corrections for the presence of additional unreconstructed 
jets (soft radiation), momentum imbalance at the particle level, and asymmetry 
bias due to non-Gaussian tails. We describe each correction 
needed to obtain the jet $p_T$ resolution.

\subsection{Dijet asymmetry}

\begin{figure}
\includegraphics[width=\columnwidth]{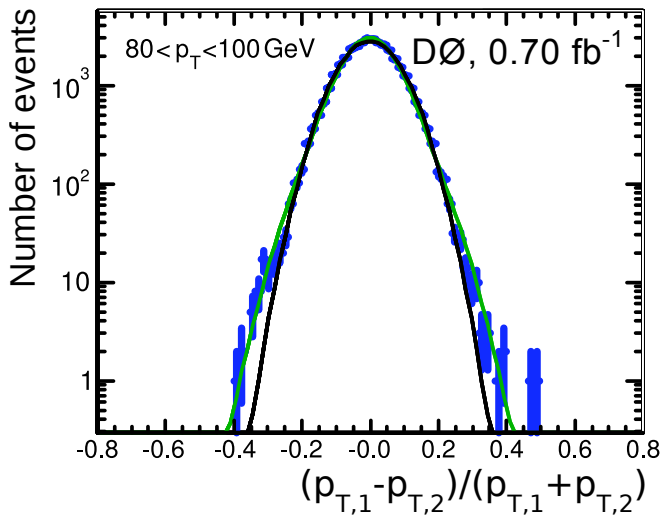}
\caption{\label{asymmetry} (color online)
Asymmetry distribution for jets in the central 
region with $80< p_T <100$~GeV. 
The probe jet is at $|y|<0.4$, the reference jet at $|y_\mathrm{ref}|<0.8$. 
The two lines display
the result of a Gaussian fit and a Gaussian with smeared exponential tails (see
Sec.~\ref{finalres}).}
\end{figure}

The jet $p_T$ resolutions are determined starting from the dijet asymmetry
\begin{equation}
A = \frac{p_{T,1} - p_{T,2}}{p_{T,1} + p_{T,2}}
\end{equation}
computed in a pure dijet sample with no additional jet identified, where
$p_{T,1}$ and $p_{T,2}$ are the $p_T$ of the leading and second-leading
jets and the two leading jets are randomly
assigned an index of 1 or 2. Both jets are 
required to be back-to-back with $\Delta\phi>3.0$ to avoid any large 
effects from QCD radiation.  The RMS of 
the asymmetry distribution is directly proportional to the jet $p_T$ 
resolution
\begin{equation}
\sigma_A = \frac{1}{\sqrt{2}}\frac{\sigma_{p_T}}{p_T},
\end{equation}
if the jets are in the same $y$ region to ensure that the $p_T$ resolution of 
both jets is the same.  To characterize the $p_T$ dependence of the 
resolution for a single jet,  $\sigma_A$ is measured in bins of 
$p_T = (p_{T,1} + p_{T,2})/2$.  This method can be used directly to 
measure the jet $p_T$ resolution in the central region where the statistics 
are high. However, in the forward region, the statistics for forward-forward 
jet pairs is small compared to central-forward jet pairs. If one of 
the jets is in the central region and the other in the forward region, it 
is possible to infer the jet $p_T$ resolution $\sigma_{p_T}$ in the 
forward region once the resolution for jets in the central reference 
region $\sigma_\mathrm{ref}$ is known
\begin{eqnarray}
\frac{\sigma_{p_T}}{p_T} &=& 
\sqrt{4\cdot \sigma_A^2 - \left(\frac{\sigma_\mathrm{ref}}{p_T}\right)^2}.
\end{eqnarray}
The central reference region used in this study is $|y_\mathrm{ref}|<0.8$, 
with the probe jet binning following the same 0.4 binning in rapidity as the 
rest of the analysis. The asymmetry distribution in the central region is 
shown in Fig.~\ref{asymmetry} for 
$80 < p_T < 100$~GeV
as an example and other $p_T^{jet}$ bins also show similarly small 
non-gaussian tails.

\subsection{Corrections to the resolution}
The jet $p_T$ resolution determined from the dijet asymmetry can be affected 
by physics and instrumental effects.  The final parameterization 
of the resolution used in this analysis includes corrections to remove 
biases in the measurement as described below.

\subsubsection{Soft radiation corrections}

\begin{figure*}
\begin{overpic}[width=0.49\textwidth]
{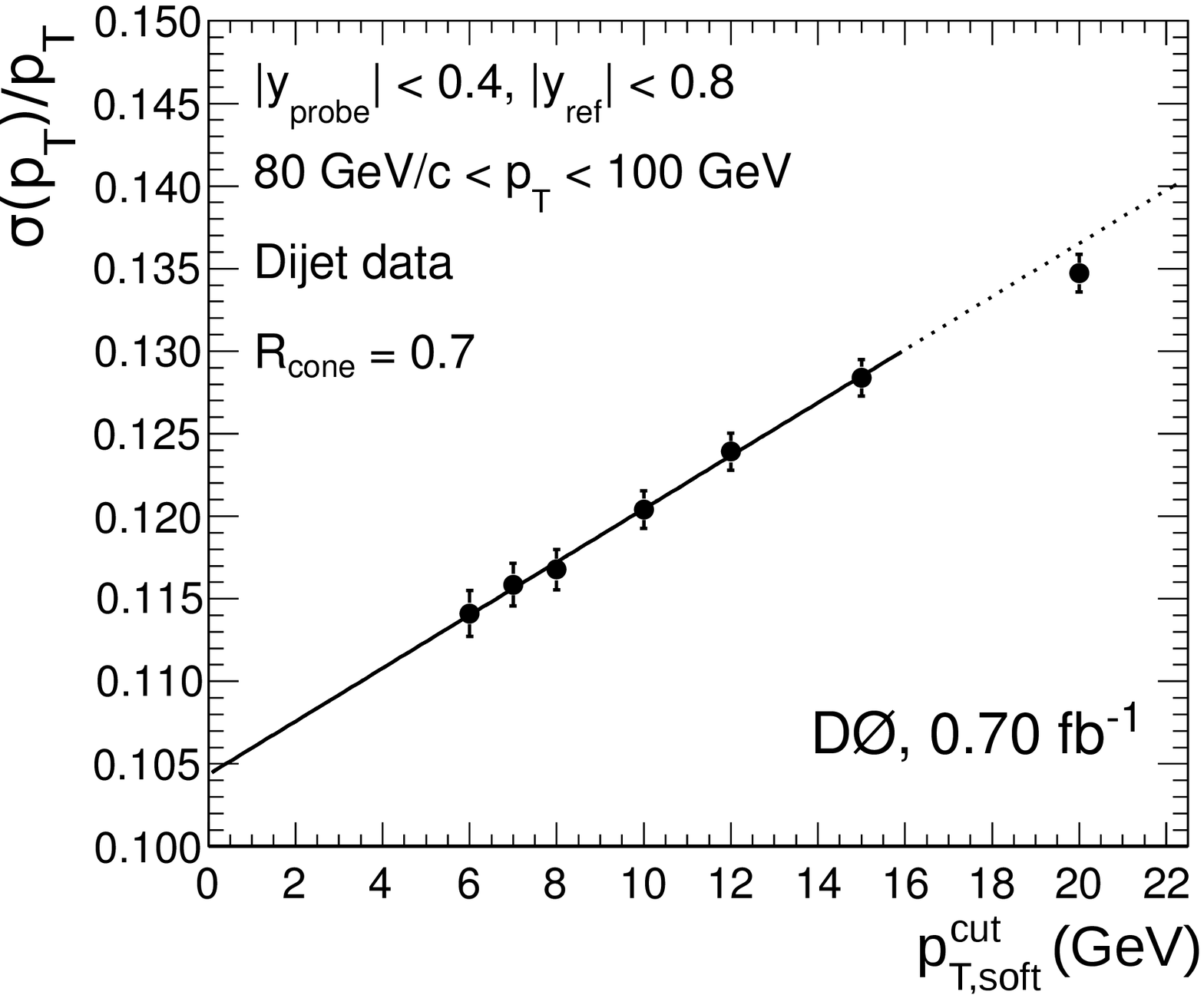}
\put(220,180){\textsf{(a)}}
\end{overpic}
\begin{overpic}[width=0.49\textwidth]
{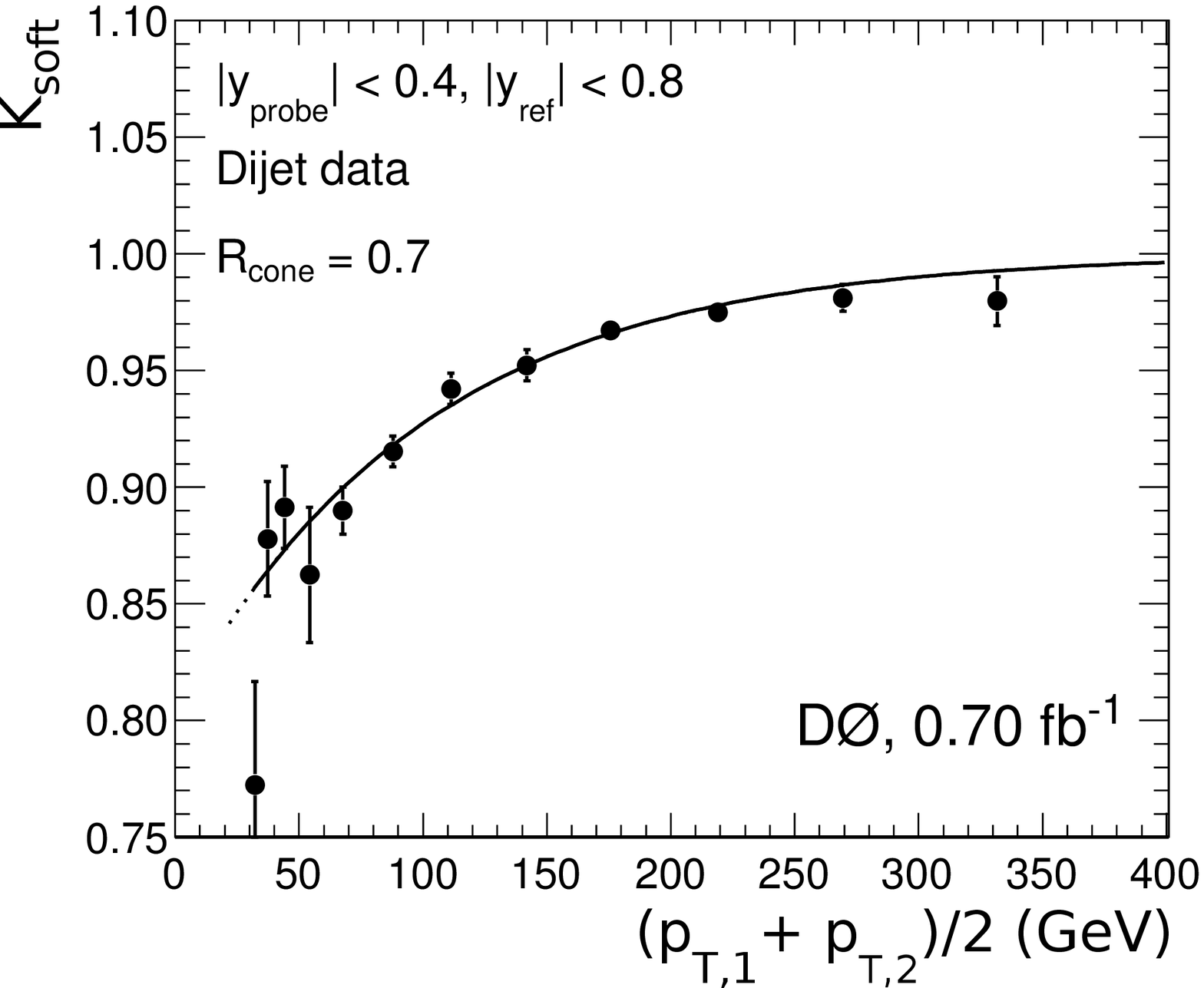}
\put(220,180){\textsf{(b)}}
\end{overpic}
\caption{\label{soft} (a) Jet $p_T$ resolution extrapolated to a jet $p_T$ 
reconstruction threshold of 0 GeV (in the $80<p_T<100$ GeV and $|y|<0.4$ 
bin).
(b) Soft radiation correction factor as a function of the average jet $p_T$ 
for the $0.4<|y|<0.8$ bin. The solid lines are the fit result and 
the dashed lines are the result of the extrapolations. }
\end{figure*}

The asymmetry method to compute the jet $p_T$ resolution is biased by the
presence of non-reconstructed jets in the sample. The $p_T$ threshold to 
reconstruct a jet is 6 GeV, and requesting the presence of only two jets in 
the sample to compute the asymmetry does not ensure the absence of
jets with $p_T$ below 6 GeV. The corrections for such soft radiation are 
determined directly in data. We compute the asymmetry and the jet 
$p_T$ resolution for different $p_T$ thresholds for jet reconstruction, 
namely 7, 8, 10, 12, 15, 20, and 40 GeV. The jet $p_T$ resolution as a 
function of the jet reconstruction threshold is shown in 
Fig.~\ref{soft} (a) for one bin in jet $p_T$ and $|y|$. A linear fit 
allows for extrapolating the jet $p_T$ resolution to a threshold $p_T$ of 0. 
The soft radiation factor, 
\begin{equation}
K_\mathrm{soft} = 
\frac{\sigma_{p_T}(p_{T,\mathrm{soft}}^\mathrm{~cut}\rightarrow 0)/p_T}
{\sigma_{p_T}(p_{T,\mathrm{soft}}^\mathrm{~cut}=6\mathrm{~GeV})/p_T},
\end{equation}
is studied as a function of the average jet $p_T$ in each $|y|$ bin as 
illustrated in Fig.~\ref{soft} (b). To better describe the low $p_T$ region 
and limit the statistical fluctuations, the dependency of $K_\mathrm{soft}$ 
versus $p_T$ is 
fitted with
\begin{equation}
K_\mathrm{soft}(p_T) = 1 - \exp(-p_0 - p_1 p_T),
\end{equation}
where $p_0$ and $p_1$ are two parameters of the fit.

\subsubsection{Particle imbalance and combined corrections}

The remaining correction needed to obtain the final jet $p_T$ resolution 
is the particle imbalance correction. Even in the ideal situation of only two 
particle jets and no soft radiation, the two jets are not necessarily perfectly 
balanced. In particular, fragmentation effects cause some energy and 
$p_T$ to be found outside the jet cone. This effect is purely related to 
QCD and is determined using a MC simulation. The particle level 
imbalance is corrected for soft radiation using the same method as introduced 
for data
\begin{eqnarray}
K_\mathrm{soft}^\mathrm{MC} &=& \frac{\sigma_{p_T}^\mathrm{ptcl}
(p_{T,\mathrm{ptcl}}^\mathrm{threshold}\rightarrow 0)/p_T}
{\sigma_{p_T}^\mathrm{ptcl}(p_{T,\mathrm{ptcl}}^\mathrm{threshold} =
6\mathrm{~GeV})/p_T},\\
\sigma_\mathrm{MC} &=& K_\mathrm{soft}^\mathrm{MC}\cdot\sigma_{p_T}^\mathrm{ptcl},
\end{eqnarray}
where $\sigma_{p_T}^\mathrm{ptcl}$ is the resolution evaluated at the 
particle level in the MC and $p_{T,\mathrm{ptcl}}^\mathrm{threshold}$ 
is the $p_T$ threshold of jet reconstruction at the particle level.

The corrected particle level imbalance $\sigma_\mathrm{MC}$ is subtracted in 
quadrature from the soft-radiation corrected resolution computed in data 
(see previous section),
\begin{equation}
\sigma_\mathrm{corr} = \sqrt{\left(K_\mathrm{soft}\sigma_{p_T}\right)^2 - 
\sigma_\mathrm{MC}^2}.
\end{equation}
The relative correction due to particle level imbalance is about (7--9)\% in 
the CC, (2--6)\% in the ICR and the EC, for $p_T>50$~GeV. The systematic
uncertainties on 
particle imbalance corrections are mainly due to the differences between the 
Gaussian one standard deviation and the RMS of the particle level 
imbalance distribution due
to non-Gaussian tails. The RMS is used for the central correction. The main 
non-Gaussian tails in particle level imbalance corrections are caused by 
muons and neutrinos, which are not included in the definition of 
D0 particle jets.

\subsection{Final jet $p_T$ resolutions{\label{finalres}}}

\begin{figure*}
\begin{overpic}[width=0.43\textwidth]
{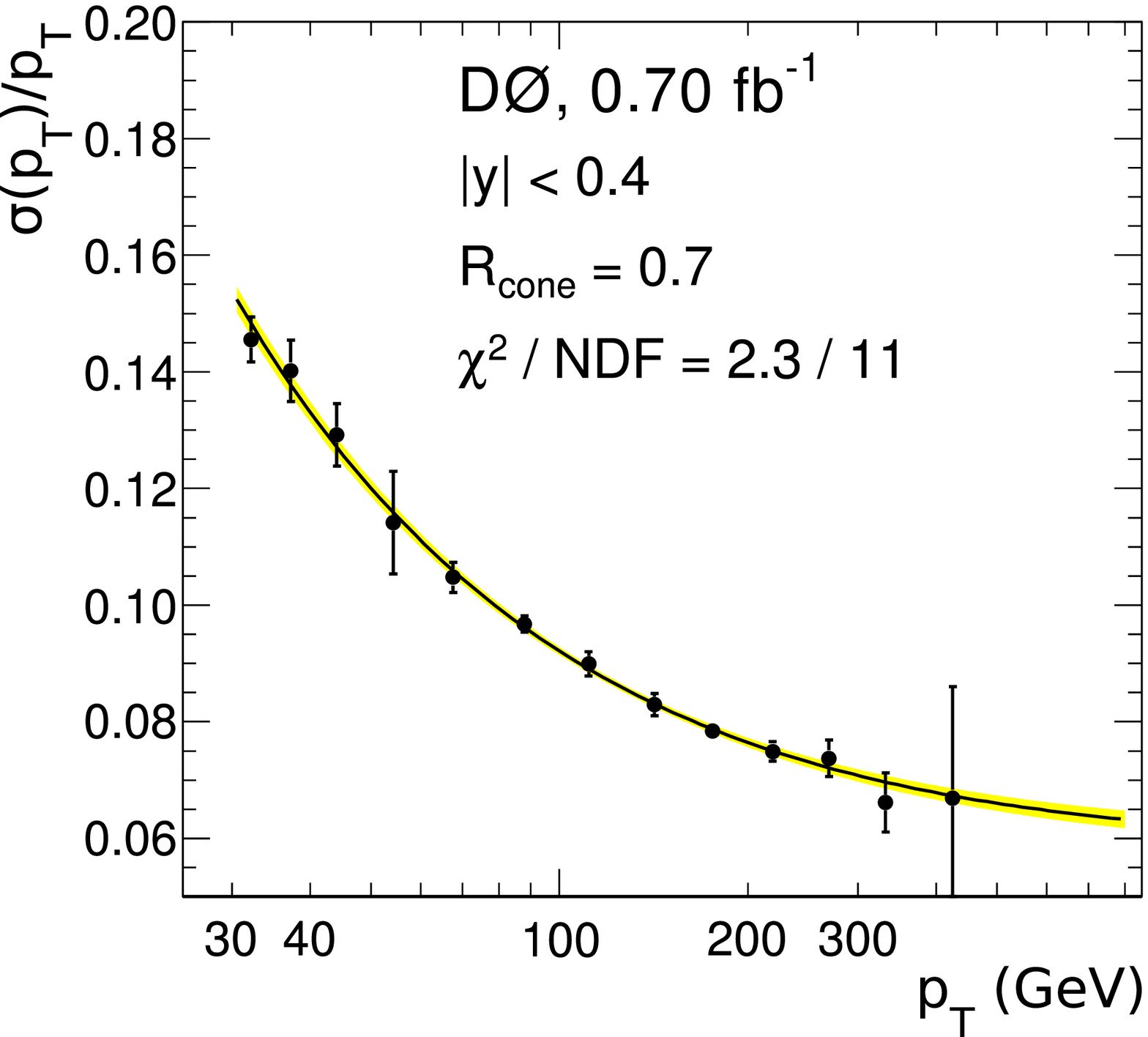}
\put(200,160){\textsf{(a)}}
\end{overpic}
\begin{overpic}[width=0.43\textwidth]
{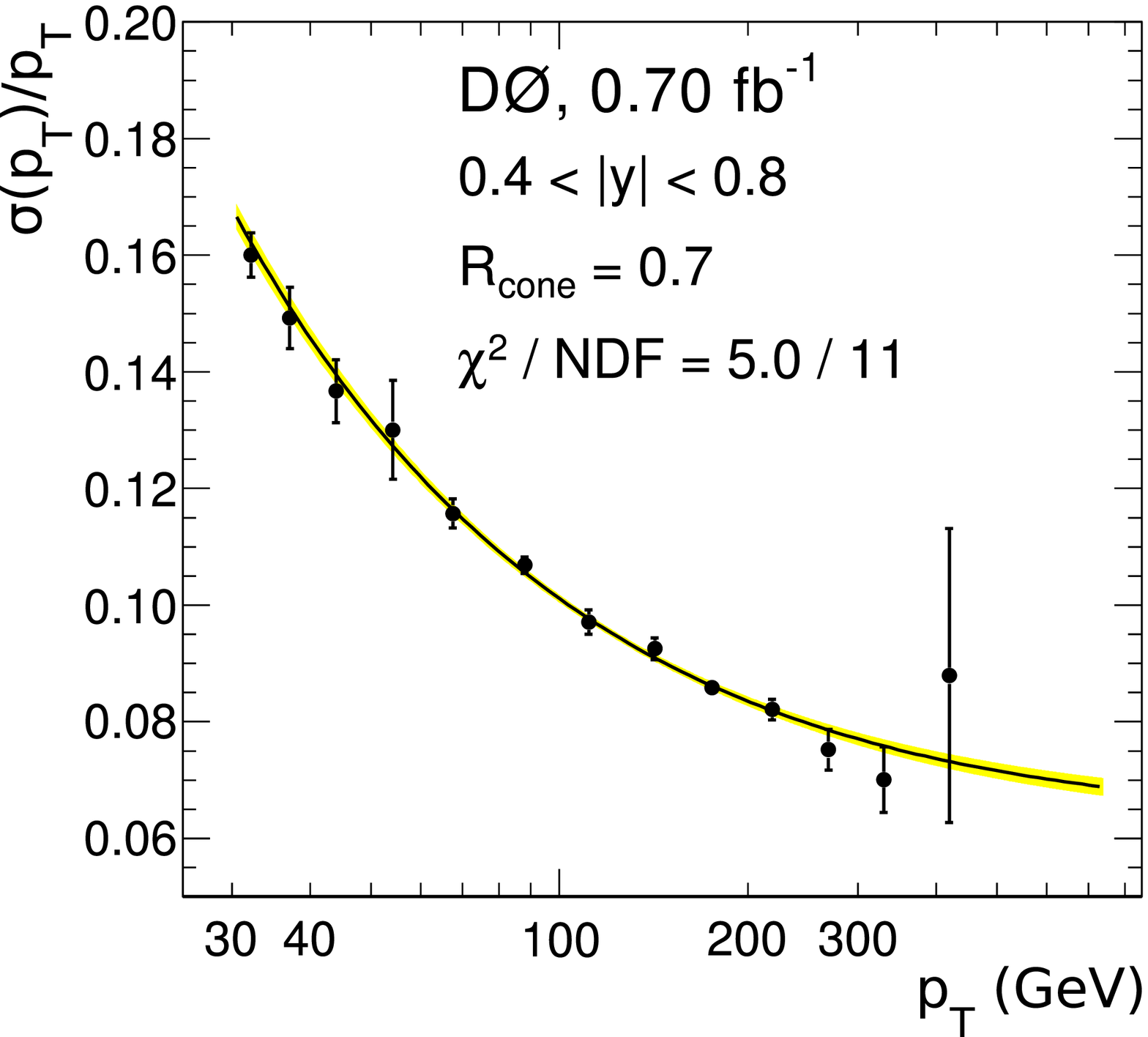}
\put(200,160){\textsf{(b)}}
\end{overpic}
\begin{overpic}[width=0.43\textwidth]
{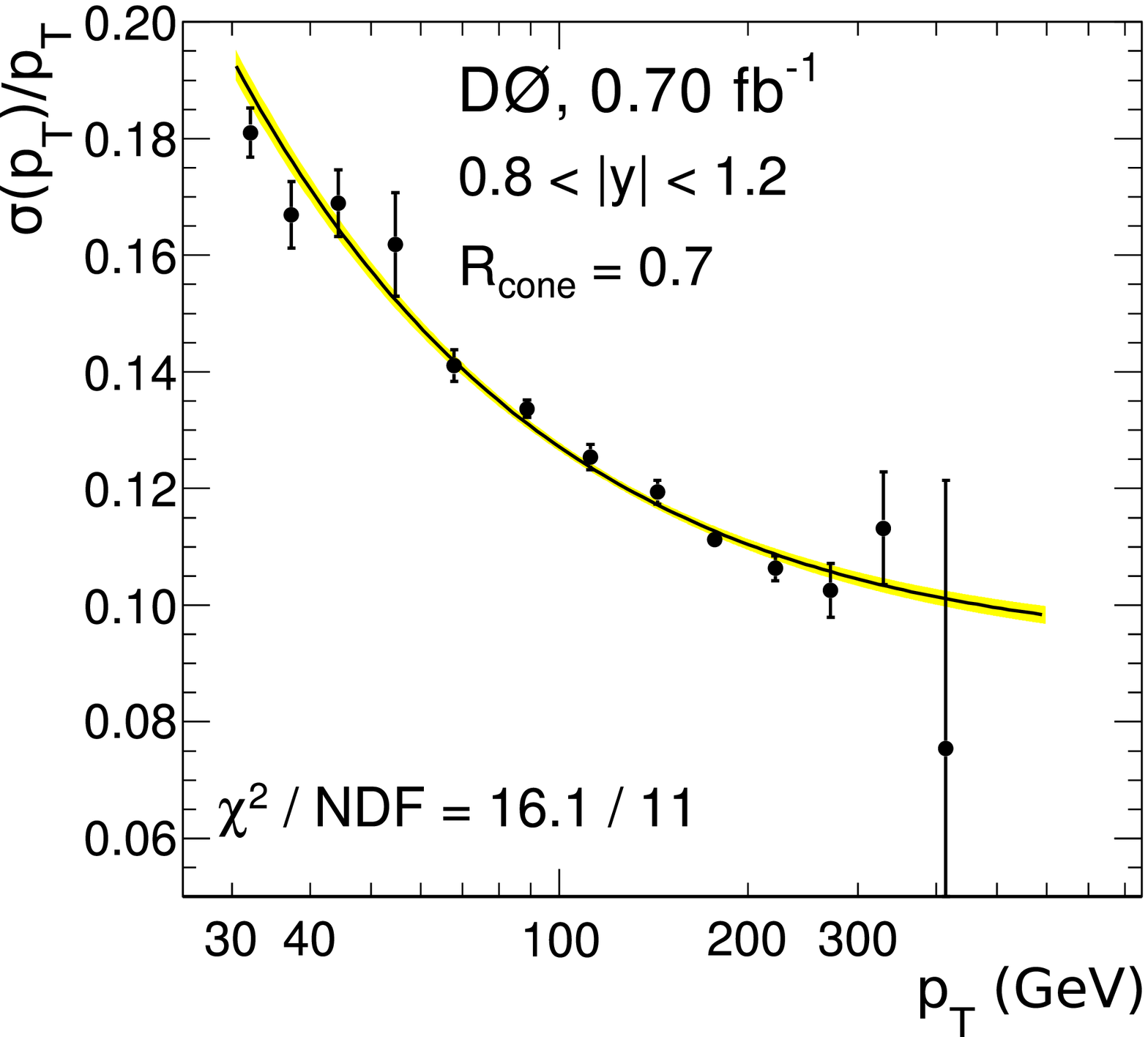}
\put(200,160){\textsf{(c)}}
\end{overpic}
\begin{overpic}[width=0.43\textwidth]
{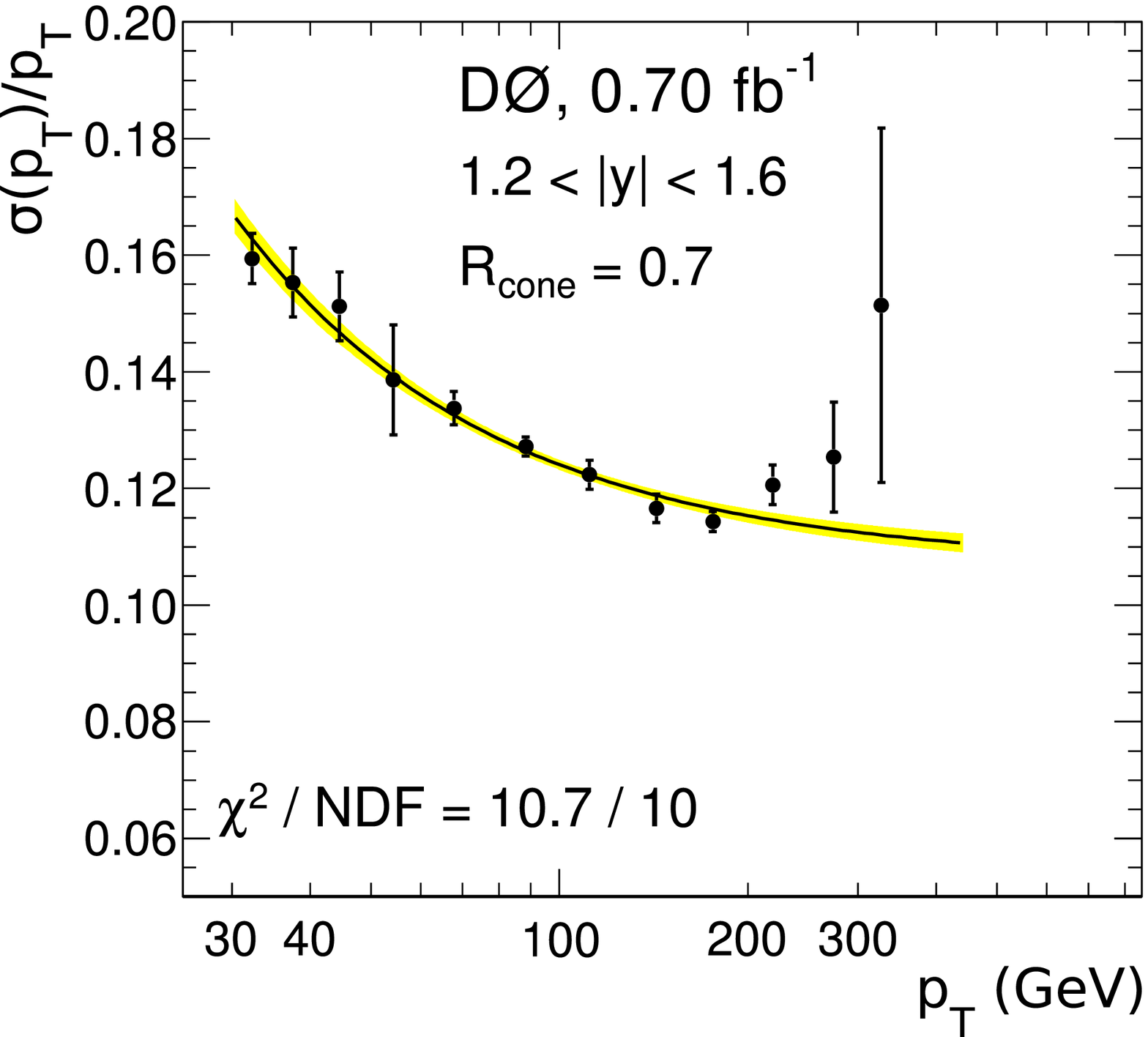}
\put(200,160){\textsf{(d)}}
\end{overpic}
\begin{overpic}[width=0.43\textwidth]
{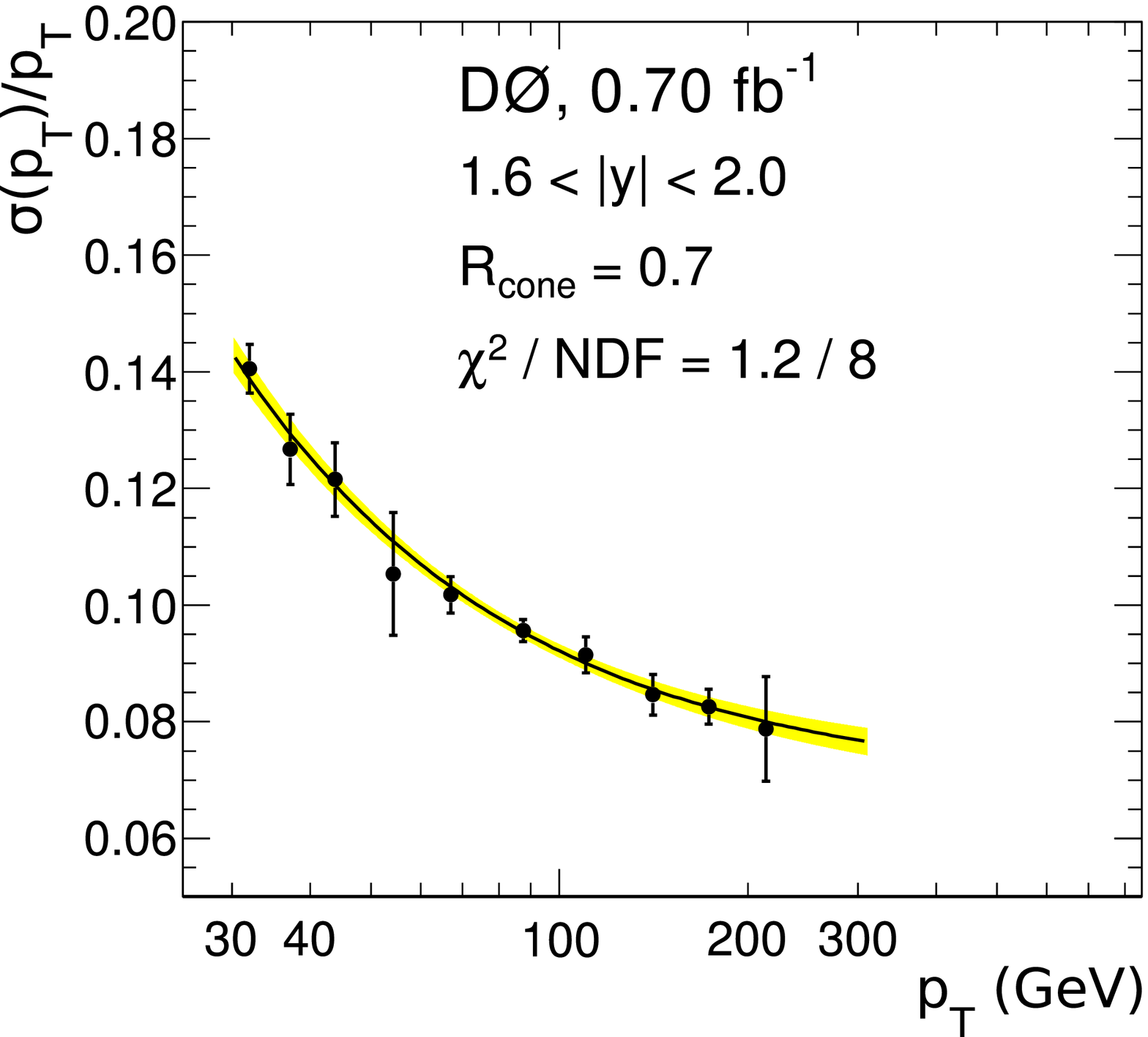}
\put(200,160){\textsf{(e)}}
\end{overpic}
\begin{overpic}[width=0.43\textwidth]
{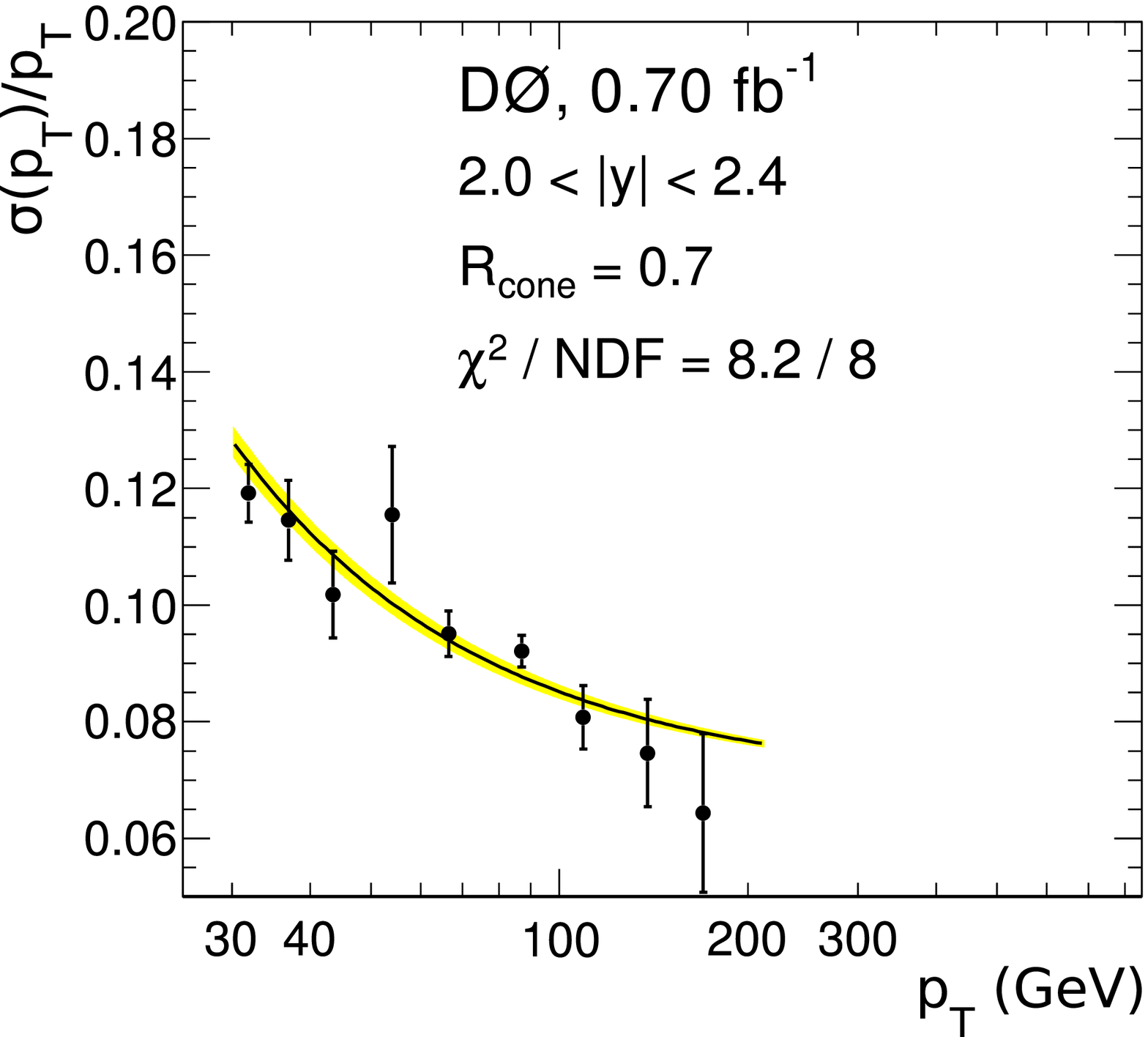}
\put(200,160){\textsf{(f)}}
\end{overpic}
\caption{\label{resolution} (color online)
Jet $p_T$ resolution determined in data for the six
rapidity regions. The solid curves
are the results of the fit. The fit uncertainty is given by the shaded band.}
\end{figure*}

Using the asymmetry method and the various corrections discussed above, we 
obtain the jet $p_T$ resolutions shown in Fig.~\ref{resolution}. The measured 
resolutions are fitted with the parameterization
\begin{equation}
\frac{\sigma_{p_T}}{p_T} = \sqrt{\frac{N^2}{p_T^2} + \frac{S^2}{p_T} + C^2},
\end{equation}
where $N$ is the noise term, $S$ the stochastic term, and $C$ the constant 
term. The values of the parameters are given in Table~\ref{resol}. 
These 
resolutions are used to obtain the inclusive jet $p_T$ cross section 
as described in the next section.

\begin{figure*}
\begin{overpic}[width=0.49\textwidth]
{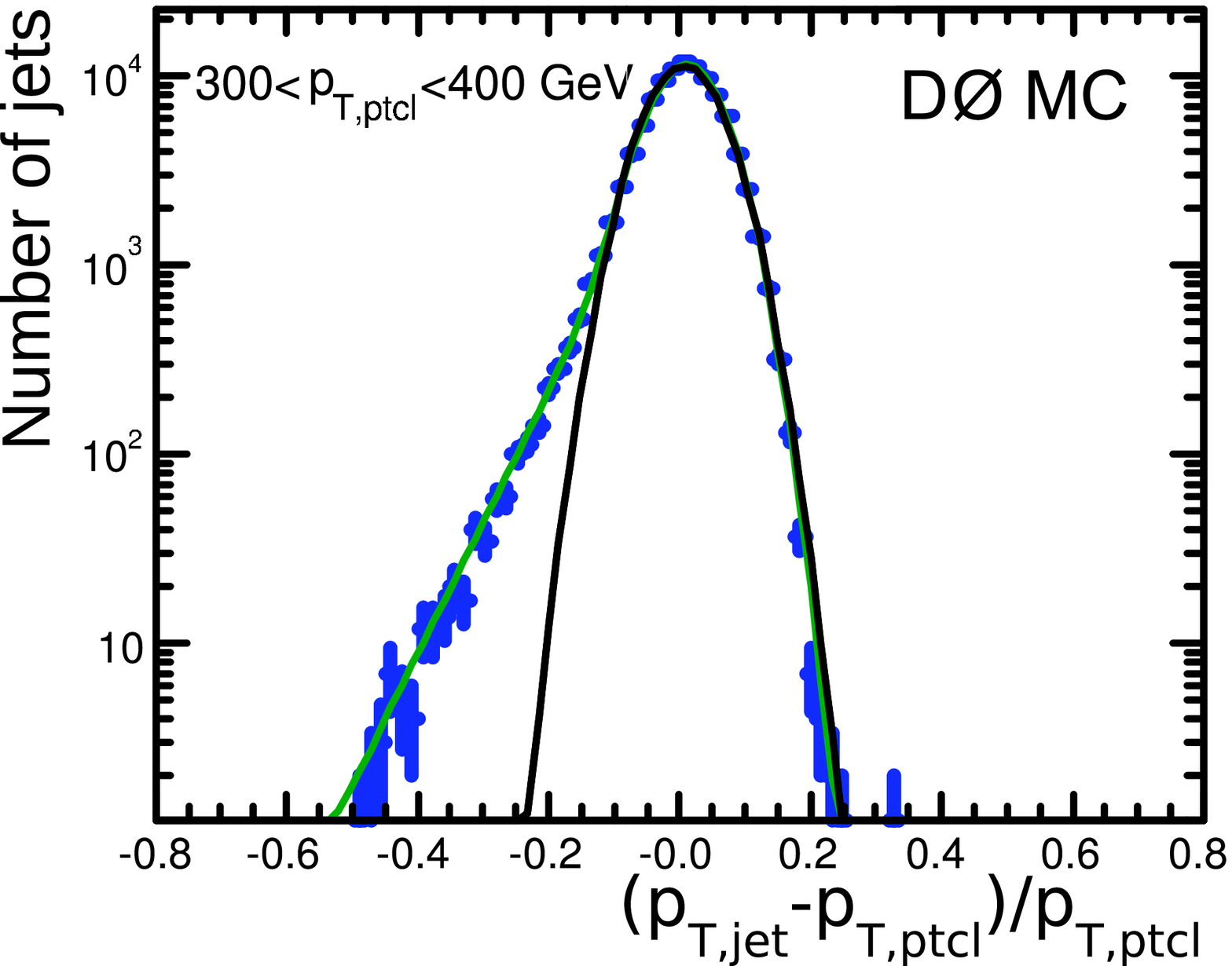}
\put(220,160){\textsf{(a)}}
\end{overpic}
\begin{overpic}[width=0.49\textwidth]
{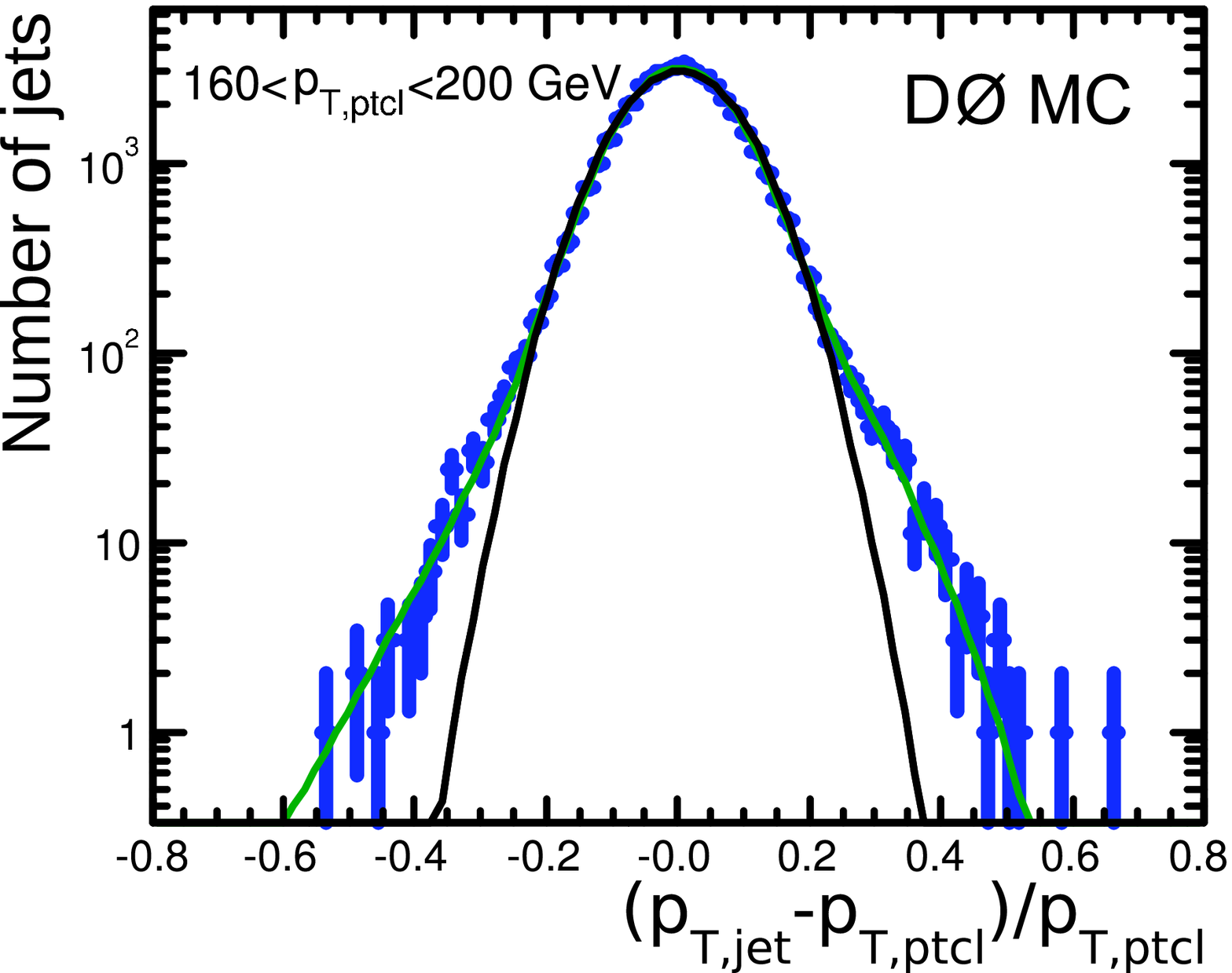}
\put(220,160){\textsf{(b)}}
\end{overpic}
\caption{\label{punch} (color online)
(a) Example of punch through for $|y|<0.4$ at high jet 
$p_T$ for $300<p_T<400$~GeV and $|y|<0.4$. (b) Example of tails of the jet 
$p_T$ resolution in the ICR for $160<p_T<200$~GeV and $0.8<|y|<1.2$. The 
two curves are the result of the Gaussian fit and of a Gaussian 
plus exponential tails.}
\end{figure*}

\begin{figure*}
\begin{overpic}[width=0.485\textwidth]
{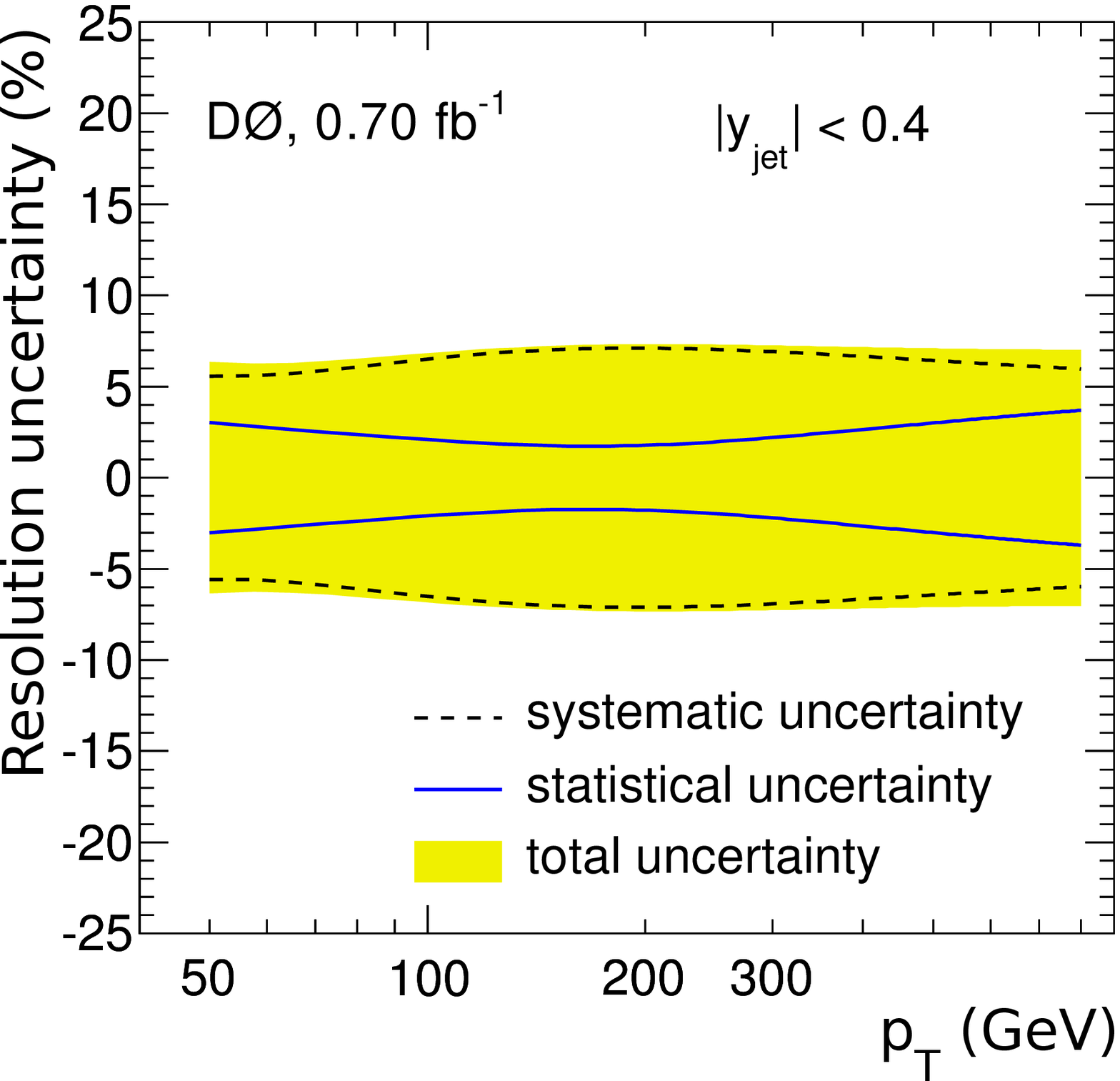}
\put(220,190){\textsf{(a)}}
\end{overpic}
\hfill
\begin{overpic}[width=0.485\textwidth]
{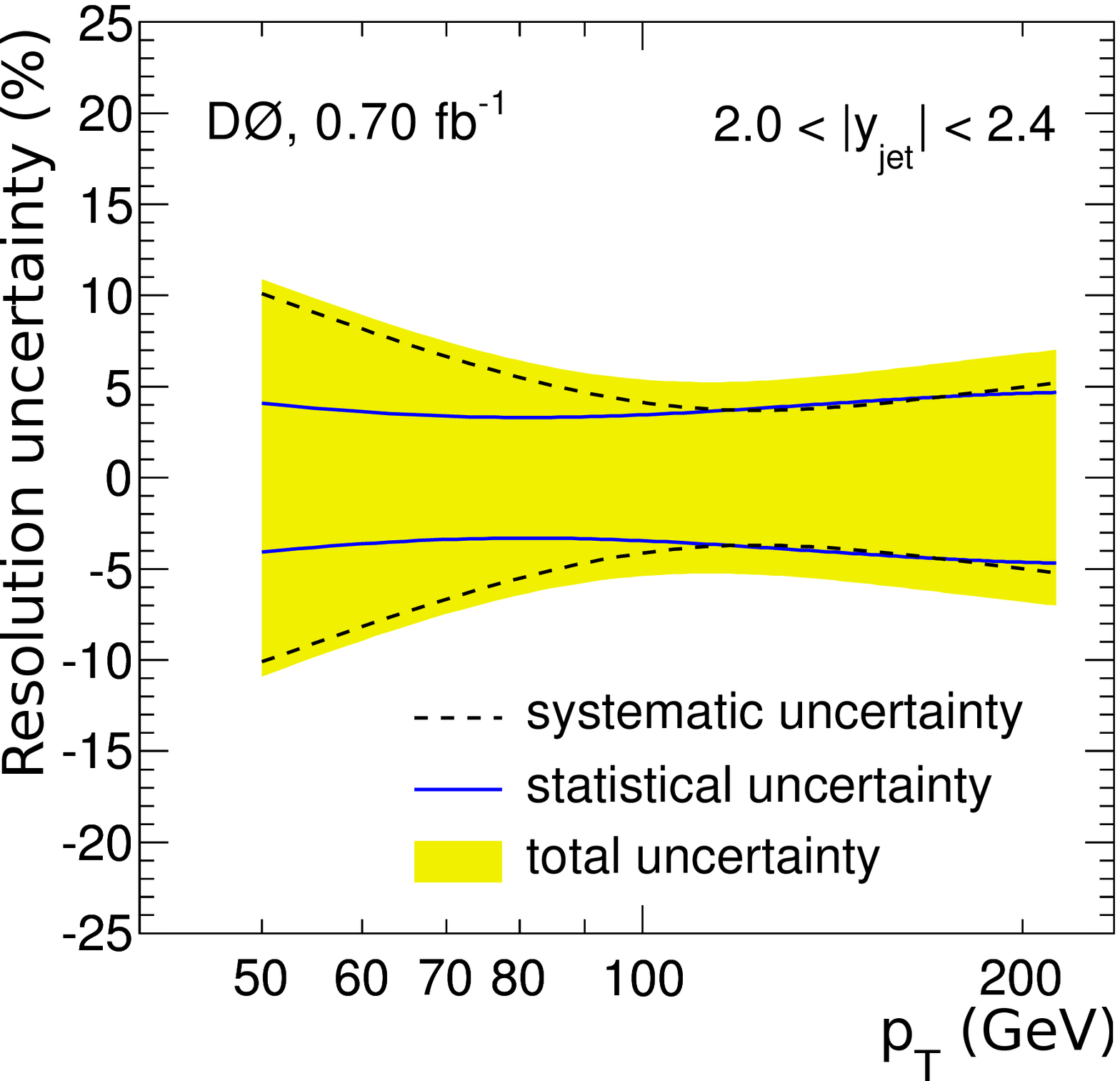}
\put(220,190){\textsf{(b)}}
\end{overpic}
\caption{\label{resoluncert} (color online)
Relative statistical and systematic uncertainties on 
jet $p_T$ resolution for (a) $|y|<0.4$ and (b) $2.0<|y_{\mathrm{jet}}|<2.4$.}
\end{figure*}

We note that the 
resolution is not Gaussian at high $p_T$ even in the central region because of 
calorimeter punch-through: jets at very high $p_T$ are not always fully 
contained in the calorimeter and can deposit energy into the muon system. In 
Fig.~\ref{punch} (a), we show the distribution of 
($p_{T,\mathrm{jet}}/p_{T,\mathrm{ptcl}}-1$) --- the ratio of the 
reconstructed to the particle level jet $p_T$ --- obtained from MC 
simulation of the detector in the central region of the calorimeter at high 
$p_T$. The ICR also exhibits non-Gaussian tails as shown in 
Fig.~\ref{punch} (b), which are explained by the changing structure of the 
calorimeter in this region. The non-Gaussian tails are modeled using a 
smeared exponential
\begin{eqnarray}
\nonumber
f(p_T,\mu,\sigma,P,\lambda) &=& (1-P)\cdot 
\frac{1}{\sqrt{2\pi}\sigma}e^{(p_T-\mu)^2/(2\sigma^2)}\\
\nonumber
&+& \frac{P\lambda}{2}\cdot \exp\left[\lambda(p_T-\mu+\frac{\lambda\sigma^2}{2})\right]\\
&&\times\mathrm{erfc}\left(\frac{p_T-\mu+\lambda\sigma^2}
{\sqrt{2}\sigma}\right),
\end{eqnarray}
with $\mu$, $\sigma$, $P$, and $\lambda$ as free parameters. 
The fitted shape from MC is scaled by varying the parameter $\sigma$ such that 
folding the distributions for the leading jets with the exponential $p_T$ 
spectrum from data results in precisely the same RMS of the jet $p_T$ 
resolution as observed in data. This method can account for any 
shaping of the non-Gaussian tails that takes place due to bin-to-bin 
migrations in data. The full MC shape with tuned $\sigma$ is later used in the 
unfolding of the data.

The uncertainties on jet $p_T$ resolution are given in Fig.~\ref{resoluncert} 
for two bins in rapidity as an example. The uncertainties come primarily from 
the statistical uncertainties in the fits.  An additional component is added 
to cover non-statistical variations between the fit model and the data.
The total uncertainty coming from the jet $p_T$ 
resolution is (5--10)\% over the full kinematic range covered by 
the inclusive jet cross section measurement ($p_T>50$ GeV). The leading 
systematic uncertainty in the central region is (4--5)\% due to the 
uncertainties on the particle level imbalance corrections. In the ICR, an 
important systematic is due to the uncertainty on the 
tails in the resolution for this region.  This systematic is estimated 
by varying the size of the tails by a factor of two, and
is not included in the RMS of the resolution, but rather 
the resulting variation in shape was used in the unfolding procedure for data. 
Another important source of uncertainty is taken from the 
following MC closure test: the 
full resolution measurement using the asymmetry is redone using a full 
simulation of the D0 detector, 
and the difference between the MC input true resolution and the 
result of the method is taken as a systematic uncertainty. This amounts 
to up to about 10\% uncertainty in the resolution at $p_T=50$~GeV in the 
forward region.

\begin{table}[htb!]

\begin{tabular}{|cccc|}
\hline \hline & $N$(oise) & $S$(tochastic) & $C$(onstant) \\
$|y|<0.4$ & 2.07 & 0.703 & 0.0577 \\
$0.4<|y|<0.8$ & 2.07 & 0.783 & 0.0615 \\
$0.8<|y|<1.2$ & 2.07 & 0.888 & 0.0915 \\
$1.2<|y|<1.6$ & 2.07 & 0.626 & 0.1053 \\
$1.6<|y|<2.0$ & 2.07 & 0.585 & 0.0706 \\
$2.0<|y|<2.4$ & 2.07 & 0.469 & 0.0713 \\
\hline \hline
\end{tabular}
\caption{\label{resol}
Parameters of the fits to the jet $p_T$ resolution versus $p_T$ for data.
The noise term is fixed to the MC value with an uncertainty of 1 GeV since 
it is not constrained by the data.}
\end{table}

\section{Jet rapidity resolution}

\begin{figure}
\includegraphics[width=0.45\textwidth]{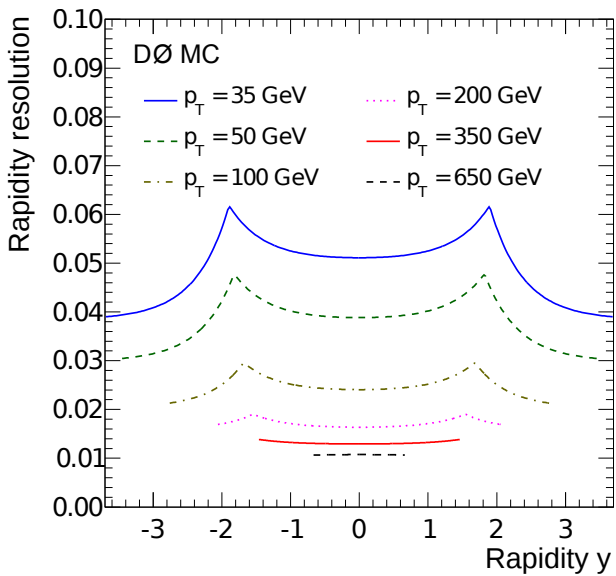}
\caption{\label{rapidity} (color online)
Rapidity resolution (RMS) as a function of $y$ in different jet $p_T$ regions.}
\end{figure}

Compared to the jet $p_T$ resolution, the rapidity resolution is a small 
effect which is determined using a MC simulation of the detector. The 
bin width in $y$ is much larger than the $y$ resolution and bin-to-bin 
migrations only occur at the bin edges. To unfold the effect of the rapidity 
resolution, a smooth parameterization of the resolution as a function of $y$ 
is used. The result of the parameterization of the $y$ resolution in 
different $p_T$ bins is shown in Fig.~\ref{rapidity}.

\section{Unfolding}\label{sec:unfolding}

In this section, we describe the method used to unfold the data
as a function of jet $p_T$ and $y$. As we 
already mentioned, the main smearing effect is due to the jet $p_T$ resolution 
while the $y$ smearing is only a second order effect. The steeply falling 
jet $p_T$ cross section convoluted with the jet $p_T$ resolution leads to an 
increase of the observed cross section as a function of the measured jet 
$p_T$. To unfold the data, we use the so-called ansatz method. We
start with a functional form for the cross section that has only a 
few parameters, smear it with the jet $p_T$ and $y$ resolutions, 
and fit the parameters so that it describes the raw 
cross section measurement before unfolding. 

The ansatz used in each rapidity bin contains a $p_T$ dependence term
and an additional rapidity dependence
\begin{eqnarray}
\label{eq:ansatz}
f(p_T,\eta)  &=&  N_0\left(\frac{p_T}{100\mathrm{~GeV}}\right)^{-\alpha}
\left[1-\frac{2p_T\cosh(|y_\mathrm{min}|)}{\sqrt{s}}\right]^{\beta} \nonumber
\\
&~& \cdot \exp \left(-\gamma p_T\right).
\end{eqnarray}
Here $\sqrt{s}=1960$~GeV is the center-of-mass energy and $|y_{\text{min}}|$ 
is the low 
edge of the bin in absolute rapidity. The ansatz is based on phenomenological 
fits and motivated by the parton model~\cite{parton}. The exponential term 
represents hydrodynamic production by freezing out particles from the quark and 
gluon sea. The value of $\gamma$ is expected to be of the order of 0.3--0.6 
GeV$^{-1}$, typical of the proton size. The first power term characterized by
$\alpha$
represents the scaling violations associated with hard production. 
Typical values of $\alpha$ are 4--6 for single particle production. The 
second power term characterized by $\beta$ represents the kinematic suppression 
effect at the edges of the phase space of particle production. 

\begin{figure*}
\includegraphics[width=0.42\textwidth]{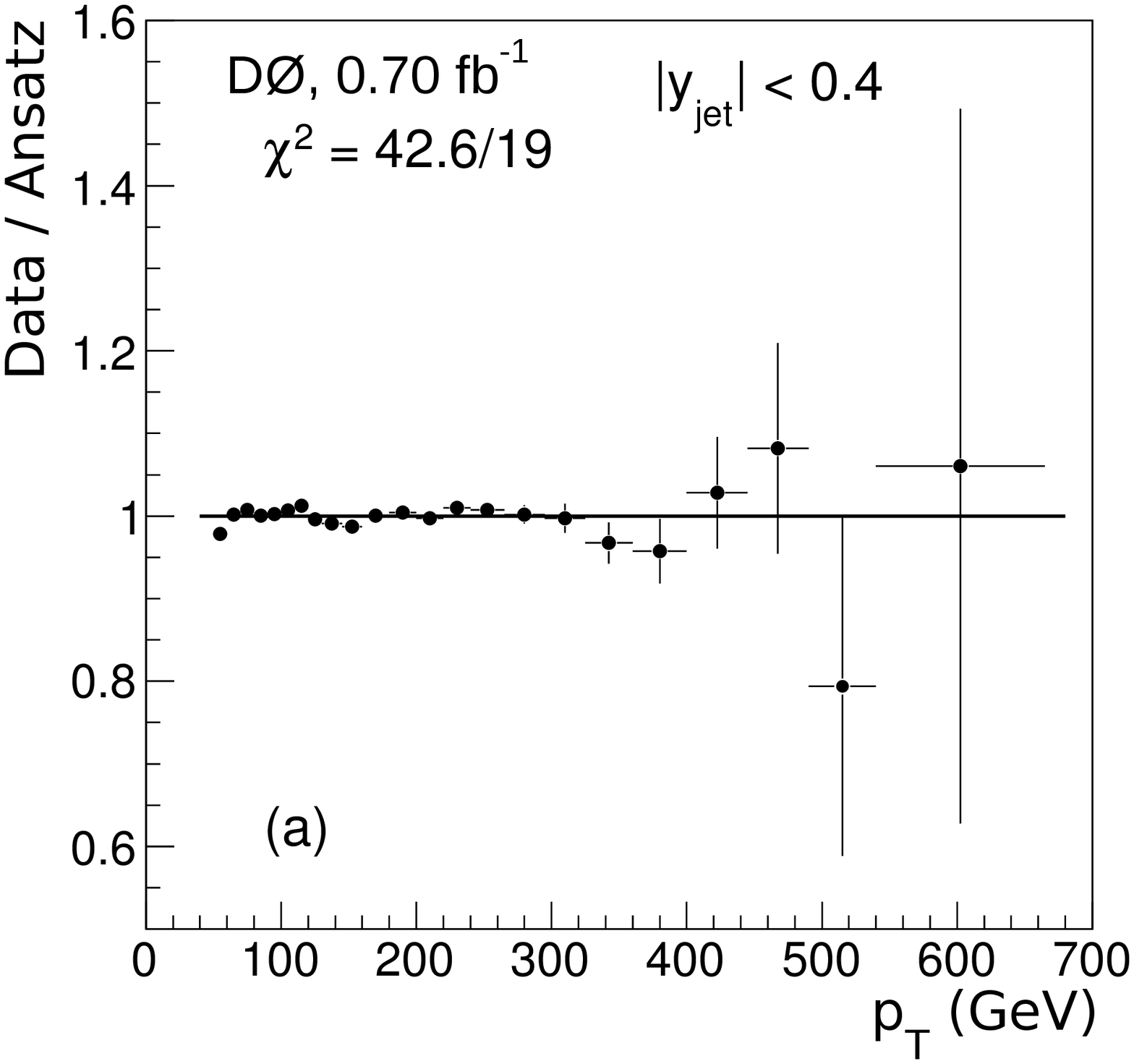}
\includegraphics[width=0.42\textwidth]{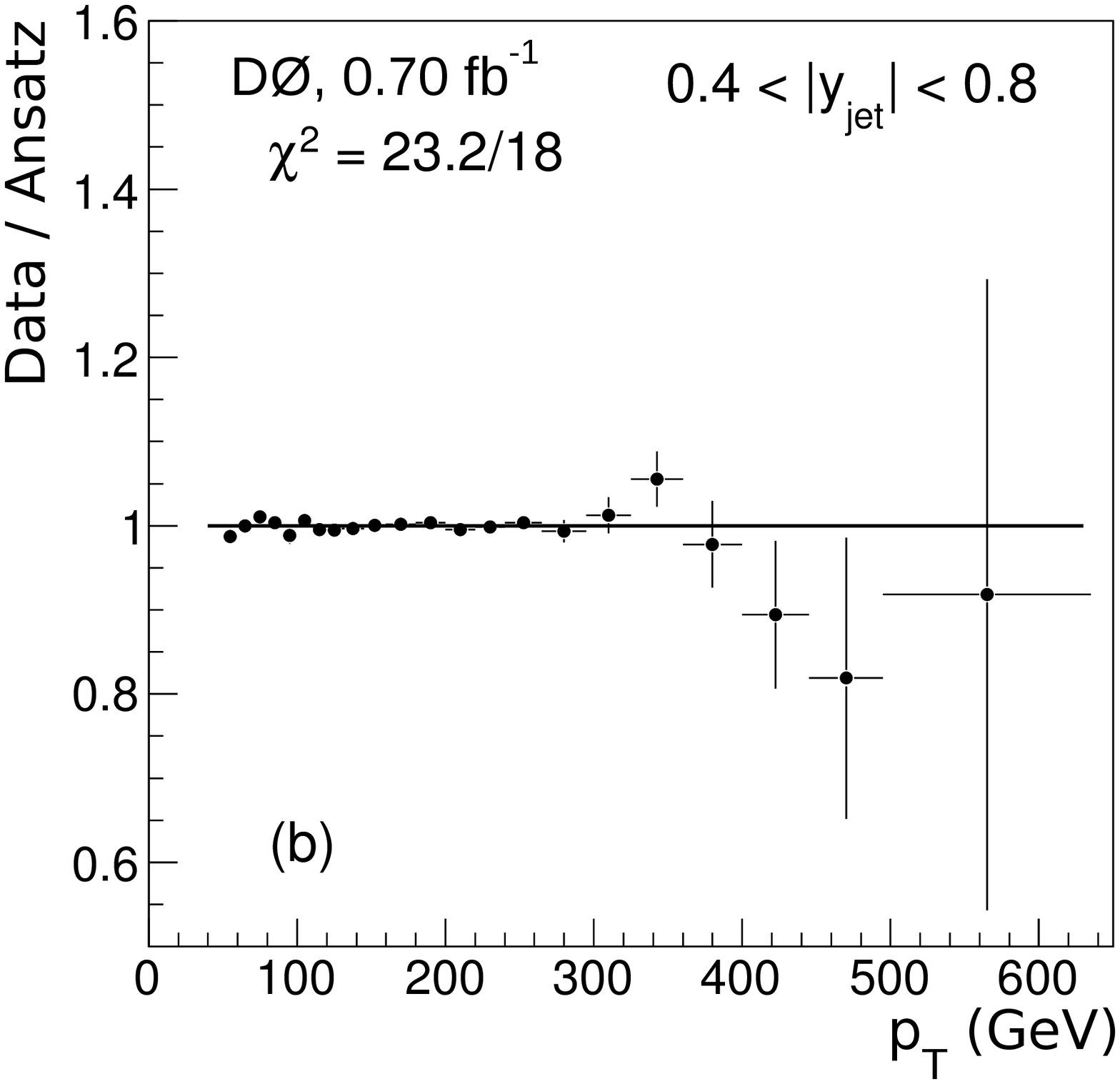}
\includegraphics[width=0.42\textwidth]{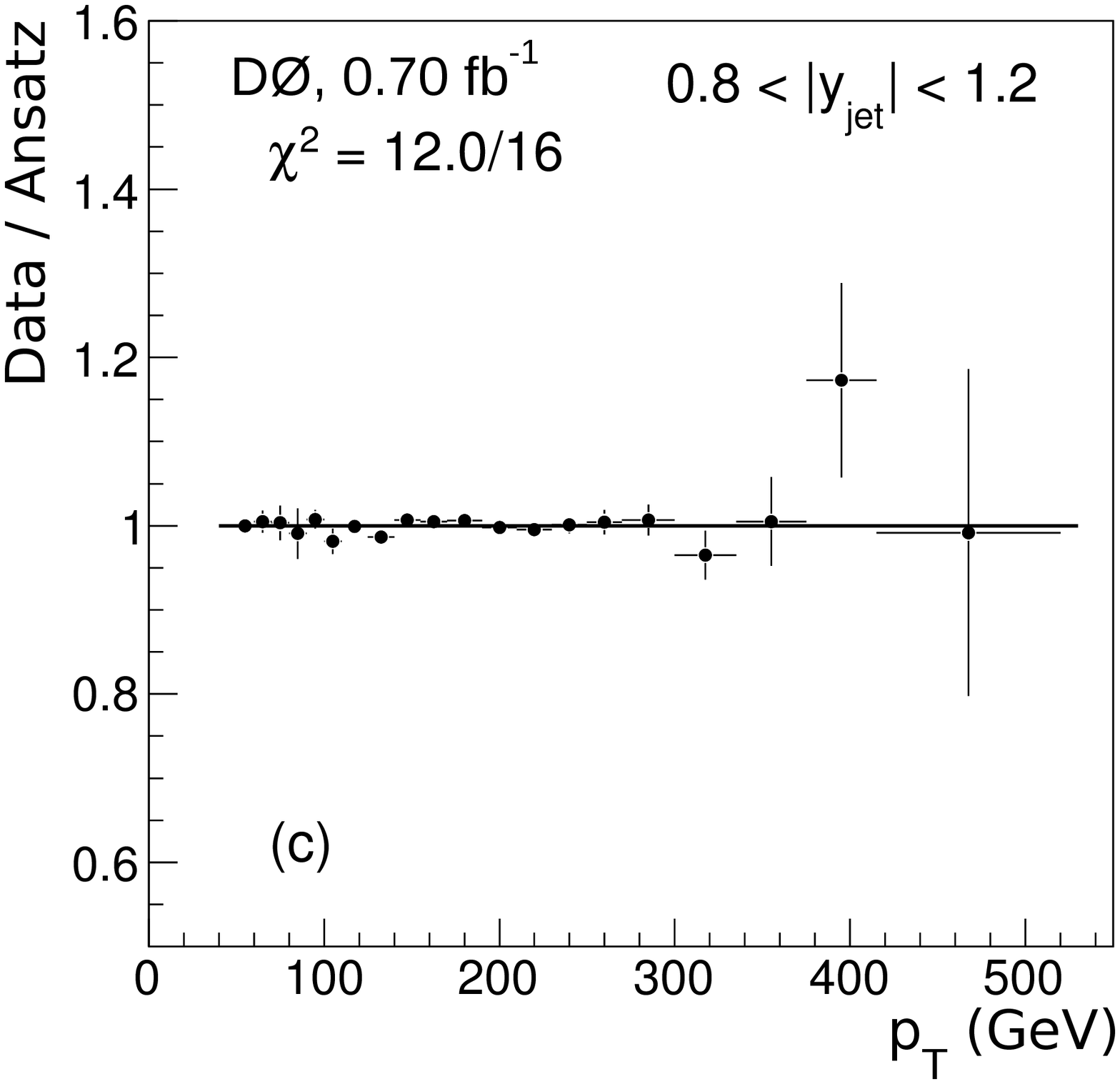}
\includegraphics[width=0.42\textwidth]{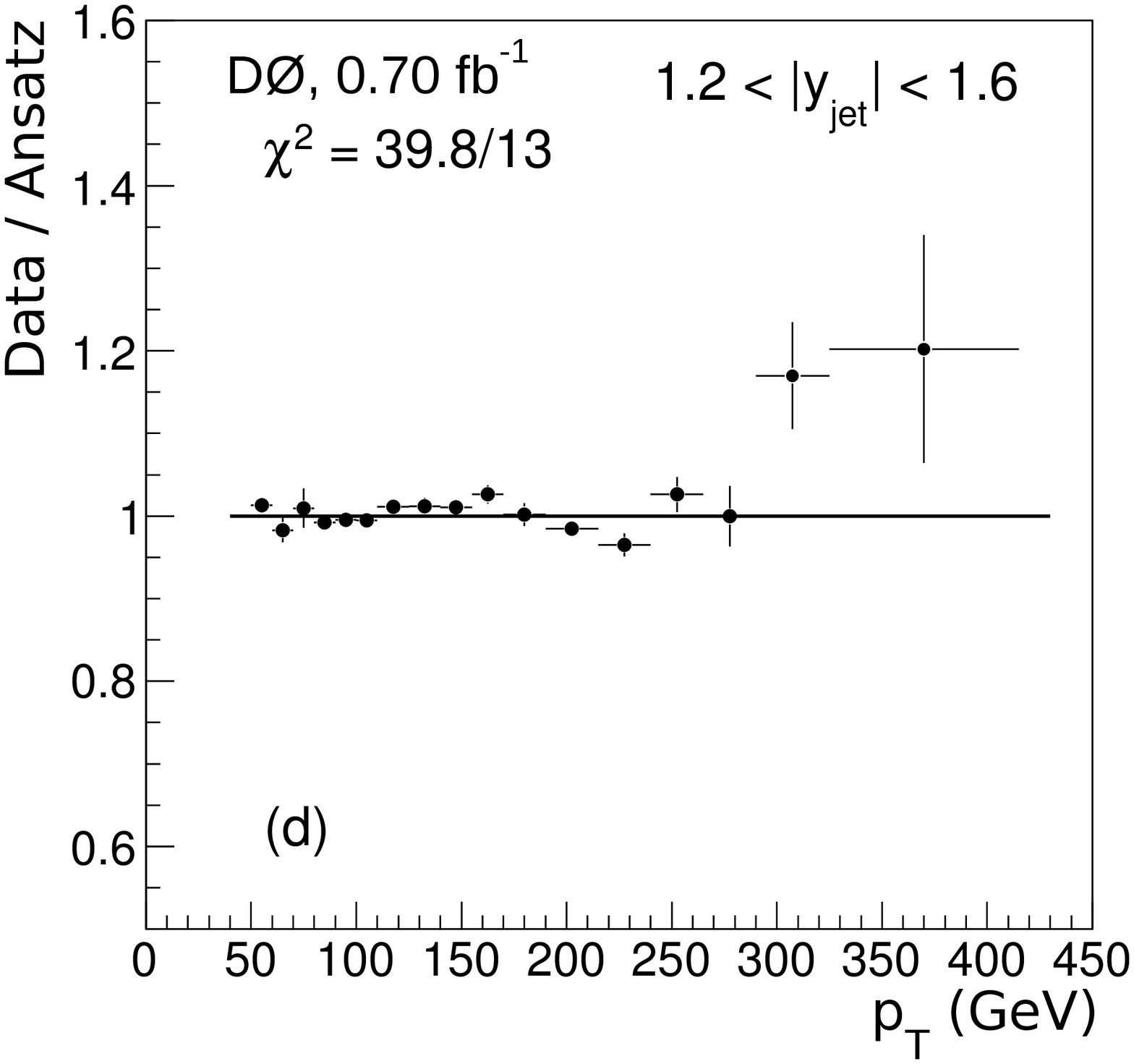}
\includegraphics[width=0.42\textwidth]{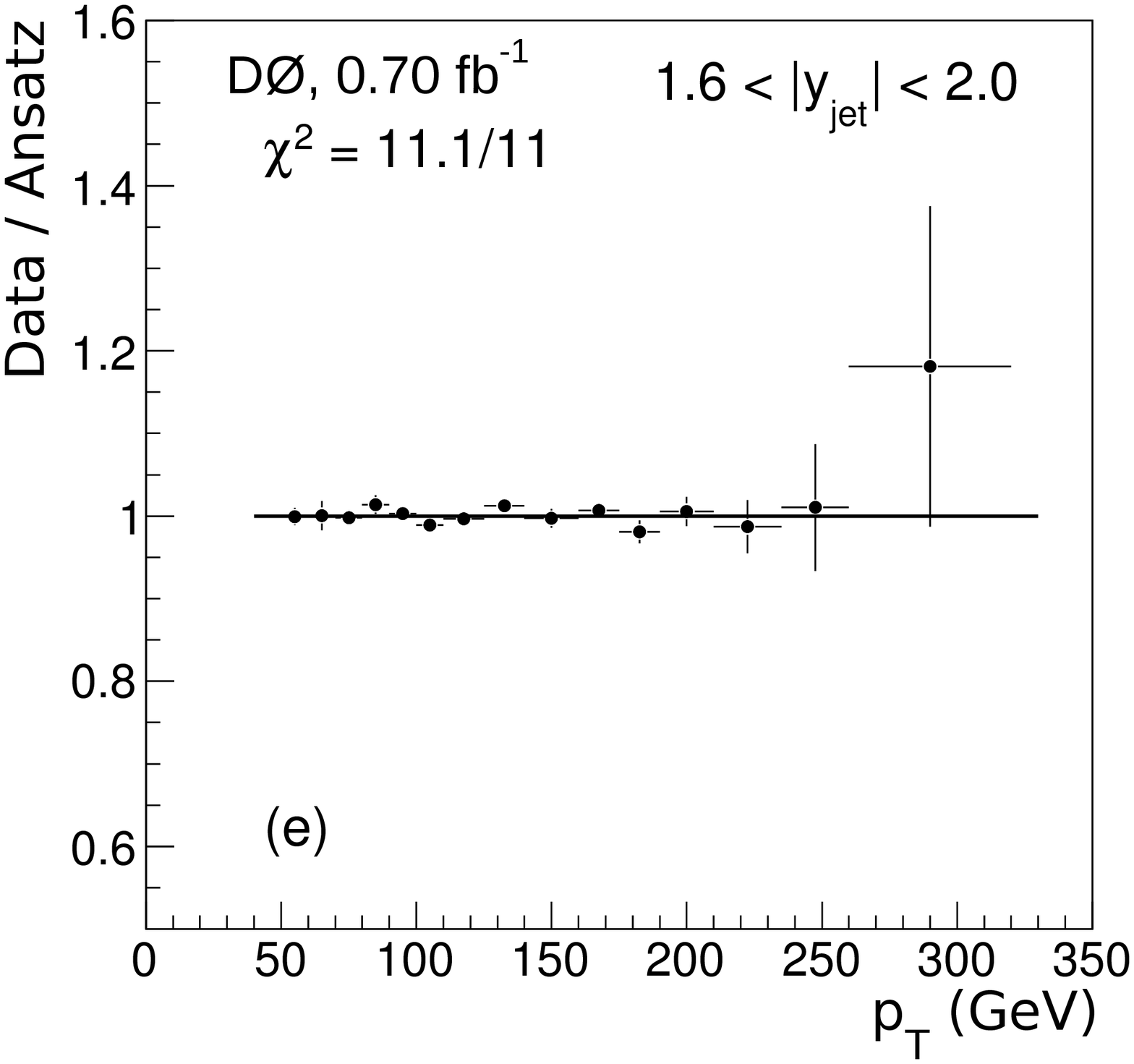}
\includegraphics[width=0.42\textwidth]{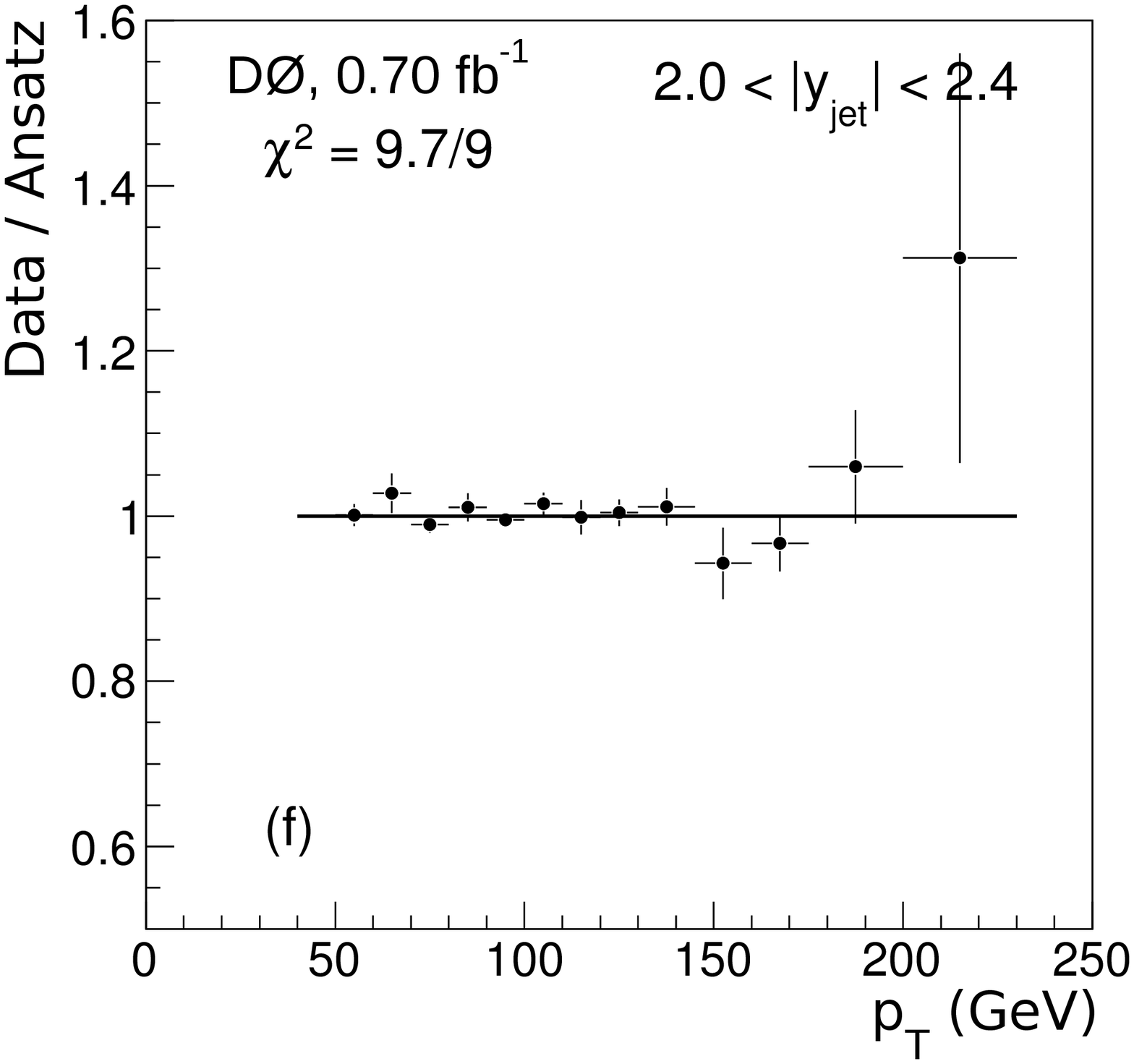}
\caption{\label{unfoldingb} Data divided by the ansatz fit with models 
for $p_T$ and $y$ smearing in the six rapidity regions.}
\end{figure*}

\begin{figure*}
\includegraphics[width=0.38\textwidth]{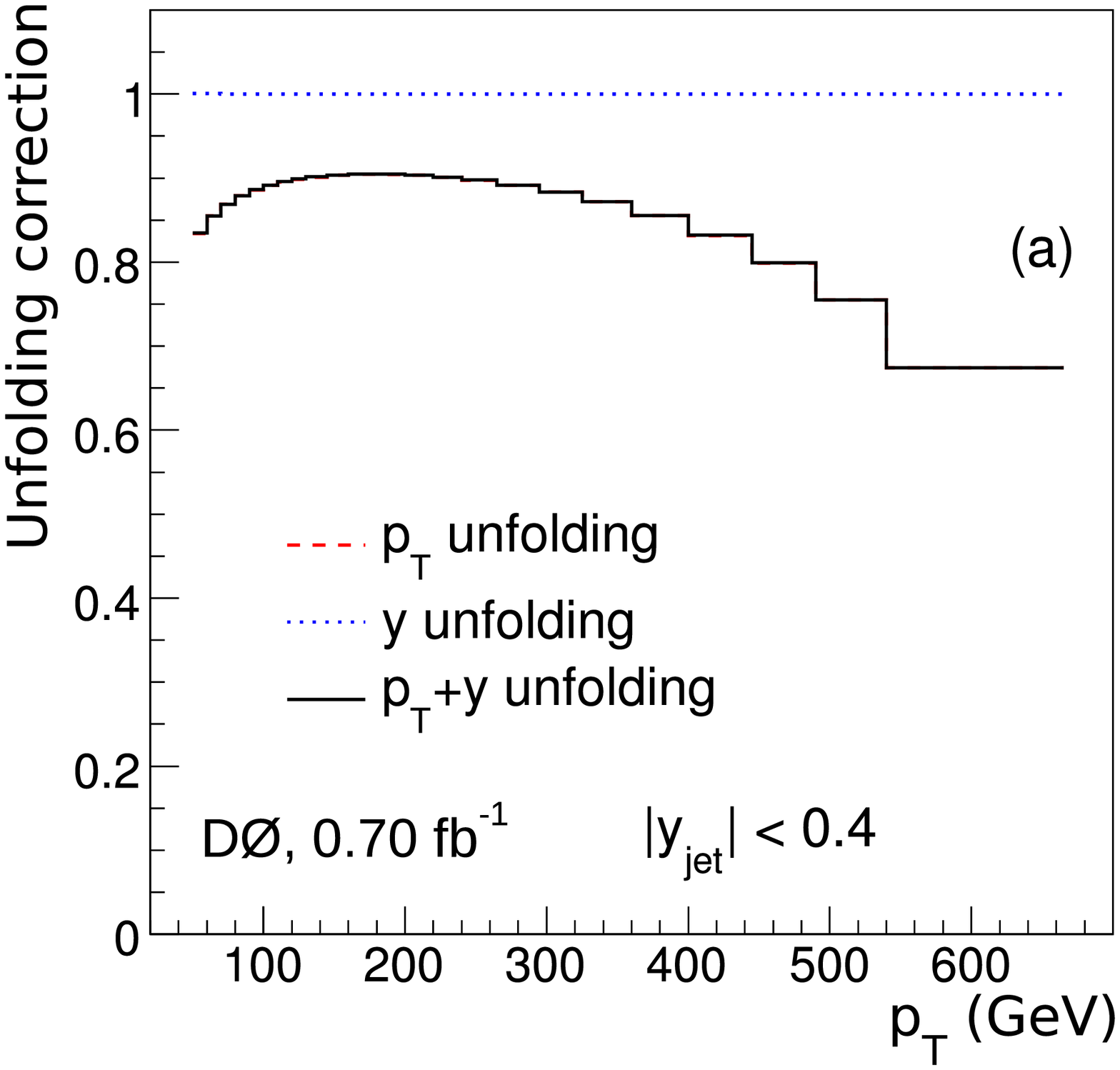}~~~
\includegraphics[width=0.38\textwidth]{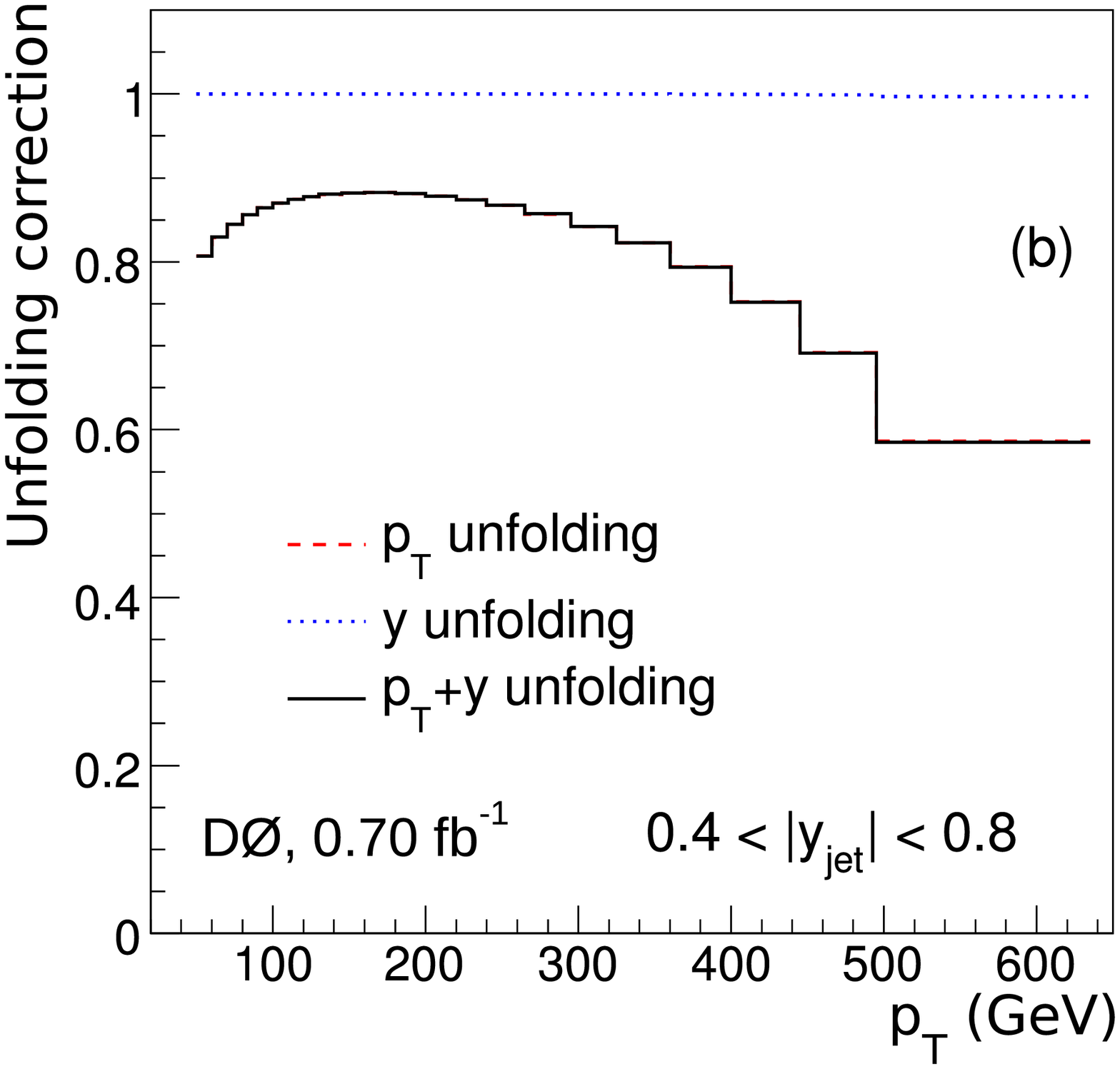} \\ \vspace{5mm}
\includegraphics[width=0.38\textwidth]{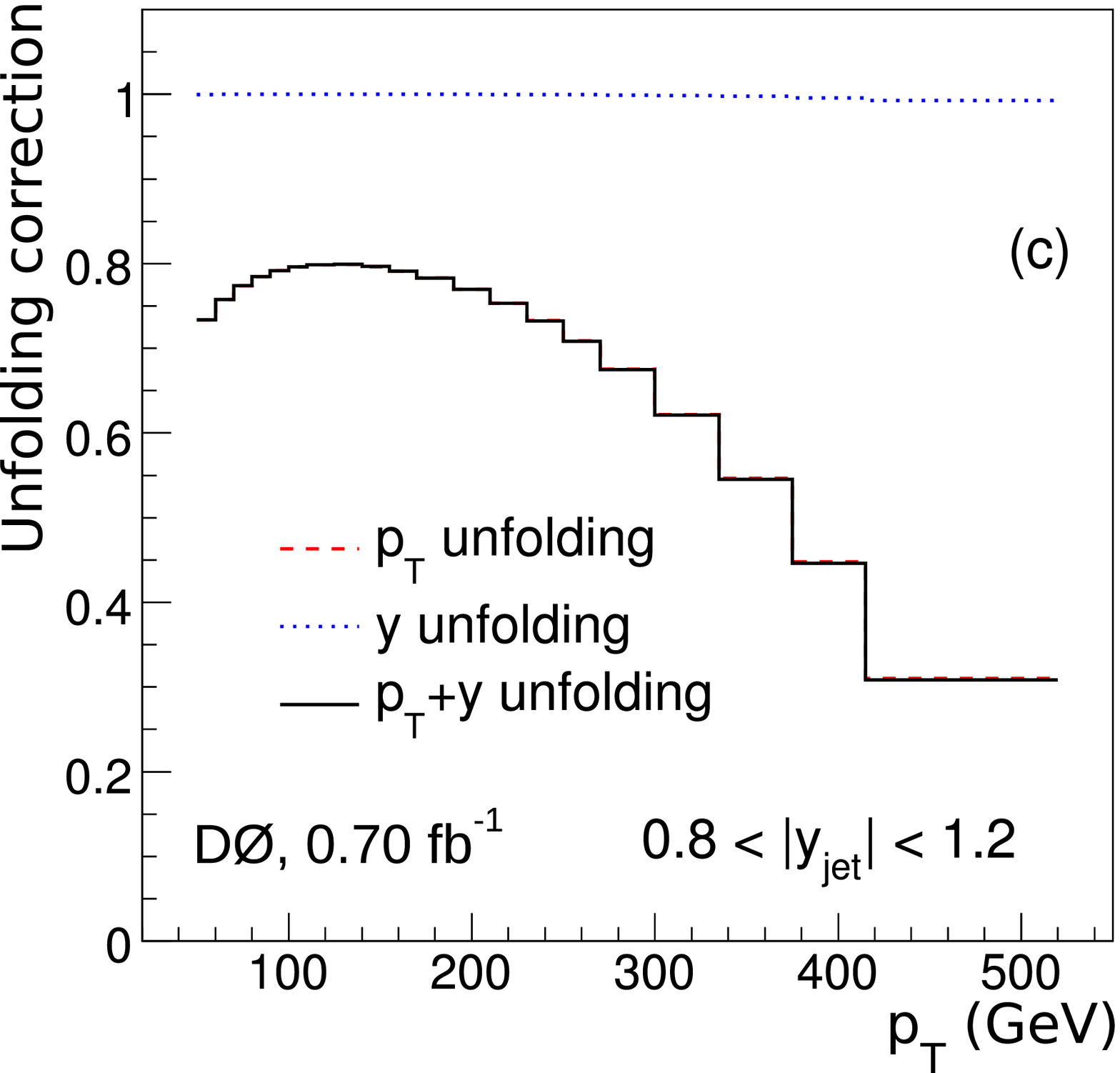}~~~
\includegraphics[width=0.38\textwidth]{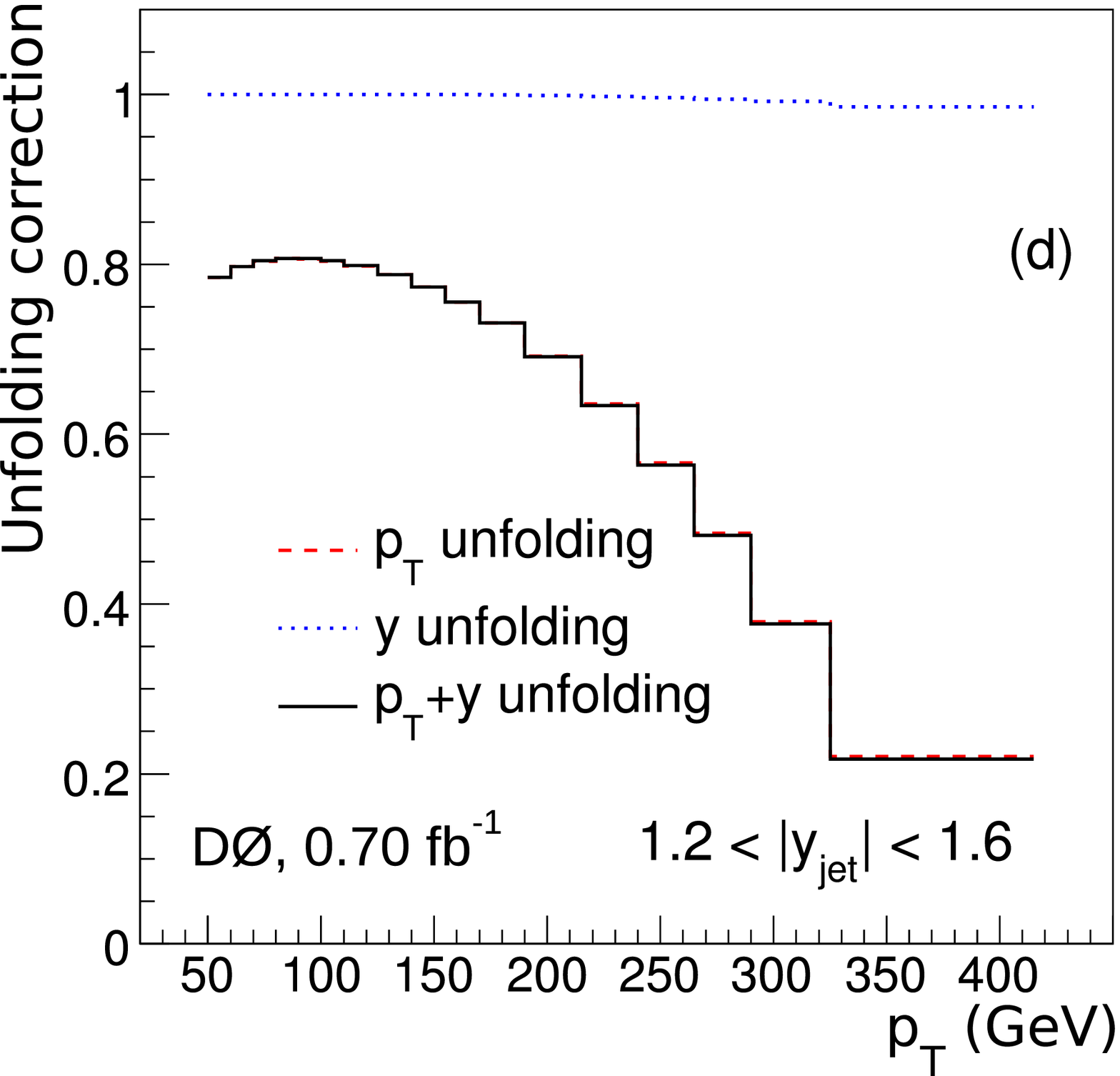} \\ \vspace{5mm}
\includegraphics[width=0.38\textwidth]{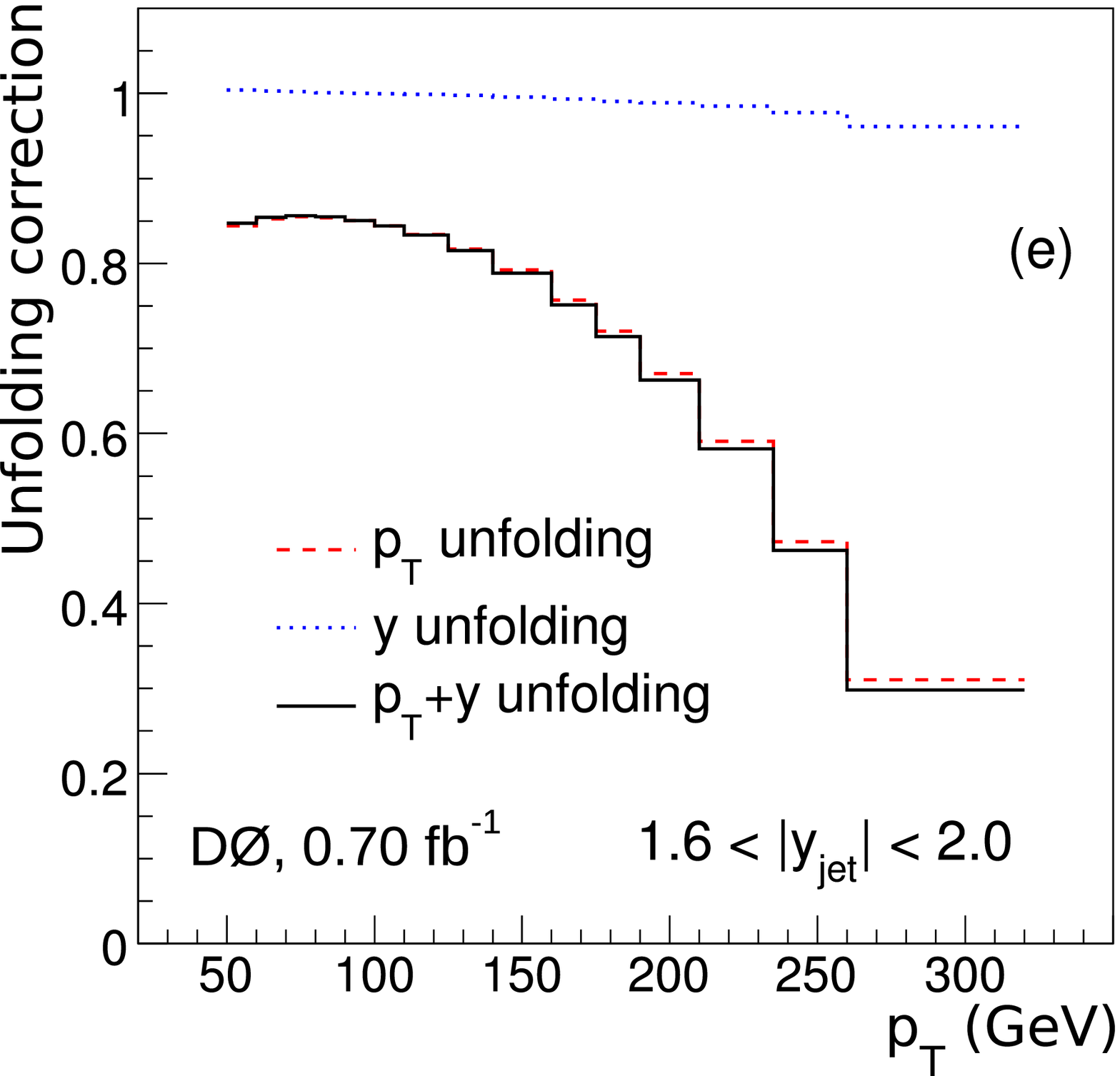}~~~
\includegraphics[width=0.38\textwidth]{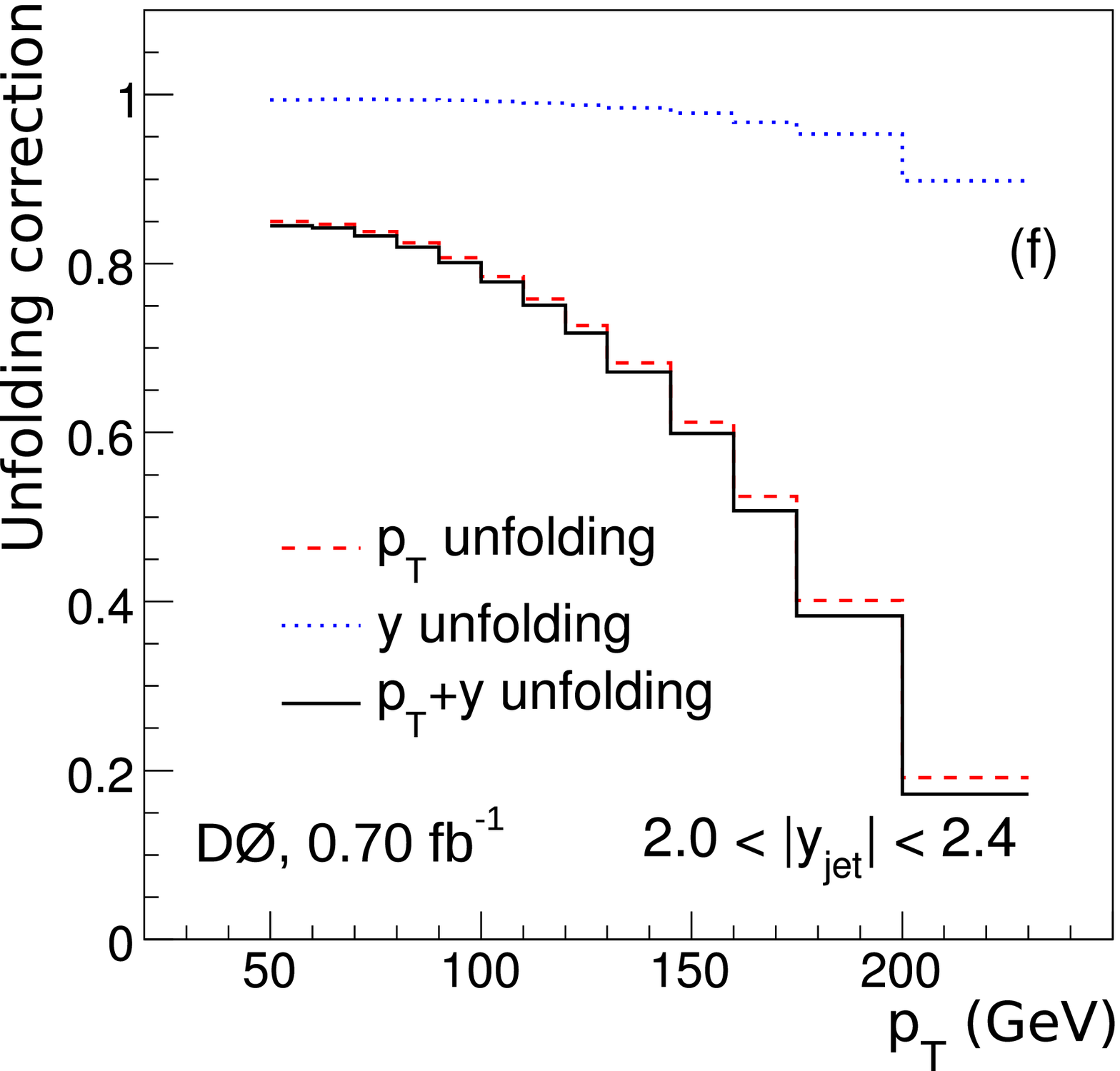}
\caption{\label{unfolding} (color online)
Unfolding corrections in the six rapidity regions 
as a function of jet $p_T$. The corrections are given for $p_T$, $y$ unfolding 
separately and combined.}
\end{figure*}

The ratios between the data and the smeared ansatz are shown in 
Fig.~\ref{unfoldingb}, where the ansatz correctly describes the data in all 
$y$ bins. The unsmearing corrections for the $p_T$ resolution effects are 
shown by the dashed lines in Fig.~\ref{unfolding}. The unfolding corrections 
are  (10--40)\% in the CC, (20--80)\% in the ICR where the jet $p_T$ 
resolution is worse, and (15--80)\% in the EC where the jet cross section 
falls steeply. The highest $p_T$ bin (where the unfolding corrections 
are the largest) where the cross section is measured is chosen so that 
the cross section measurement is still meaningful; 
the number of events should still be sufficient to give a lower 
limit on the measured cross section at the 95\% C.L.
($N_\mathrm{theory}/\sqrt{N_\mathrm{smeared~theory}} \ge 1.645$).
Although in some bins most 
of the events migrate from lower $p_T$, the migrations are well understood and 
result in a relatively small uncertainty compared to the uncertainty from
the jet energy scale. 
The ansatz unfolding is found to be in good agreement with the results using 
the {\sc{Pythia}} MC where the cross section is rescaled to data and the 
jets at particle level are smeared according to the $p_T$ resolutions obtained 
in Sec.~\ref{sec:unfolding}.

The same ansatz unfolding method can be used to unfold the cross section for 
effects of the resolution for resolving rapidity,
assuming the $p_T$ and $y$ resolutions are 
uncorrelated. Since the $y$ resolution is much better than the 
$p_T$ resolution, the effects of the $y$ resolution are a small 
perturbation on top of the $p_T$ smearing. The fits to 
the unfolded $p_T$ spectra (unfolded for $p_T$ resolution effects only) in 
neighboring rapidity bins are interpolated with respect to rapidity to 
produce a smooth, continuous two dimensional spectrum in $p_T$ and $y$.  A
final unfolding is performed to correct for events that migrate into
neighboring rapidity regions due to effects of the $y$ resolution.
The results of the $y$ unfolding as a function of jet $p_T$ in the 
different $y$ bins are given in 
Fig.~\ref{unfolding}, together with the results of the global unfolding 
corrections in jet $p_T$ and $y$. As expected, the effects of $y$ unfolding 
are very small with respect to the effects of the $p_T$ unfolding.

\section{\boldmath Inclusive jet $p_T$ cross section measurement}\label{sec:results}

\begin{figure}
\includegraphics[width=\columnwidth]{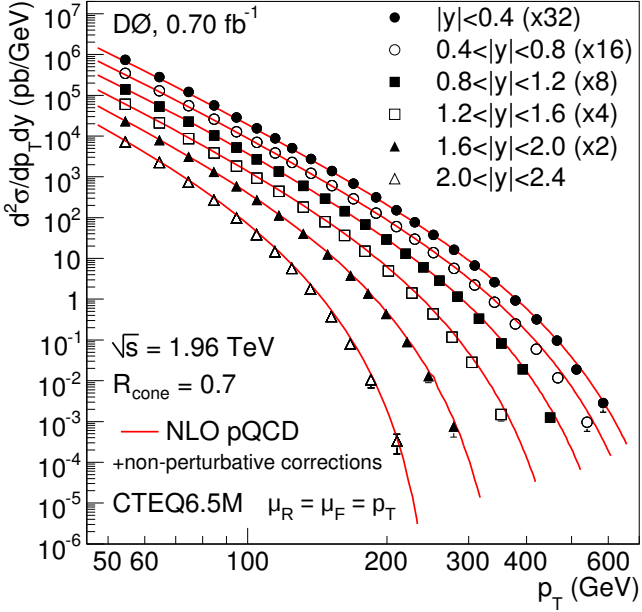}
\caption{\label{fig1} (color online)
Inclusive jet cross section measurements as a function of jet $p_T$ 
in six $|y|$ bins. 
The data points are multiplied by 2, 4, 8, 16, and 32 for the bins 
$1.6<|y|<2.0$, $1.2<|y|<1.6$, $0.8<|y|<1.2$, $0.4<|y|<0.8$, and $|y|<0.4$, 
respectively.}
\end{figure}

In this section, we describe the final result on the inclusive jet $p_T$ 
cross section measurement applying the corrections defined in the 
previous sections: jet energy scale, efficiencies, and unfolding, used in
order to compute the true number of events observed in each $p_T$ and $y$ bin.
The cross section results are given in Fig.~\ref{fig1} in the six $y$ bins as a 
function of jet $p_T$.   The data points are plotted according 
to the prescription described in~\cite{plotting} and the tabulated data are 
available from Ref.~\cite{table}.

\begin{figure}
\vspace{1.3mm} 
\includegraphics[width=\columnwidth]{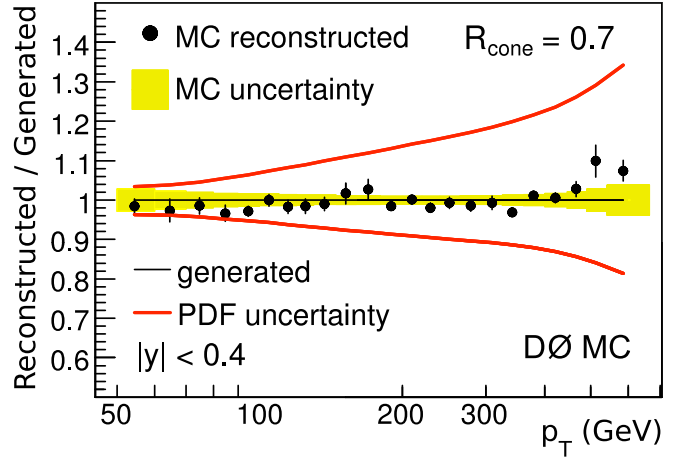}
\caption{\label{closure} (color online)
MC closure test of the method used to extract the inclusive jet 
$p_T$ cross section for the jet $|y|<0.4$ bin. The full analysis 
was repeated treating MC events as data and comparing the result to the input
cross section.
Good agreement is found within the statistical uncertainties of fits to jet 
energy scale and resolution, and unfolding present in MC (shaded band), which 
are much smaller than the systematic uncertainties in data.}
\end{figure}

The method used to extract the cross section is repeated and cross checked 
using a MC simulation of the detector. 
Events are generated using {\sc pythia} and weighted to match the 
NLO prediction calculated using the CTEQ6.5M PDFs
and including nonperturbative corrections.
The MC events are treated in the same way as data, 
all corrections are rederived using MC events, 
and the derived cross section is compared to 
the input cross section to perform a closure test of the measurement. 
The results given in Fig.~\ref{closure} show 
that the method used to extract the 
cross section works well within the statistical uncertainties of the fits to 
the jet response, jet $p_T$ resolution and $p_T$ spectrum. These MC 
uncertainties are significantly smaller than the systematic uncertainties 
present in data.

In Fig.~\ref{fig1} the measurement is compared to the prediction of NLO QCD 
using the CTEQ6.5M PDF parameterization computed using the {\sc{nlojet++}} 
program and {\sc{FastNLO}}. The central CTEQ6.5M prediction uses the 
factorization and renormalization scales $\mu_F = \mu_R = p_T$. The 
alternative scale choices $\mu_F = \mu_R = 0.5p_T$ and $\mu_F = \mu_R = 2p_T$ 
are used to estimate the theoretical uncertainty on the higher order 
corrections. 

\begin{figure*}
\includegraphics[width=0.8\textwidth]{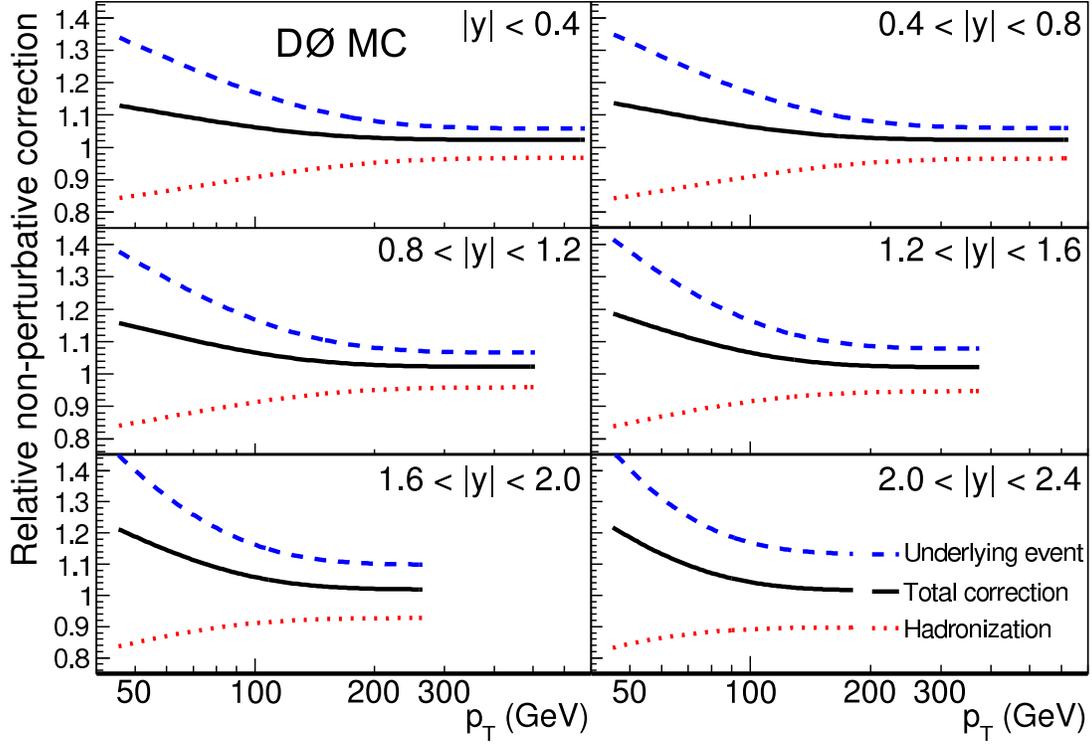}
\caption{\label{nonpert} (color online)
Hadronization (dashed line) and underlying event (dotted line) corrections 
for inclusive jet cross section and the product of both corrections 
(solid line). The 
uncertainty on the theory is estimated as 50\% of the individual corrections 
added in quadrature.}
\end{figure*}

The NLO PQCD prediction is corrected for non-perturbative effects to connect 
the parton level jets predicted by theory to the measured particle level jets. 
The leading non-perturbative corrections are hadronization and underlying 
event which partially cancel. Another small correction is the exclusion 
of muons and neutrinos from the definition of the particle jets. 
The muon/neutrino energy loss is not corrected by the JES procedure
using the MPF method in $\gamma$+jet events. 
The MC corrections 
have been obtained using {\sc{Pythia}} v6.412 with parameters for tune 
QW~\cite{qwtune} obtained by tuning {\sc{Pythia}} to reproduce CDF 
Run II data. The strong coupling constant is fixed to 
$\alpha_s(M_Z)=0.118$ at the $Z$ boson mass and uses the 2-loop formula 
for the $Q^2$ evolution of $\alpha_s$. The {\sc{Pythia}} 
cross section is reweighted in $\hat{s}$ so that the {\sc{Pythia}} 
parton shower prediction agrees with NLO pQCD. The correction factors 
for hadronization and the underlying event are shown in 
Fig.~\ref{nonpert}. As shown in Fig.~\ref{fig1}, 
the measurement is well described by NLO QCD over eight orders of 
magnitude in the six $y$ bins.

\begin{figure*}
\includegraphics[width=1.0\textwidth]{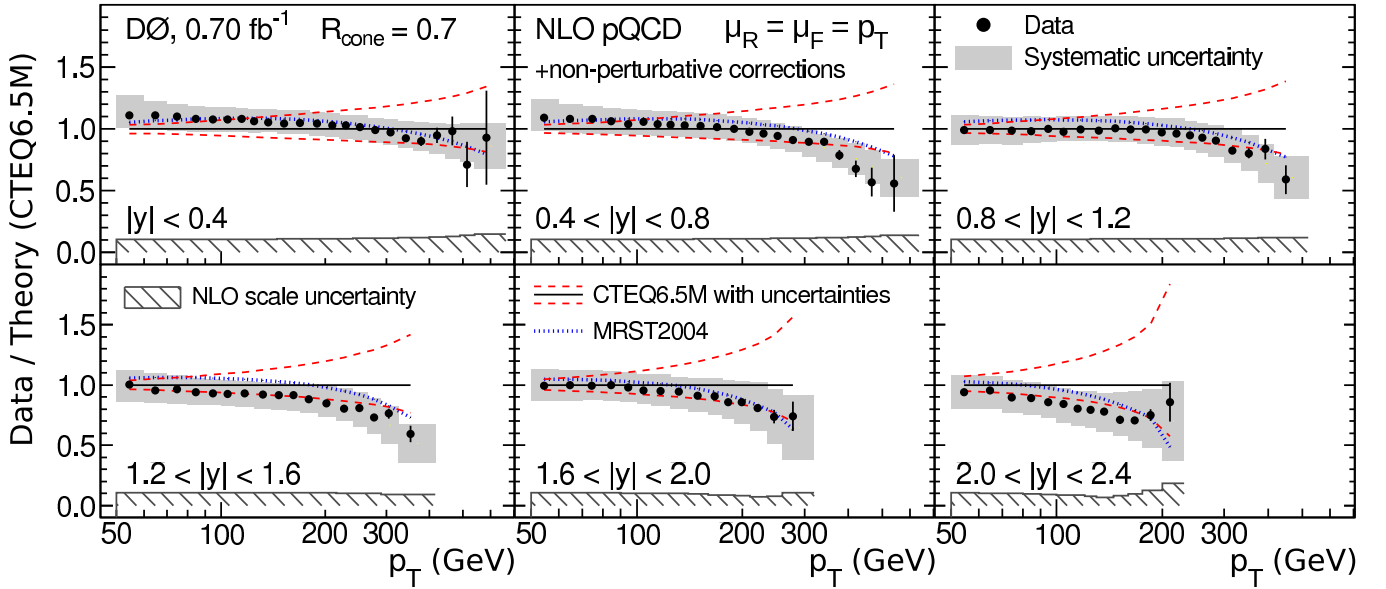}
\caption{\label{fig2} (color online)
Measured data divided by theory for the inclusive jet cross section as a 
function of jet $p_T$ in the six $|y|$ bins. The data systematic 
uncertainties are 
displayed by the shaded band. NLO pQCD calculations, with renormalization 
and factorization scales set to jet $p_T$ using the CTEQ6.5M PDFs and 
including non-perturbative corrections, are compared to the data. The 
CTEQ6.5 PDF uncertainties are shown as dashed lines and the predictions with 
MRST2004 PDFs as dotted lines. The theoretical uncertainty, determined by 
changing the renormalization and factorization scales between $p_T/2$ and 
$2 p_T$, is shown at the bottom of each figure.}
\end{figure*}

To check more precisely how well the measurement is described by the 
NLO QCD theory, we 
display the ratio of data over theory in Fig.~\ref{fig2}, where the theory is 
calculated using the CTEQ6.5M PDF parameterization. 
The PDF uncertainties represented 
as dashed lines are calculated using the set of 20 eigenvectors 
provided by the CTEQ Collaboration for the CTEQ6.5M PDF fits. Data and theory 
agree within experimental and theoretical uncertainties, but data seems to 
favor the lower end of the CTEQ6.5M PDF 90\% confidence level 
uncertainty band. Data are also 
compared to the NLO QCD calculations using the MRST2004 PDF 
parameterization and our agreement in shape is good.  
The experimental uncertainties are smaller than the present PDF uncertainties, 
so these data further constrain the PDFs.

\begin{figure*}
\includegraphics[width=1.0\textwidth]{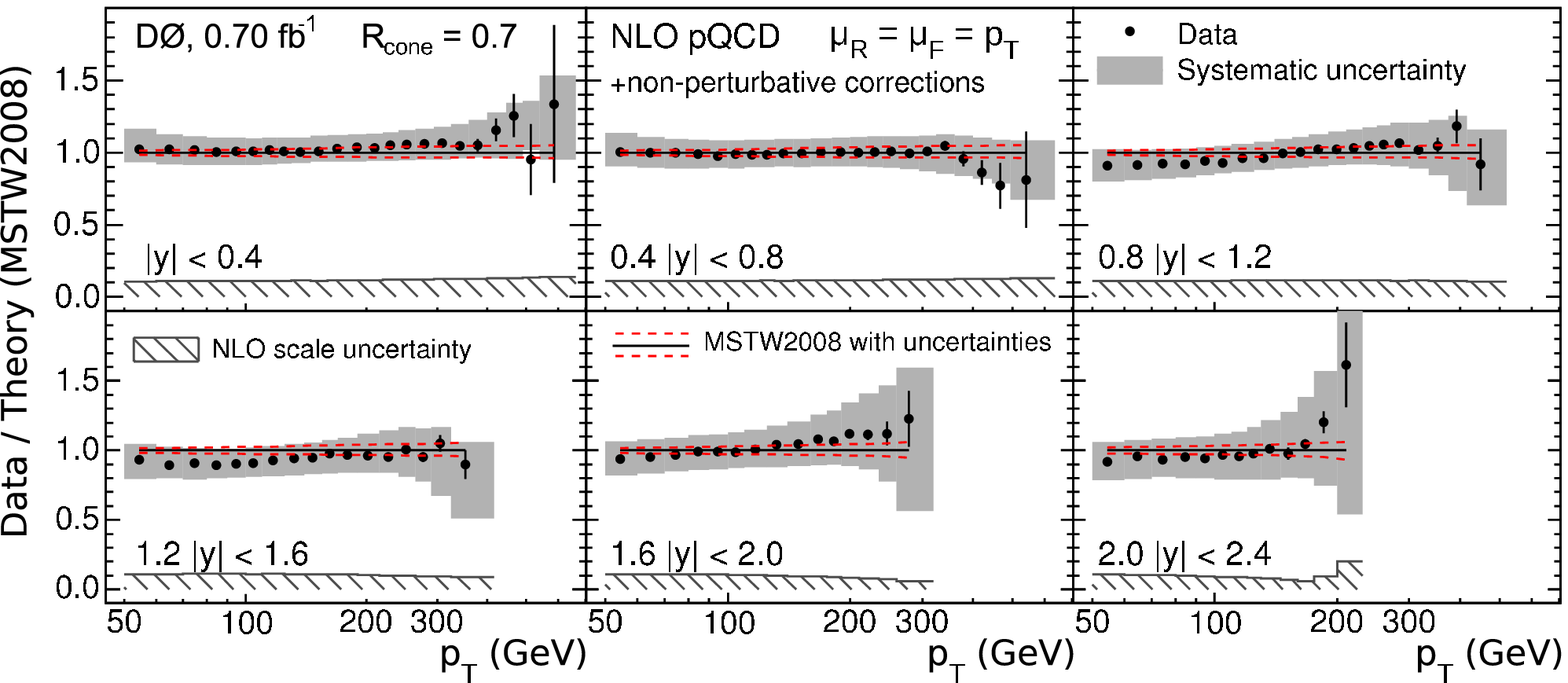}
\caption{\label{fig2b} (color online)
Measured data divided by theory for the inclusive jet cross section as a 
function of jet $p_T$ in the six $|y|$ bins using the MSTW2008 
parameterization. The data systematic uncertainties are 
displayed by the shaded band. The theoretical uncertainty, determined by 
changing the renormalization and factorization scales between $p_T/2$ and 
$2 p_T$, is shown at the bottom of each figure.}
\end{figure*}

\begin{figure*}
\includegraphics[width=1.0\textwidth]{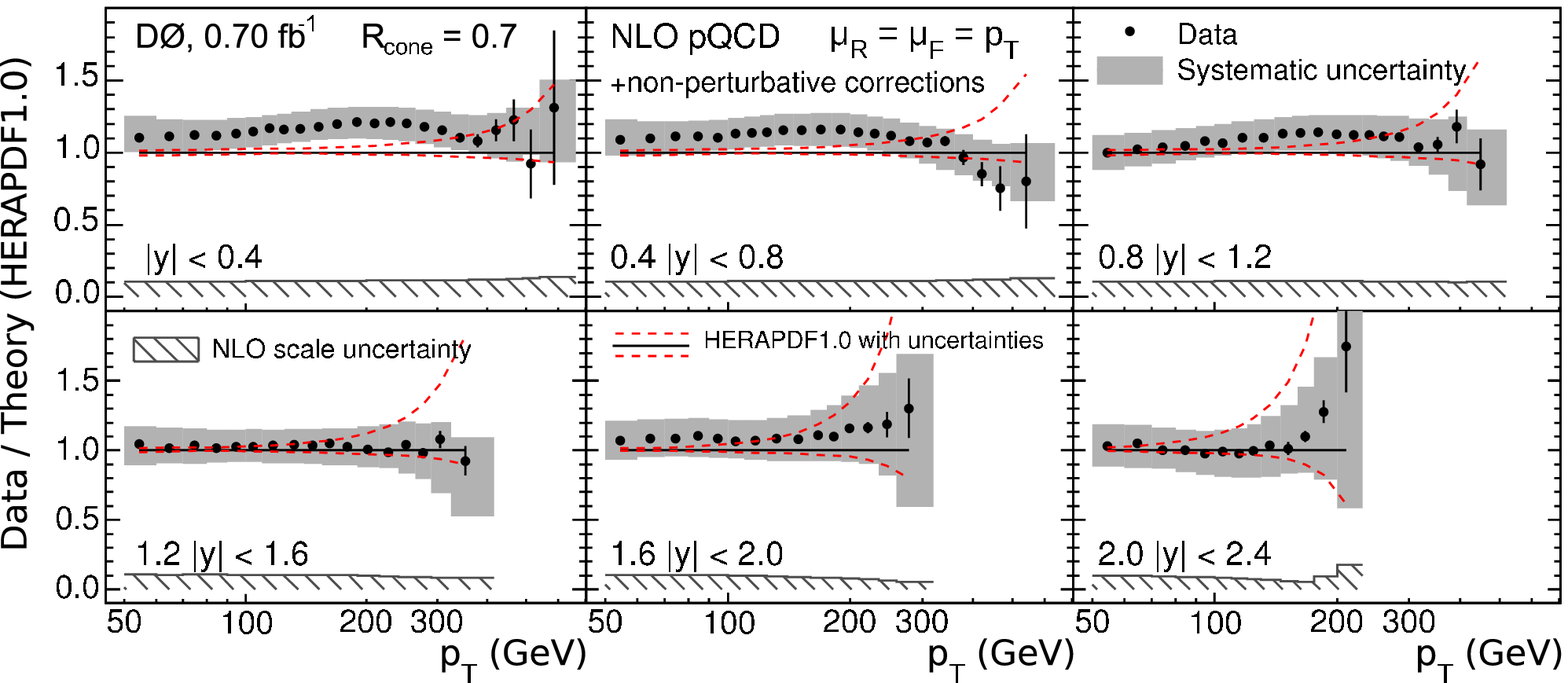}
\caption{\label{fig2c} (color online)
Measured data divided by theory for the inclusive jet cross section as a 
function of jet $p_T$ in the six $|y|$ bins using the HERAPDF1.0 
parameterization. The data systematic uncertainties are 
displayed by the shaded band. The theoretical uncertainty, determined by 
changing the renormalization and factorization scales between $p_T/2$ and 
$2 p_T$, is shown at the bottom of each figure.}
\end{figure*}

\begin{figure*}
\includegraphics[width=1.0\textwidth]{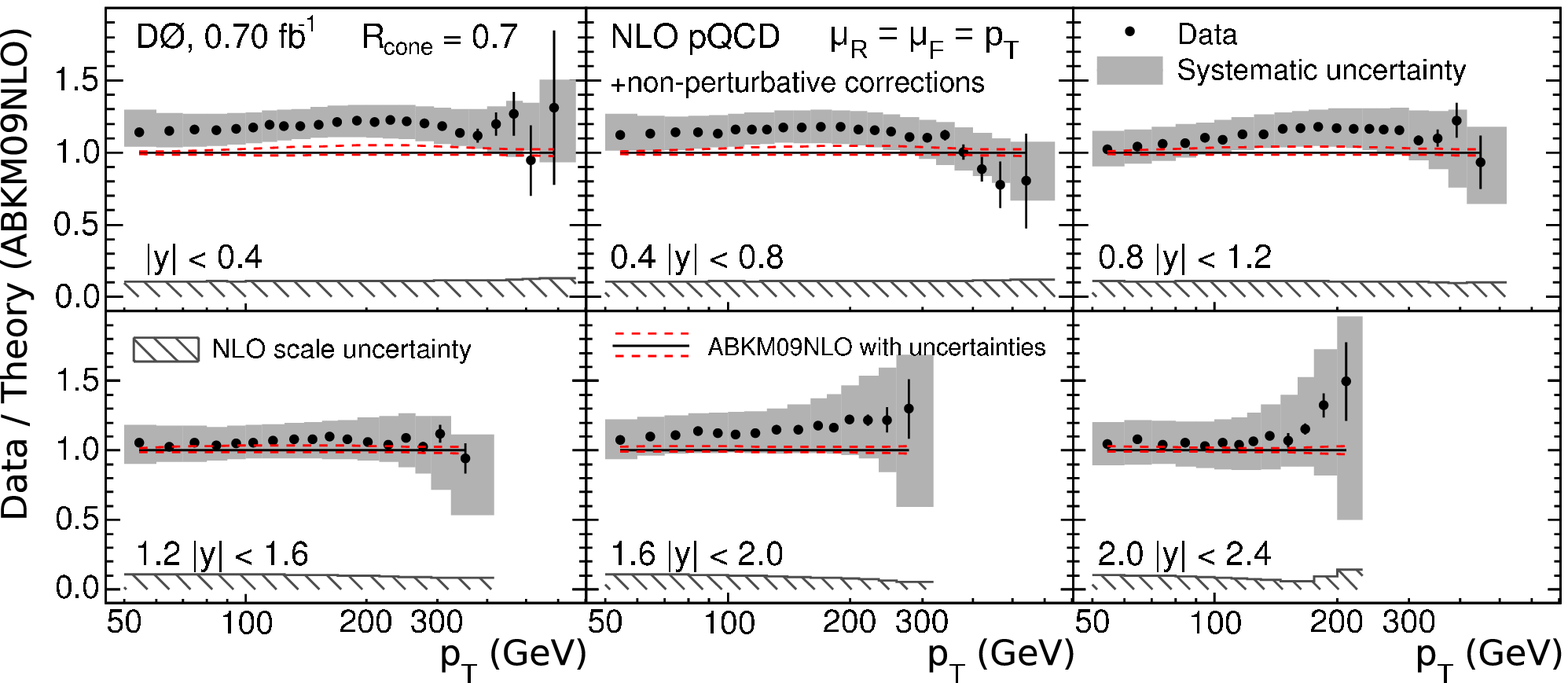}
\caption{\label{fig42} (color online)
Measured data divided by theory for the inclusive jet cross section as a 
function of jet $p_T$ in the six $|y|$ bins using the ABKM09 parameterization. 
The data systematic uncertainties are 
displayed by the shaded band. The theoretical uncertainty, determined by 
changing the renormalization and factorization scales between $p_T/2$ and 
$2 p_T$, is shown at the bottom of each figure.}
\end{figure*}

\begin{figure*}
\includegraphics[width=1.0\textwidth]{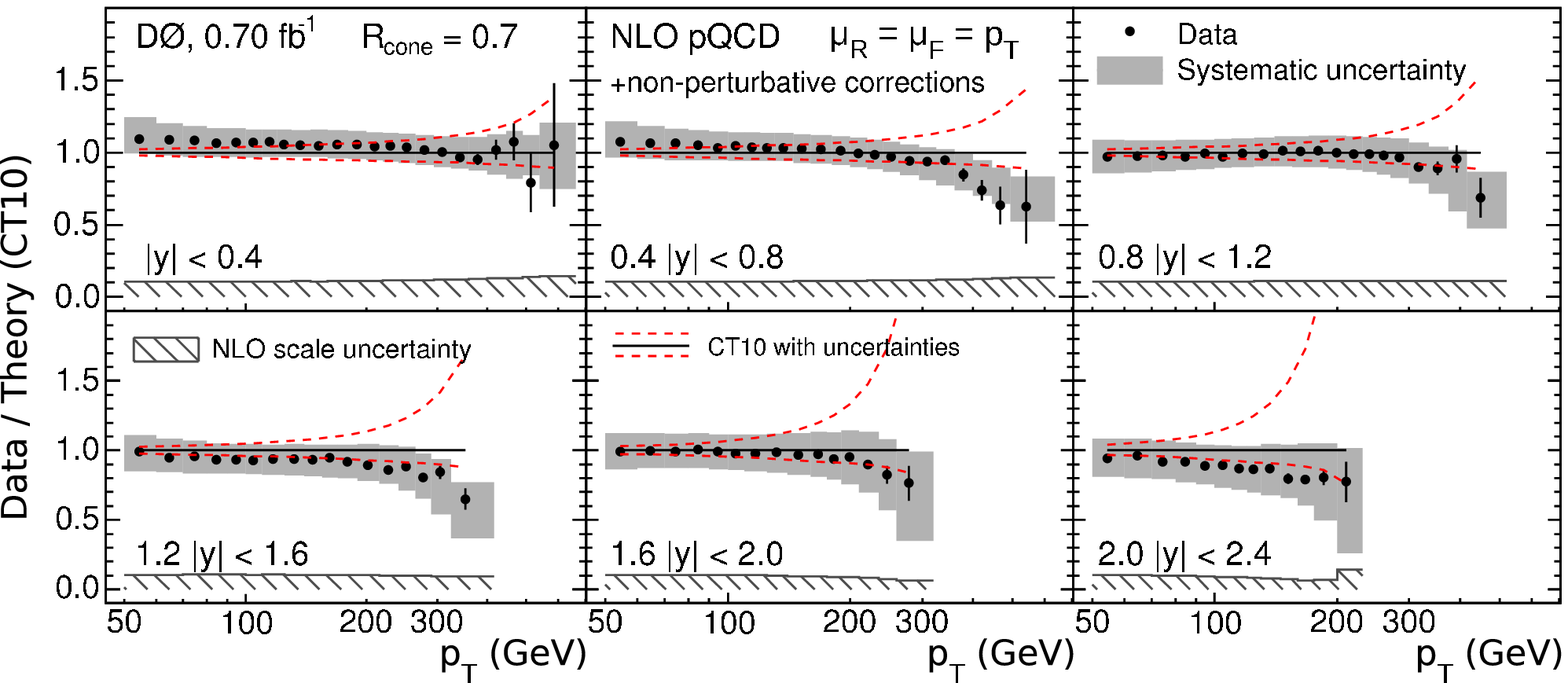}
\caption{\label{fig43} (color online)
Measured data divided by theory for the inclusive jet cross section as a 
function of jet $p_T$ in the six $|y|$ bins using the CT10 parameterization. 
The data systematic uncertainties are 
displayed by the shaded band. The theoretical uncertainty, determined by 
changing the renormalization and factorization scales between $p_T/2$ and 
$2 p_T$, is shown at the bottom of each figure.}
\end{figure*}

\begin{figure*}
\includegraphics[width=1.0\textwidth]{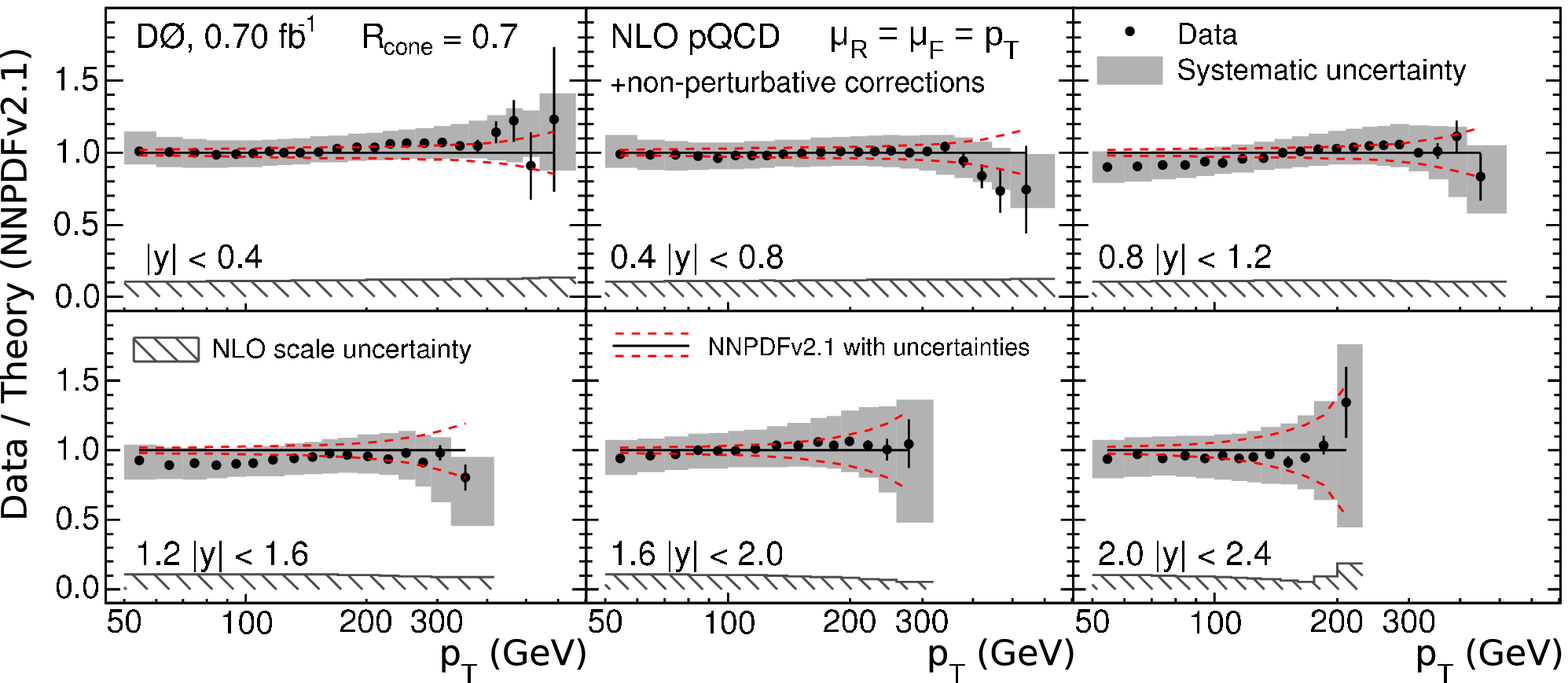}
\caption{\label{fig44} (color online)
Measured data divided by theory for the inclusive jet cross section as a 
function of jet $p_T$ in the six $|y|$ bins using the NNPDFv2.1 
parameterization. The data systematic uncertainties are 
displayed by the shaded band. The theoretical uncertainty, determined by 
changing the renormalization and factorization scales between $p_T/2$ and 
$2 p_T$, is shown at the bottom of each figure.}
\end{figure*}

Some recent parameterizations have already used our measured jet cross sections 
described here to further constrain
the PDFs. As an example, we display in Fig.~\ref{fig2b} the ratio data over
theory, where the NLO theory is calculated using the MSTW08 NLO 
PDF~\cite{mstw08} which displays
good agreement between our measurement and this parameterization, with a
tendency to be slightly different at high jet $p_T$ where the uncertainties are
larger. For reference, we also display in Fig.~\ref{fig2c} the ratio of data 
over theory where the theory uses the recent HERAPDFv1.0 PDF~\cite{hera01}, 
which uses only HERA data to constrain PDFs. We notice some discrepancies 
between our measurement and the HERAPDFv1.0 PDF at medium jet $p_T$ especially 
in the central region, and at high $p_T$ in the forward region. We also 
compare our data with the ABKM09NLO~\cite{abkm} parameterization in 
Fig.~\ref{fig42} and we notice some disagreement between
our data and the predictions in particular on the
normalization. This shows the capability of our data to constrain 
further the PDFs.  Furthermore, we compare our measurements to
the recent CT10~\cite{ct10} parameterization in Fig.~\ref{fig43}. There is 
a good agreement with data with the tendency of the CT10 parameterization 
to be higher at large $p_T$ in all $|y|$ bins. Finally, we compare our 
measurement with the predictions from the NNPDFv2.1~\cite{nnpdf} 
parameterization in Fig.~\ref{fig44} and again good agreement is found 
with our data.

\begin{figure*}
\begin{overpic}[width=0.43\textwidth]
{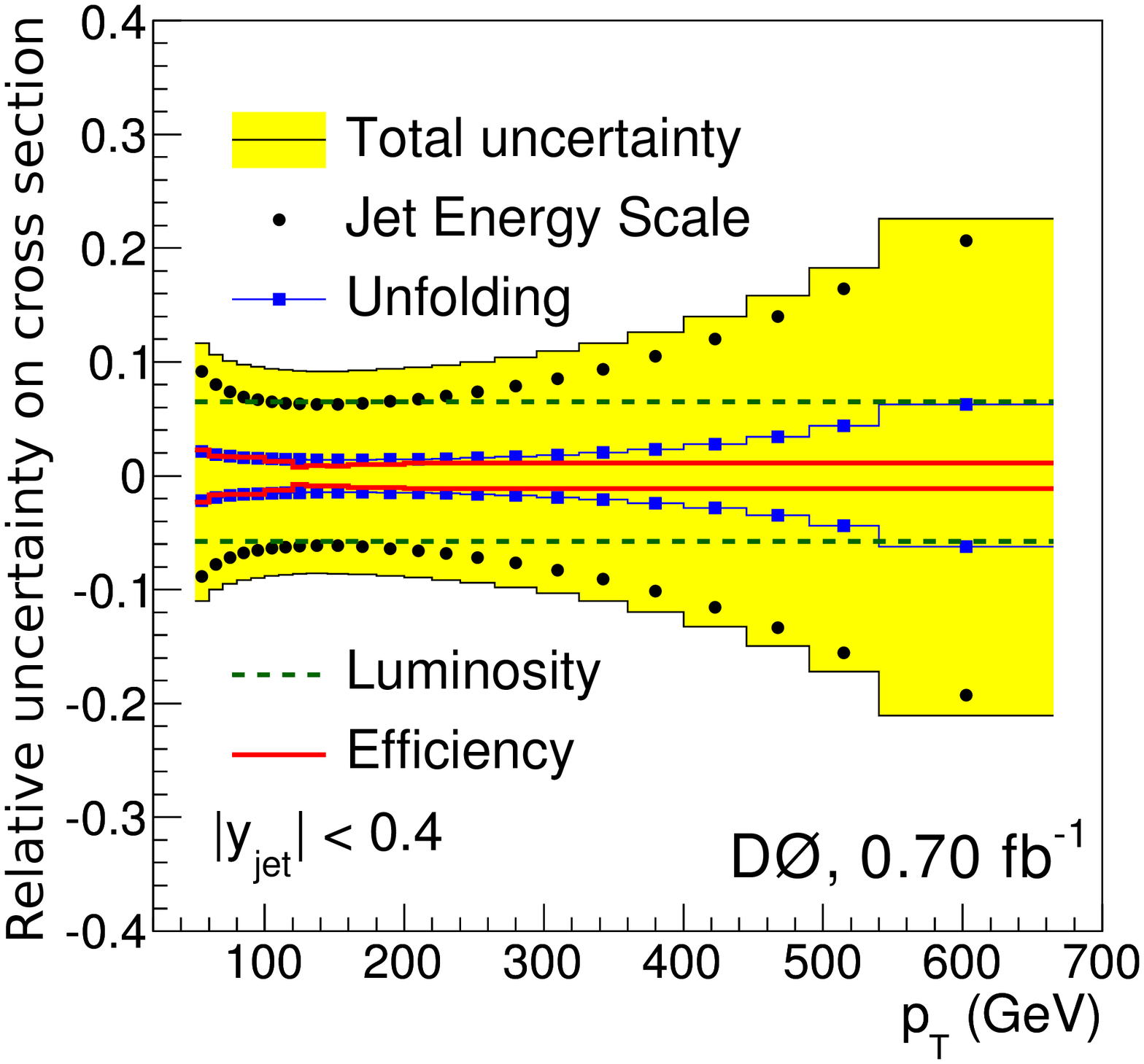}
\put(185,175){\textsf{(a)}}
\end{overpic}
\begin{overpic}[width=0.43\textwidth]
{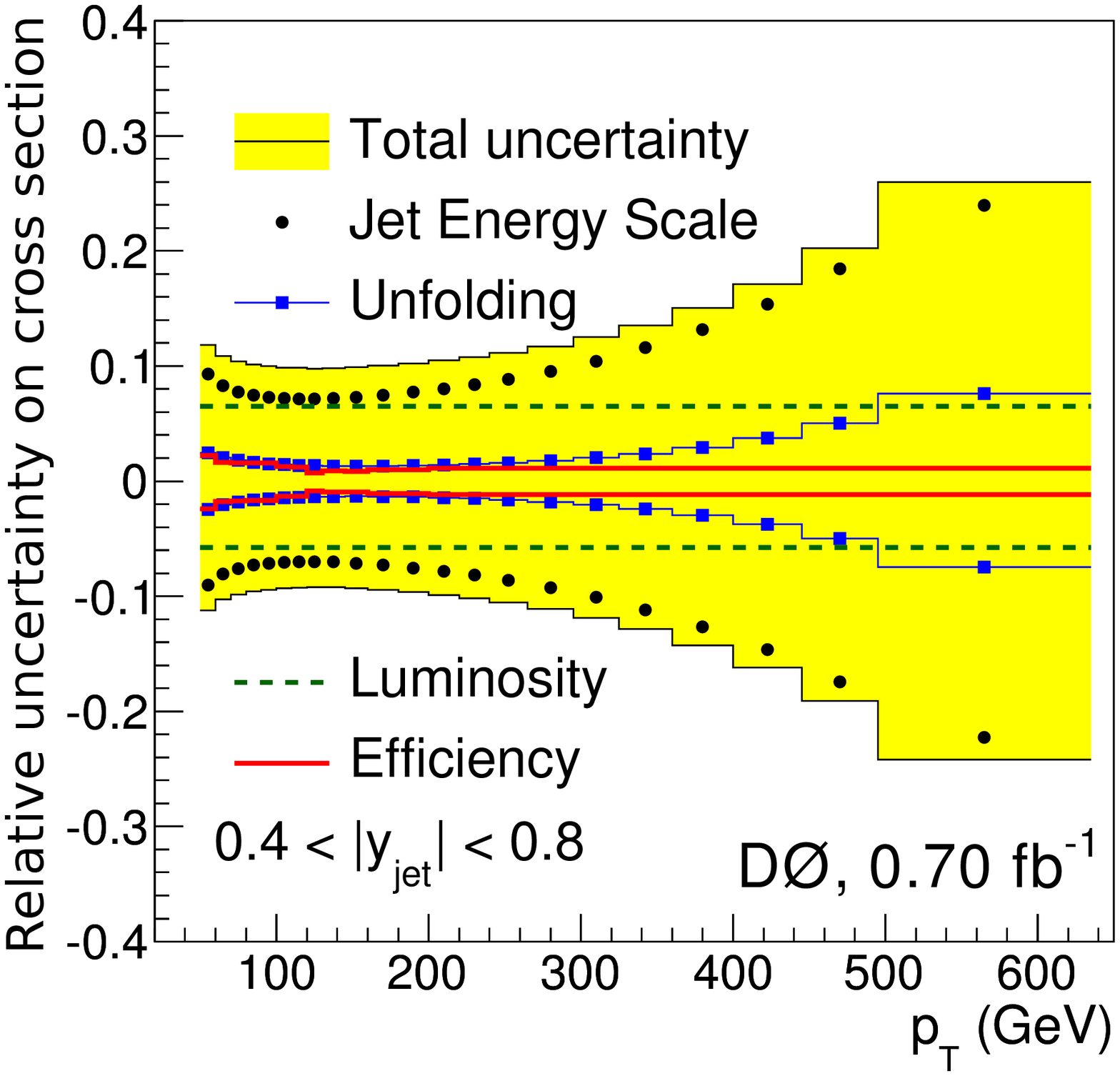}
\put(185,175){\textsf{(b)}}
\end{overpic}
\begin{overpic}[width=0.43\textwidth]
{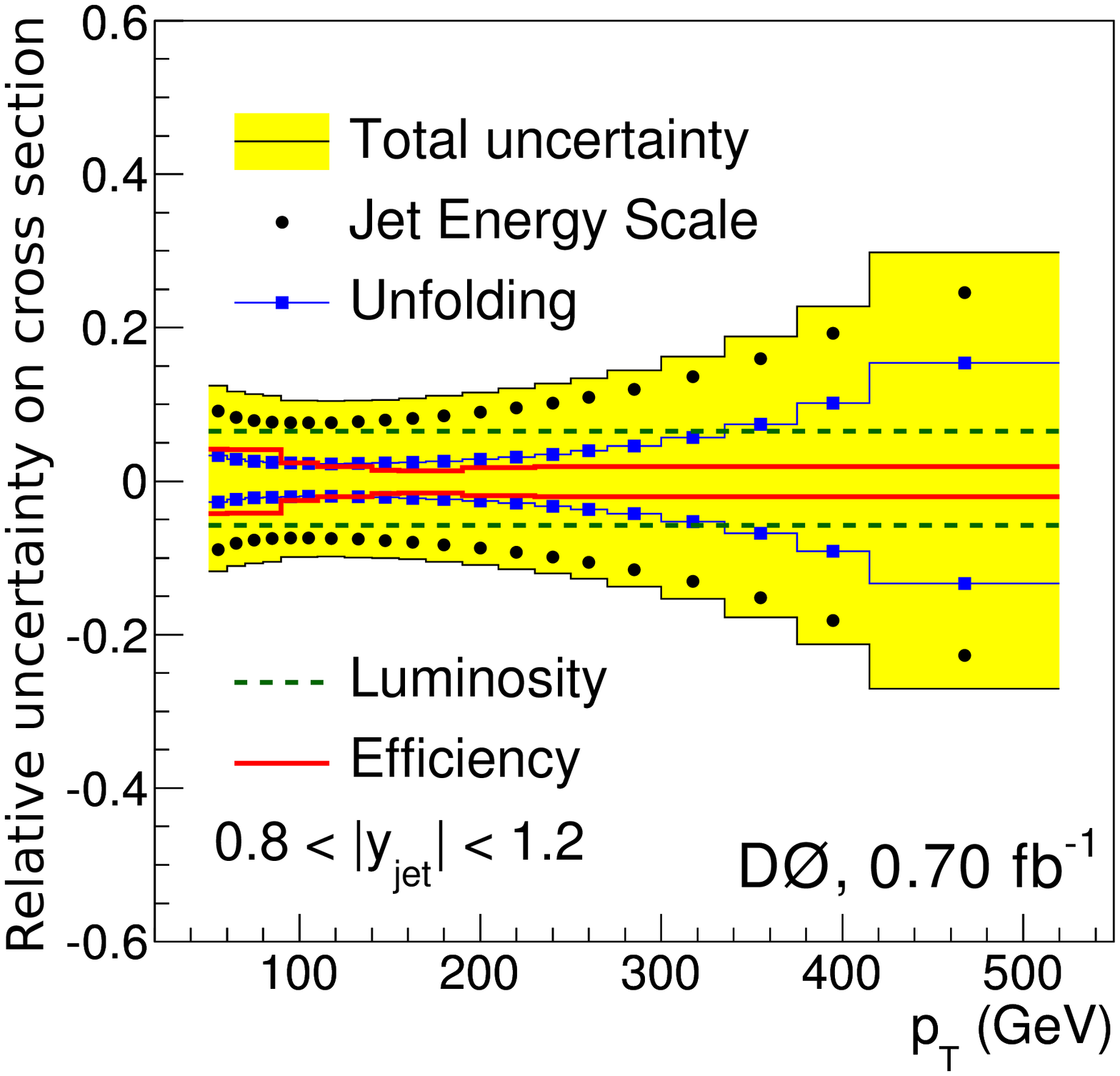}
\put(185,175){\textsf{(c)}}
\end{overpic}
\begin{overpic}[width=0.43\textwidth]
{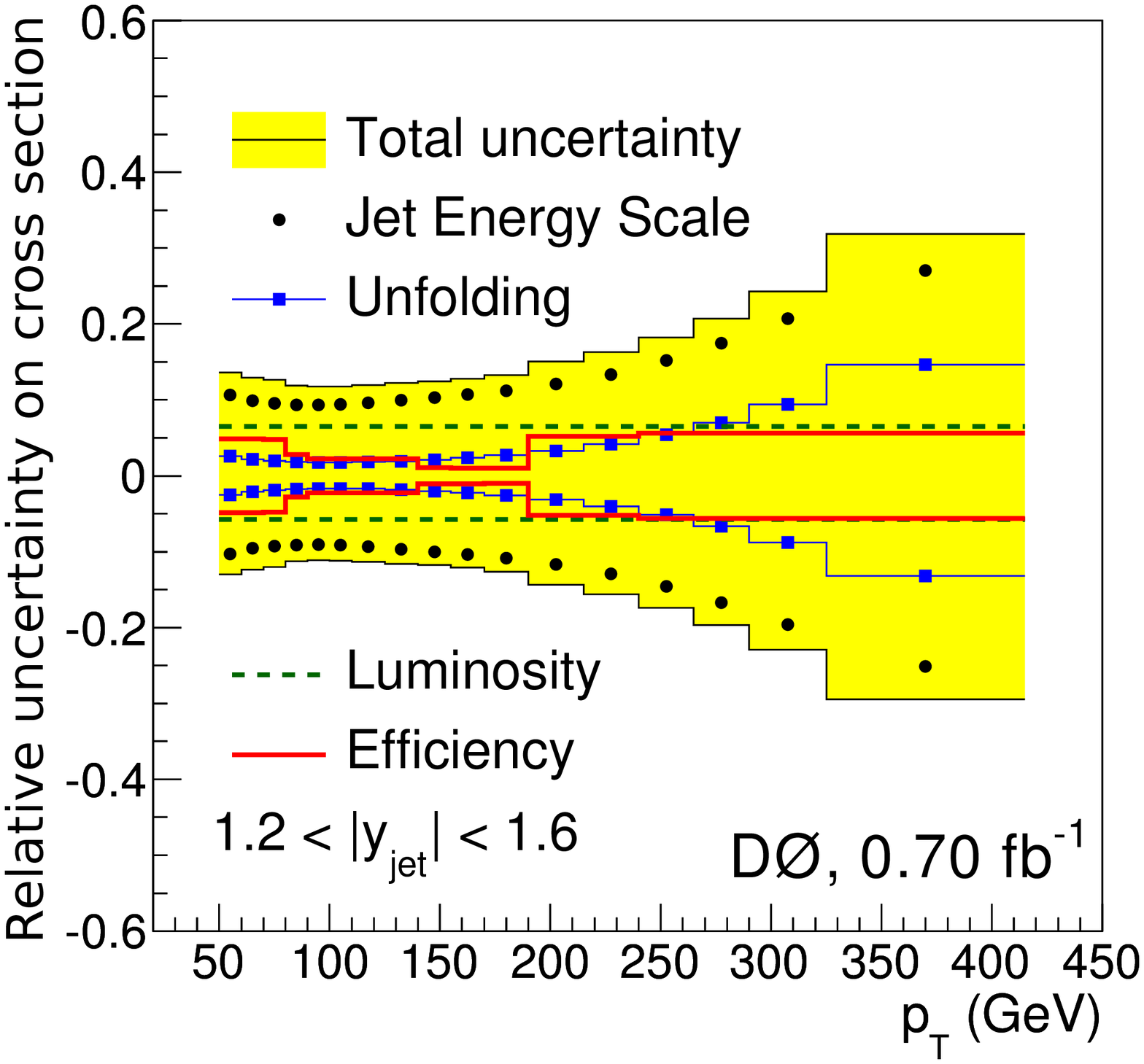}
\put(185,175){\textsf{(d)}}
\end{overpic}
\begin{overpic}[width=0.43\textwidth]
{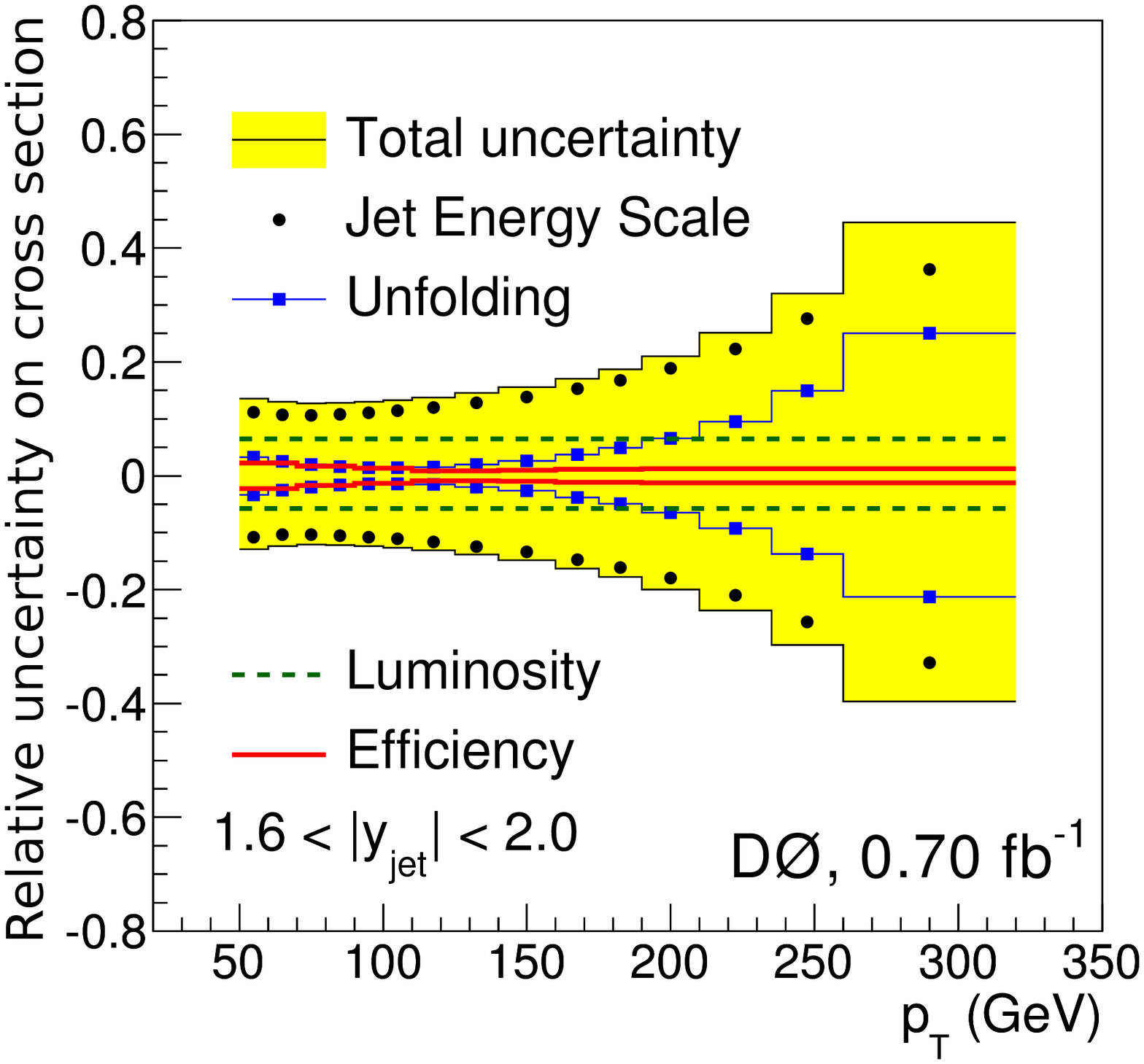}
\put(185,175){\textsf{(e)}}
\end{overpic}
\begin{overpic}[width=0.43\textwidth]
{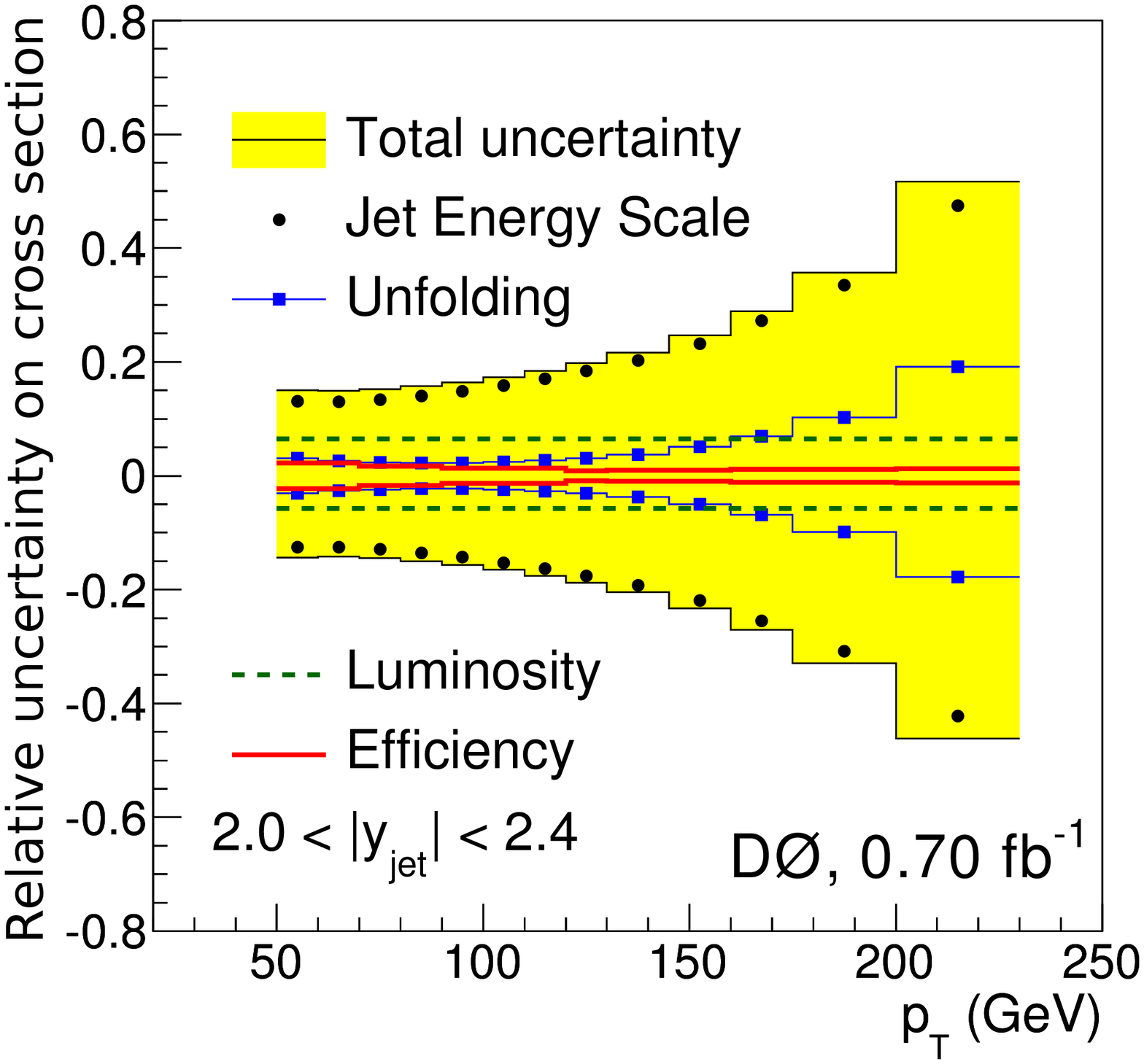}
\put(185,175){\textsf{(f)}}
\end{overpic}
\caption{\label{uncertainties}(color online)
Different components of the systematic 
uncertainty as a function of jet $p_T$ in the six $y$ bins.}
\end{figure*}

The details of the uncertainties on the inclusive jet $p_T$ cross section are 
given in Fig.~\ref{uncertainties}. The dominant uncertainty is due to the 
systematic uncertainties on the jet energy scale, but the unfolding and the 
uncertainties related to the resolution in jet $p_T$ are also important, 
especially at high $p_T$ and high $|y|$. The 6.1\% luminosity uncertainty is 
the second largest uncertainty at low $p_T$ and the third largest at high 
$p_T$, and leads to significant uncertainty in the overall normalization 
of the cross section. For a jet $p_T ~\approx 150$~GeV, it is similar to 
the jet energy scale uncertainty. The uncertainties related to 
efficiencies are small everywhere.

\section{Uncertainty correlations}

Correlations between systematic uncertainties are studied in detail to 
increase the value of these data in future fits to model parton
distributions and their impact on LHC physics predictions in particular.  
In total, there are 91 independent sources of systematic uncertainty, 
and in this section we describe the method we use to group those 
with similar impact on the shape of the cross section to
find the principal components of the uncertainty without significantly 
impacting the overall quality of the data. Many of the systematic sources 
we describe above are small in magnitude and highly correlated in shape 
with other sources.

The traditional interpretation of uncertainties to be independent 
requires that at each point the sum of all sources in quadrature must equal 
the total systematic uncertainty. In practice, adding in quadrature sources 
with similar shapes whose orthogonal components (defined later) are small will 
lose very little information compared to the full information given in 
the 91 different systematic uncertainties.

We combine uncertainties that are
correlated and of similar shape to reduce the number of components
in the covariance matrix. 
We develop a robust systematic approach for regrouping the sources based on
the notions of source size, shape 
similarity, and orthogonality. The natural measure for the size of 
a source is the impact it has on the overall $\chi^2$ in the fit with the
ansatz function when shifted by 
one standard deviation around the minimum. To assess the similarity in shape
between different systematic uncertainties, we
define the inner product for sources $h$ and $g$ as
\begin{equation}
\label{eq:innerproduct}
\left<h\cdot g\right> = \sum_{i\in\mathrm{bins}} \frac{h_i\cdot g_i}
{\sigma_{\mathrm{stat},i}^2},
\end{equation}
where $h_i$ and $g_i$ are the values of two systematic uncertainties and
the sum is over the $p_T$ and $|y|$ bins.
The size, or magnitude, of a source $h$ can be written using this notation as
\begin{equation}
||h|| = \sqrt{\left<h\cdot h\right>}.
\end{equation}
The shape similarity of two sources $h$ and $g$ can be quantified by 
calculating their correlation, which is written in the notation of 
Eq.~\ref{eq:innerproduct} as
\begin{equation}
\rho = \frac{\left<h\cdot g\right>}{||h||\cdot||g||},
\end{equation}
which varies between -1.0 and 1.0. When $\rho=$1.0,
the sources are fully correlated, $-1.0$ fully 
anti-correlated and 0.0 completely uncorrelated. The source $g$ can 
be broken into a 
component that is fully correlated with source $h$ and another component that 
is fully uncorrelated by considering a linear transformation
\begin{equation}
\label{eq:lintransform_g}
g' = g - \alpha h.
\end{equation}
When the orthogonality of $h$ and $g'$ is defined in terms of the inner 
product,
\begin{equation}
\label{eq:ortho}
h \perp g' \equiv \left<h\cdot g'\right> = 0,
\end{equation}
Eq.~\ref{eq:lintransform_g} and Eq.~\ref{eq:ortho} together yield
\begin{equation}
\alpha = \frac{\left<h\cdot g\right>}{\left<h\cdot h\right>},
\end{equation}
defining $g'$ as the orthogonal component that is fully uncorrelated with 
source $h$. The value $g'$ has the property $\left<g'\cdot g'\right> 
\le \left<g\cdot g\right>$, $\left<g'\cdot g'\right> = 
\left<g\cdot g\right> $ is equivalent to $h$ being orthogonal to $g$, and $
\left<g'\cdot g'\right>=0$ to $h$ being parallel to $g$. Small values of $||g'||$ 
indicate that the sources can be combined with little impact on the freedom of 
the fit to the ansatz.

The sources due to statistical uncertainties in fits are 
first assigned as uncorrelated. The remaining 
sources are sorted by size and are then iteratively recombined with 
other sources most similar in shape and having the smallest 
orthogonal components. The sources are combined when their correlation 
is greater than about 85\% and the orthogonal components have a 
magnitude smaller than 10\% of the 
largest individual systematic $\epsilon_\mathrm{max}$. At the end of the 
iterative procedure, the remaining set of sources no longer has any pairings 
with an orthogonal component less than $0.1\epsilon_\mathrm{max}$.  
The smallest remaining sources with magnitude less than 
$0.1\epsilon_\mathrm{max}$ are added in quadrature to the uncorrelated 
uncertainty. The final reduced set of uncertainties has 23 correlated 
sources (principal 
components) and one fully uncorrelated uncertainty, which is a significant 
reduction compared to the original 91 sources. The reduced set of 23 
correlated sources and the total uncorrelated uncertainty are provided in 
Ref.~\cite{table}.

\begin{figure*}
\includegraphics[width=0.49\textwidth]{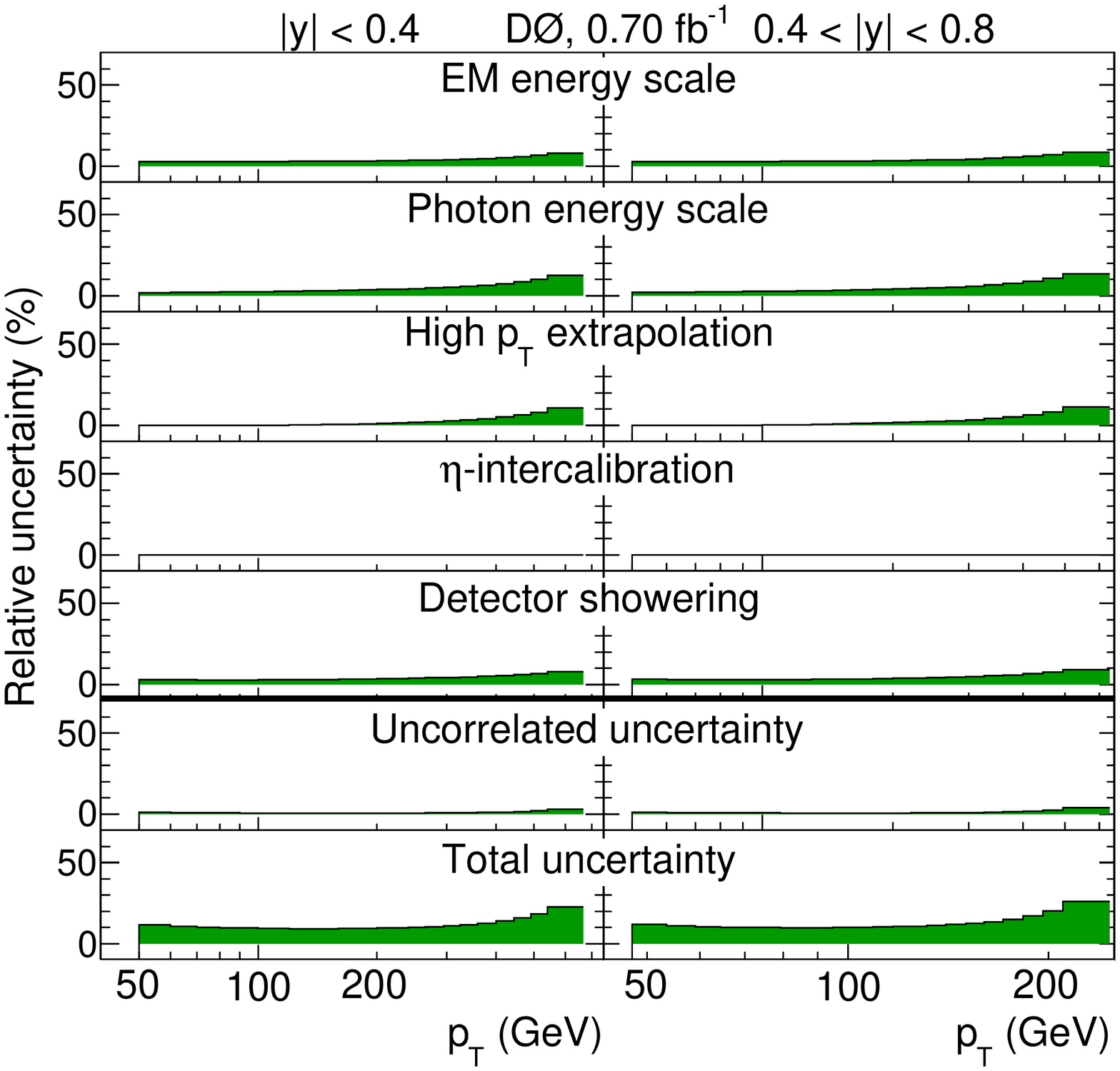}
\includegraphics[width=0.49\textwidth]{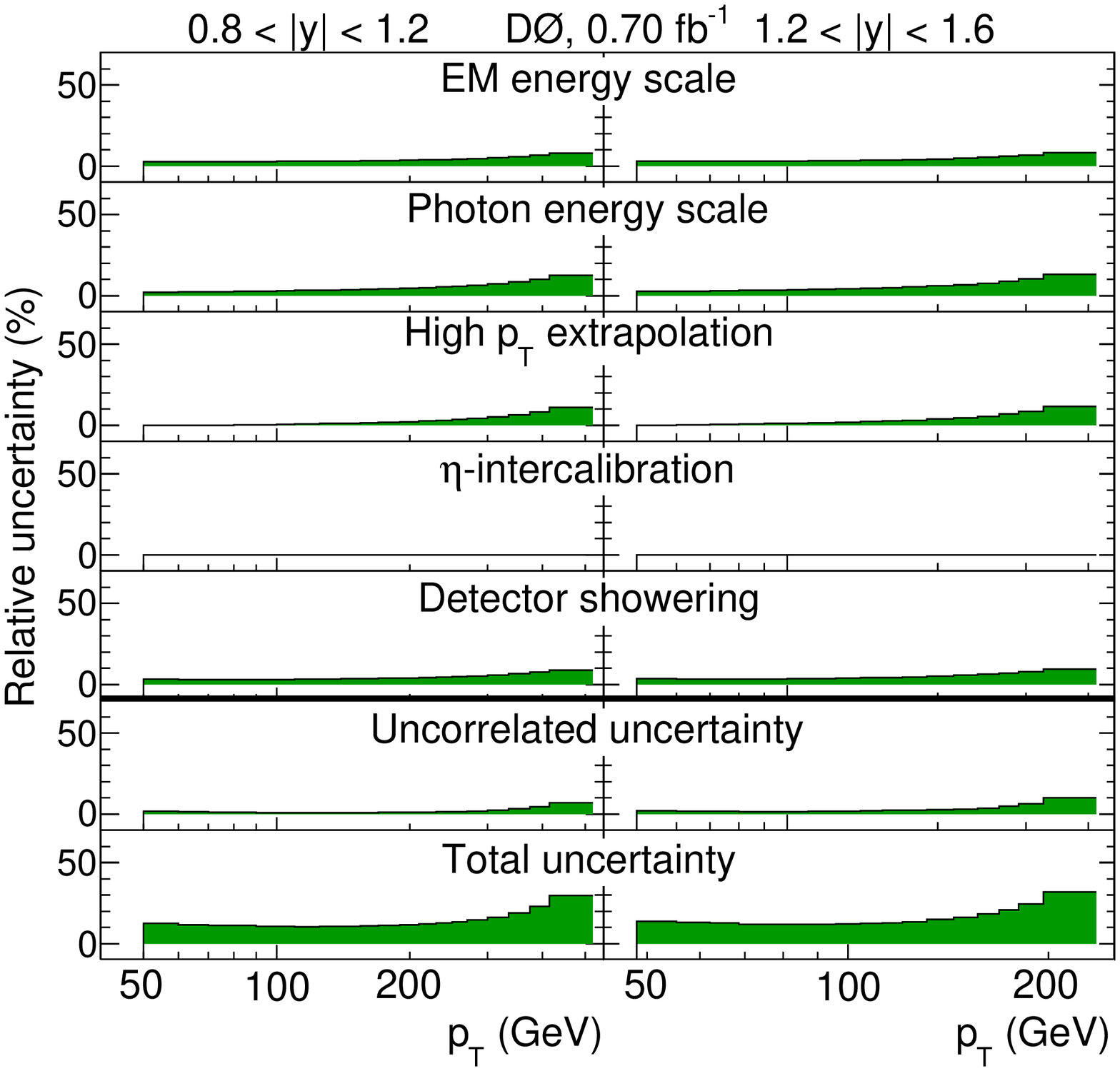}
\caption{\label{source01} (color online)
Correlated uncertainties for all central regions and the ICR as a 
function of jet $p_T$ for four $|y|$ bins, $|y|<0.4$, $0.4<|y|<0.8$, 
$0.8<|y|<1.2$, and $1.2<|y|<1.6$ . The five largest systematic 
uncertainties are shown together with uncorrelated and total 
uncertainties, computed as the sum in quadrature of all 
sources.}
\end{figure*}

\begin{figure}
\includegraphics[width=\columnwidth]{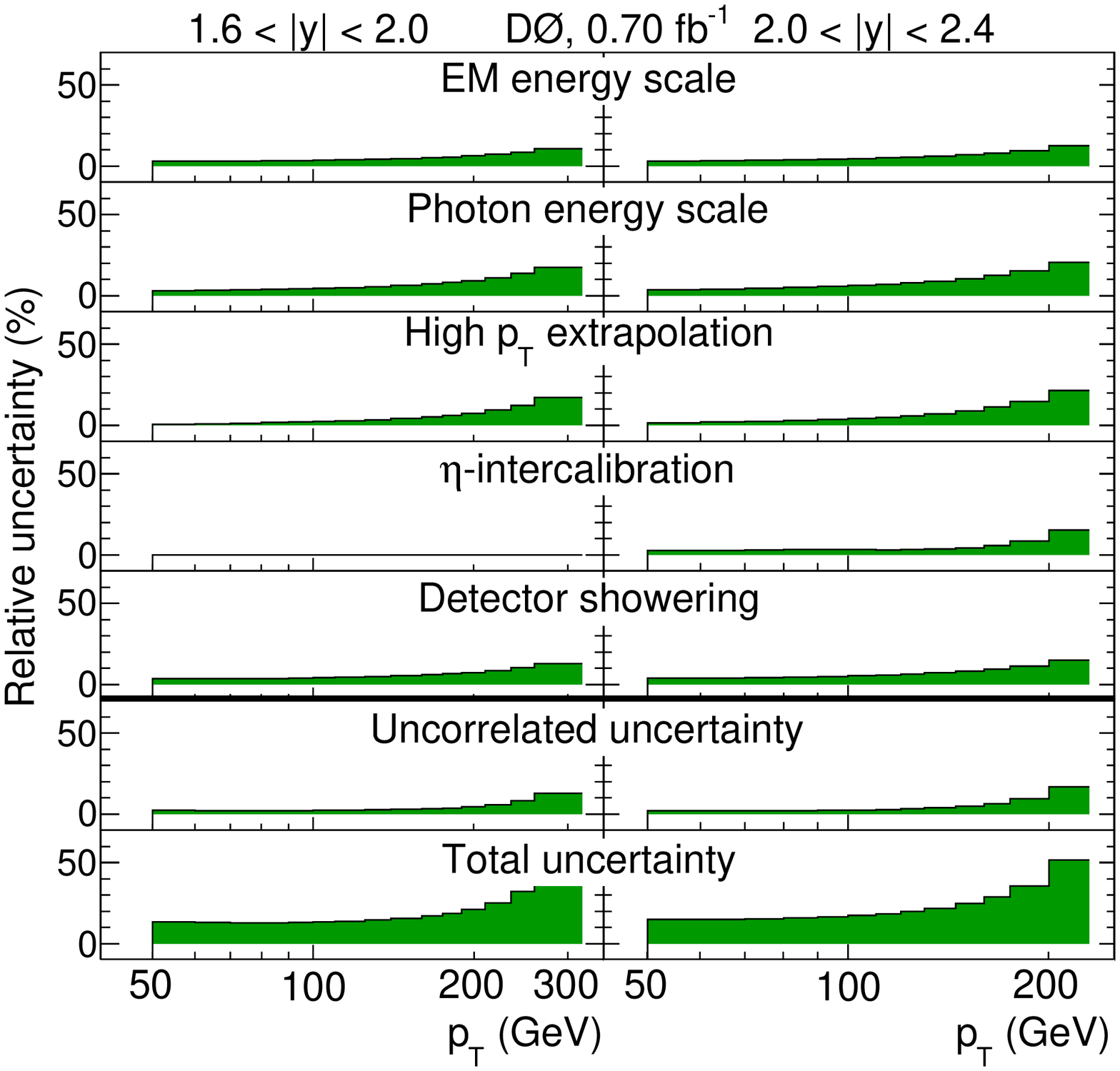}
\caption{\label{source45} (color online)
Correlated uncertainties for $1.6<|y|<2.0$ and $2.0<|y|<2.4$ as a function of 
jet $p_T$. The five largest systematic uncertainties are shown together with 
uncorrelated and total uncertainties, computed as the sum in quadrature of all 
sources.}
\end{figure}

The five leading sources from the reduced set of combined systematic
uncertainties, the 
total uncorrelated uncertainty, and the total uncertainty are shown in 
Figs.~\ref{source01} and \ref{source45} in the six $|y|$ bins. These sources 
summarize the leading systematic uncertainties for the measurement. 
The EM scale 
uncertainty comes from the calibration of the EM calorimeter using 
$Z \rightarrow e^+e^-$ events.
The photon energy scale includes the uncertainty in the MC description of the 
difference in the electron and photon responses and the uncertainty in the 
amount of passive material in front of the calorimeter, which affects the 
response difference as a function of photon $p_T$. 
The uncertainty in the high $p_T$ extrapolation is due to
differences in fragmentation models of {\sc{Pythia}} and {\sc{Herwig}}, which 
lead to an additional uncertainty in the high $p_T$ extrapolation of the 
central response. The rapidity-intercalibration uncertainty summarizes the 
uncertainty in the relative response calibration between calorimeter regions. 
The detector showering uncertainty includes the uncertainties on showering, 
but also additional significant contributions from other uncertainties such as 
sample purity and the difference between alternate tunes of 
{\sc{Pythia}} (tunes A and QW).

\section{Conclusion}\label{sec:conclusions}

In this paper, we described the measurement of the 
inclusive jet cross section by the D0 experiment. The measured inclusive 
jet cross section corrected for experimental effects to the particle level in 
$p\bar{p}$ collisions at $\sqrt{s}=$ 1.96 TeV with ${\cal L}=0.70$~fb$^{-1}$ 
was presented for six $|y|$ bins as a function of jet $p_T$. The precision 
reached in this measurement is unprecedented for results from a hadron collider,
particularly for processes dependent on gluons at high-$x$. The measurement 
was found to be in good agreement with 
NLO QCD calculations with CTEQ6.5M and MRST2004 PDFs. 
These results will also be 
useful for any experiment at a hadron collider such as the LHC where the same 
techniques can be used to extract the jet energy scale with high precision and 
to measure the inclusive jet cross section. In addition, a full analysis of 
correlations between sources of systematic uncertainty was performed, 
demonstrating a useful method to reduce the complexities
of describing numerous sources of uncertainties in the cross section, and
increasing the potential impact of these data in global PDF fits.

%
We thank the staffs at Fermilab and collaborating institutions,
and acknowledge support from the
DOE and NSF (USA);
CEA and CNRS/IN2P3 (France);
FASI, Rosatom and RFBR (Russia);
CNPq, FAPERJ, FAPESP and FUNDUNESP (Brazil);
DAE and DST (India);
Colciencias (Colombia);
CONACyT (Mexico);
KRF and KOSEF (Korea);
CONICET and UBACyT (Argentina);
FOM (The Netherlands);
STFC and the Royal Society (United Kingdom);
MSMT and GACR (Czech Republic);
CRC Program and NSERC (Canada);
BMBF and DFG (Germany);
SFI (Ireland);
The Swedish Research Council (Sweden);
and
CAS and CNSF (China).

\end{document}